\documentclass[preprint,12pt,authoryear]{elsarticle}

\usepackage{geometry}
\usepackage{hyperref}       
\usepackage{url}            
\usepackage{booktabs}       
\usepackage{amsfonts}       
\usepackage{nicefrac}       
\usepackage{microtype}      
\usepackage{lipsum}
\usepackage{fancyhdr}       
\usepackage{graphicx}       
\graphicspath{{./frames/}}
\usepackage{amsmath}  
\usepackage{multirow}
\usepackage{bm}
\usepackage{subfigure}
\usepackage{subcaption}
\usepackage{float}
\journal{Computational Statistics and Data Analysis}

\begin{document}

\begin{frontmatter}



\title{Expectile Periodogram} 


\author{Tianbo Chen$^a$}
\author{Ta-Hsin Li$^b$}
\author{Hanbing Zhu$^a$}
\author{Wenwu Gao$^a$ }\fntext[myfootnote]{Corresponding author. Email: wenwugao528@163.com.}

\affiliation{organization={School of Big Data and Statistics, Anhui University},
            country={China}}
\affiliation{organization={IBM Watson Research Center},
	country={US}}
\begin{abstract}
	This paper introduces a novel periodogram-like function, called the expectile periodogram (EP), for modeling spectral features of time series and detecting hidden periodicities. The EP is constructed from trigonometric expectile regression (ER), in which a specially designed loss function is used to substitute the squared $\ell_{2}$ norm that leads to the ordinary periodogram. The EP retains the key properties of the ordinary periodogram as a frequency-domain representation of serial dependence in time series, while offering a more comprehensive understanding by examining the data across the entire range of expectile levels. The asymptotic theory is established to investigate the relationship between the EP and the so-called expectile spectrum. Simulations demonstrate the efficiency of the EP in the presence of hidden periodicities. In addition, by leveraging the inherent two-dimensional nature of the EP, we train a deep learning model to classify earthquake waveform data. Notably, our approach outperforms alternative periodogram-based methods in terms of classification accuracy.
\end{abstract}


\end{frontmatter}

\section{Introduction}
Spectral analysis plays a crucial role in time series analysis, where data are analyzed in the frequency domain. The ordinary periodogram (PG), a raw non-parametric estimator of the power spectrum, is widely applied across various fields  \citep{caiado2006periodogram,polat2007classification,baud2018multi,euan2018hierarchical,maadooliat2018nonparametric,martinez2019periodogram,chen2021clustering}. The PG is constructed by the ordinary least squares (OLS) regression on trigonometric regressors. However, OLS exhibits limitations in terms of robustness and effectiveness, particularly in handling data with asymmetric or heavy-tailed distributions. Moreover, OLS regression focuses primarily on the conditional mean, lacking the ability to provide a comprehensive view of the conditional distribution of the data, which subsequently affects the performance of the PG \citep{bloomfield2004fourier}.

An alternative regression approach is quantile regression (QR), which evaluates the variation of conditional quantiles with respect to the response variable. The pioneering work of \cite{koenker1978regression} introduced the concept of quantile regression, and it has been comprehensively extended by  \cite{koenker2001quantile}. For a detailed and systematic introduction to quantile regression and its extensions, we refer to \cite{kouretas2005conditional,cai2008nonparametric,cai2012semiparametric,koenker2017quantile}.  Equipped with a specially designed check function, quantile regression provides a more complete picture of the relationship between the response variable and the covariates, while also demonstrating strong robustness against outliers and heavy-tailed distributions. As a result, quantile regression has been applied in various fields \citep{garcia2001wide,machado2005counterfactual,sharif2021disaggregated}. One innovative development in quantile regression is the quantile periodogram (QP) \citep{li2012quantile} constructed by trigonometric quantile regression, which demonstrates its ability to detect hidden periodicities in time series. Similar to the behavior of the power spectrum and the PG, the QP is an unbiased estimator of the so-called quantile spectrum, which is a scaled version of the ordinary power spectrum of the level-crossing process. Notably, the Laplace periodogram  \citep{li2008laplace} represents a specialized case of the QP when the quantile is set to 0.5. Related works on the QP include \cite{li2012detection,li2014quantile,dette2015copulas,kley2016quantile,birr2017quantile,meziani2020penalised,chen2021semi,li2023quantile}. 

However, a distinctive feature of quantile regression is that its objective function is not differentiable, which presents a computational challenge. Asymmetric least squares (ALS) regression, also known as ER, was proposed by \cite{newey1987asymmetric} to address this issue. ER can capture the entire distribution of the data and is much easier to compute using weighted least squares estimators. Furthermore, consistent estimation of the joint asymptotic covariance matrix of several ALS estimators does not require estimation of the density function of the error terms. In addition, ER exhibits high-normal efficiency.  A comprehensive comparative analysis of quantiles and expectiles is presented in \cite{waltrup2015expectile}, wherein the relationships between these two approaches are thoroughly examined. In addition, \cite{jones1994expectiles} provided a mathematical proof that expectiles indeed correspond to quantiles of a distribution function uniquely associated with the underlying data distribution. Building on this foundation, \cite{yao1996asymmetric} established the existence of a bijective function that directly relates expectiles to quantiles, thereby facilitating the calculation of one from the other. Alternative approaches for estimating quantiles from expectiles are introduced in \cite{efron1991regression,  schnabel2009non}, offering methodologies for estimating the density (and also, quantiles) from a set of expectiles by using penalized least squares. As a generalization of both quantile and ER, \cite{jiang2021k} introduced the $k$-th power ER with $1\textless k \leq 2$.    
 One commonly noted disadvantage of the expectile framework is its less direct interpretability.  \cite{philipps2022interpreting} provided nine perspectives for interpreting expectiles, which make expectiles more intuitive and accessible.

Expectile techniques are applied across diverse domains. One notable field is financial time series analysis, where the desirable properties of expectiles make them an alternative to popular risk measures such as value at risk (VaR) and expected shortfall (ES) \citep{ziegel2016coherence}. In contrast to VaR and ES, which only concern the lower tail, the expectile relies on both tails of the distribution to measure risk. \cite{bellini2017risk} reviewed the properties and financial interpretations of expectiles, VaR, and ES, along with their asymptotic behaviors.  \cite{daouia2018estimation} used tail expectiles to estimate alternative measures for the VaR, ES and marginal ES. \cite{xu2020mixed} developed a novel mixed data sampling ER (MIDAS-ER) model to measure financial risk and demonstrated its exceptional performance when applied to VaR and ES.  \cite{jiang2017expectile} introduced an ER neural network (ERNN) model, which incorporates a neural network structure into ER, thereby facilitating the exploration of potential nonlinear relationships between covariates and the expectiles of the response variable. \cite{xu2021elastic} involved the elastic-net penalty into ER, and applied the model to two real-world applications: relative location of  computed tomography (CT) slices on the axial axis and metabolism of tacrolimus drug.

Inspired by the notable success of ER and the foundational work of the QP, we propose a novel spectral estimator termed the EP, for spectral analysis of time series data. We demonstrate that the EP not only shares similar properties with the PG as a frequency-domain representation of serial dependence in time series, but also provides a more comprehensive analysis by exploring the full range of expectile levels. The remainder of the paper is organized as follows. In Section \ref{EP}, we define the EP and the expectile spectrum. We establish the asymptotic analysis in Section \ref{aa}. In Section \ref{NR}, we present comparative studies evaluating the performance of different periodograms by simulations. In Section \ref{EC}, we apply our method to two real-world examples and an earthquake data classification task, where we leverage the two-dimensional nature of the EP and train a deep learning model to classify the earthquake events. Conclusions and future work are discussed in Section \ref{CO}.

\section{Expectile Periodogram and Expectile Spectrum}\label{EP}
\subsection{Ordinary Periodogram}
Let $\omega_{\nu} := 2\pi \nu /n$ for $\nu =0,  1, \ldots, n - 1$ denote the Fourier frequency and let  $\{y_t : t = 1, \ldots, n\}$ be a real-valued time series. The raw PG is defined as 
\begin{equation}\label{pg1}
I_{n}(\omega_{\nu}):=n^{-1}|z_{n}(\omega_{\nu})|^{2}:=n^{-1}\left|\sum_{t=1}^{n}y_{t}\exp(-i\omega_{\nu}t)\right|^{2},
\end{equation}
where $i:=\sqrt{-1}$ and $z_{n}(\omega_{\nu})$ denotes the discrete Fourier Transform (DFT) of $\{y_t\}$. This Fourier representation (\ref{pg1}) can also be expressed in an equivalent form via  OLS regression.  Let $
\tilde{\bm{\beta}}_n(\omega_{\nu}) := [ \tilde{\beta}_1(\omega_{\nu}) , \tilde{\beta}_2(\omega_{\nu}) , \tilde{\beta}_3(\omega_{\nu}) ] ^ {\top}
$ be the minimizer of
\begin{equation}\label{ols}
\tilde{\bm{\beta}}_n(\omega_{\nu}) := \arg\min_{\bm{\beta} \in \mathbb{R}^3} \sum_{t=1}^{n} \left|y_t - \mathbf{x}_{nt}^{\top} \bm{\beta}\right|^2,
\end{equation}
where $\mathbf{x}_{nt}$ denotes the trigonometric regressor
\begin{equation}\label{xt}
\mathbf{x}_{nt}(\omega_{\nu}) :=\left[ 1 , \cos(\omega_{\nu} t) , \sin(\omega_{\nu} t) \right]^{\top}.
\end{equation}
Then, it can be verified that
$$I_{n}(\omega_{\nu}) = \frac{1}{4}n\left( \tilde{\beta}_2(\omega_{\nu} )^2 + \tilde{\beta}_3(\omega_{\nu} )^2 \right).$$
It is  convenient to define $I_{n}(0) = 0$, this is  computed from the demeaned series $\{y_{t}-\bar{y}:t=1,\ldots,n\}$, where $\bar{y}:=n^{-1}\sum_{t=1}^{n}y_{t}$.

\subsection{Expectile Periodogram}
\begin{figure}[ht]
\centering
\subfigure[]{\includegraphics[width=0.314\textwidth]{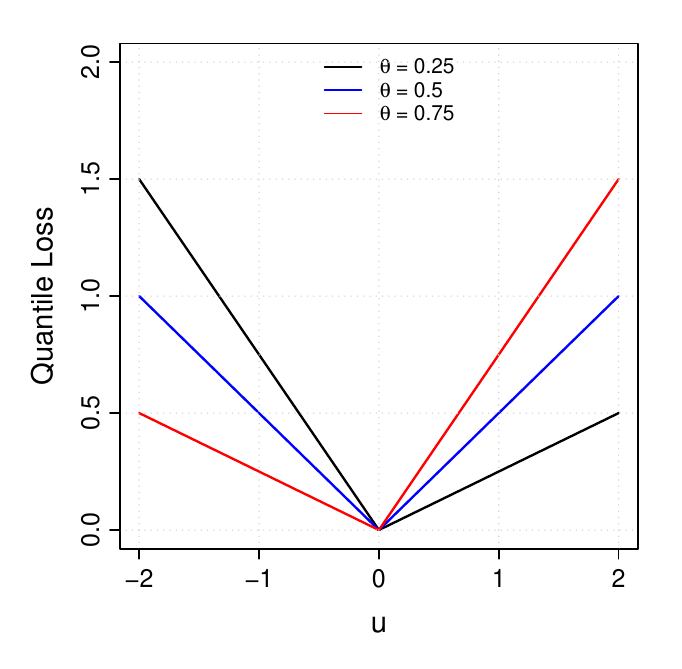}}
\subfigure[]{\includegraphics[width=0.314\textwidth]{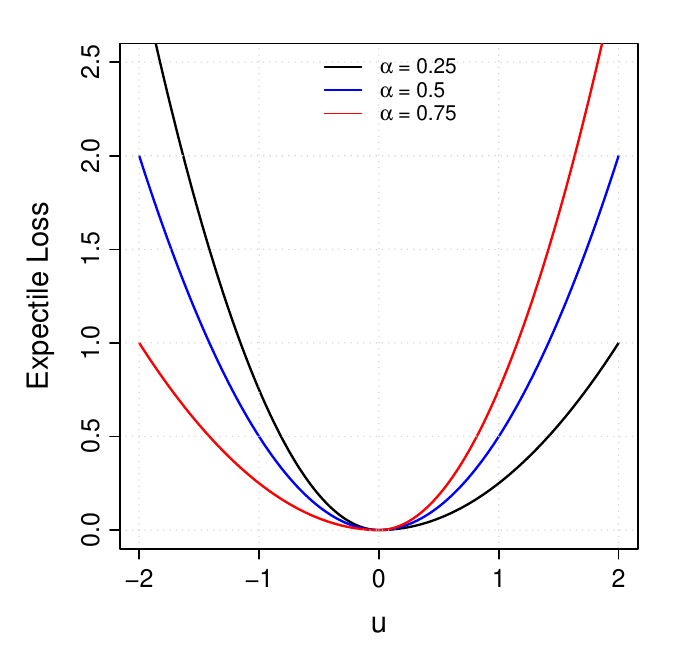}} 
\subfigure[]{\includegraphics[width=0.314\textwidth]{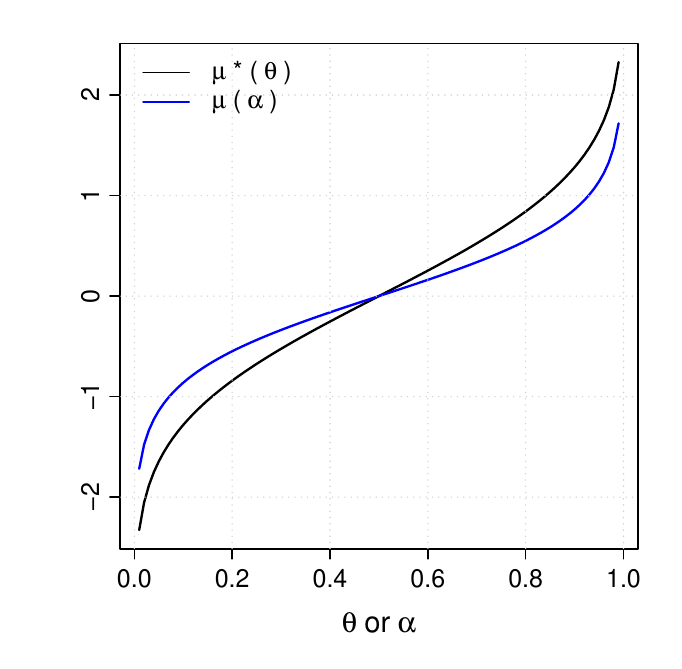}} 
\caption{(a) Loss functions for the QR at $\theta =0.25,0.5,0.75$. (b) Loss functions for the ER at $\alpha =0.25,0.5,0.75$. (c) The quantile function $\mu^*(\theta)$ and the expectile function $\mu(\alpha)$ for the standard Gaussian distribution.}
\label{loss}
\end{figure}

For any expectile $\alpha \in (0,1)$, define the expectile loss as
\begin{equation}\label{eloss}
\rho_{\alpha}(u) := u^2|\alpha - I(u<0)| =
\begin{cases}
	\alpha u^2 & \text{if } u \geq 0, \\
	(1-\alpha)u^2  & \text{otherwise,} 
\end{cases}
\end{equation}
where $I(\cdot)$ denotes the indicator function.
When $\alpha=0.5$, $\rho_{\alpha}(u) $ reduces to the quadratic loss used in OLS. The $\alpha$-expectile of $\{y_t\}$ is defined as
$$
\mu(\alpha) := \arg\min_{\mu \in \mathbb{R}} \mathrm{E} \left\{ \rho_\alpha(y_t - \mu) \right\},
$$
which satisfies the normal equation
\begin{equation}\label{ne}
\mathrm{E} \left\{ \dot{\rho}_\alpha(y_t - \mu(\alpha)) \right\} = 0,
\end{equation}
where  
$$
\dot{\rho}_\alpha(u) :=
\begin{cases}
2\alpha u, & \text{if } u \geq 0, \\
2(1 - \alpha) u & \text{otherwise}.
\end{cases}
$$ 
Figure \ref{loss} (a) and (b) compare the expectile loss and the quantile loss at different quantile or expectile levels. Let $\mu^*(\theta)$ denote the $\theta$-th quantile of $\{y_t\}$, where $\theta \in (0,1)$. Figure \ref{loss} (c) compares the quantile function $\mu^*(\theta)$  and the expectile function $\mu(\alpha)$ for the standard Gaussian distribution, where $\mu(\alpha)$ has a smaller slope than $\mu^*(\theta)$ near $\theta \text{ or } \alpha =0.5$ and a larger slope near $\theta \text{ or } \alpha = 0 \text{ or } 1$. 

Substituting the expectile loss (\ref{eloss}) into (\ref{ols}), we obtain
\begin{equation}\label{er}
\hat{\bm{\beta}}_n(\omega_{\nu}, \alpha) := \arg\min_{\bm{\beta} \in \mathbb{R}^3} \sum_{t=1}^{n}\rho_{\alpha}\left( y_t - \mathbf{x}_{nt}^{\top} \bm{\beta}\right),
\end{equation}
where  $
\hat{\bm{\beta}}_n(\omega_{\nu},\alpha) := [ \hat{\beta}_1(\omega_{\nu},\alpha) , \hat{\beta}_2(\omega_{\nu},\alpha) , \hat{\beta}_3(\omega_{\nu},\alpha) ] ^ {\top}.
$
The ER problem (\ref{er}) reduces to the OLS regression when $\alpha = 0.5$. 
Then, the raw EP at expectile level $\alpha$ is defined as
\begin{equation}\label{ep}
{\rm EP}_{n}(\omega_{\nu},\alpha):=\frac{1}{4}n \left( \hat{\beta}_2(\omega_{\nu} ,\alpha)^2 +\hat{\beta}_3(\omega_{\nu},\alpha )^2 \right), \quad (\nu=1,\ldots,n-1).
\end{equation}
When $\omega_{\nu}$ takes values $0$ and $\pi$, we modify the regressors accordingly. For $\omega_{\nu} = \pi$, define  
$
\mathbf{x}_{nt}(\pi) := [ 1 , \cos(\pi t)] ^{\top}, 
$
and let the resulting ER solution be  
$
\hat{\bm{\beta}}_n(\pi, \alpha) :=[ \hat{\beta}_1(\pi, \alpha) , \hat{\beta}_2(\pi, \alpha) ]^{\top}.
$
For $\omega_{\nu} = 0$, let $\hat{\beta}_1(0, \alpha)$ denote the ER solution with the regressor $\mathbf{x}_{nt} := 1$. Then,
$$
{\rm EP}_{n}(\omega_{\nu},\alpha) := \left\{
\begin{array}{ll}
	n|\hat{\beta}_{1}(0,\alpha)|^{2} & \omega_{\nu}=0, \\
	n|\hat{\beta}_{2}(\pi,\alpha)|^{2} & \omega_{\nu}=\pi \text{ (if $n$ is even)}.
\end{array}
\right. 
$$
In the special case of $\alpha=0.5$, the EP coincides with the PG.

\subsection{Expectile Discrete Fourier Transform}
The EP can also be derived by the Fourier transform representation. Given the ER solution $	\hat{\bm{\beta}}_n(\omega_{\nu},\alpha)$, the expectile discrete Fourier transform (EDFT) of $\{y_{t}\}$ is defined as
$$
z_{n}(\omega_{\nu},\alpha):=\left\{
\begin{array}{ll}
	n\hat{\beta}_{1}(0,\alpha) & \omega_{\nu}=0, \\
	n\hat{\beta}_{2}(\pi,\alpha) & \omega_{\nu}=\pi \text{ (if $n$ is even)}, \\
	(n/2)\{\hat{\beta}_{2}(\omega_{\nu},\alpha)-i\hat{\beta}_{3}(\omega_{\nu},\alpha)\} & \omega_{\nu} \in (0, \pi).
\end{array}
\right. 
$$
The EDFT at $\alpha=0.5$ coincides with the ordinary DFT, i.e., $z_{n}(\omega_{\nu},0.5)=z_{n}(\omega_{\nu})$. Then, it can be verified that the EP defined in (\ref{ep}) satisfies
\begin{equation}\label{def2}	{\rm EP}_{n}(\omega_{\nu},\alpha)=n^{-1}|z_{n}(\omega_{\nu},\alpha)|^{2}.
\end{equation}	
With (\ref{ep}) and (\ref{def2}), both the PG and EP have parallel representations: Fourier and regression.

\noindent
{\bf Remark 1:} 
It is also convenient to define ${\rm EP}_{n}(0,\alpha)=0$ instead of $n|\hat{\beta}_{1}(0,\alpha)|^{2}$. This is analogous to the property of the PG: $I_n(0) = 0$.  In the rest of this paper, we use $f=\omega/2\pi \in [0, 1/2]$, which represents the number of cycles per unit time in all figures.  

\noindent
{\bf Remark 2:} 
In Sections \ref{NR} and \ref{EC}, 
our primary focus is on serial dependence of the time series, with amplitude considerations being secondary. Accordingly, we normalize the EP such that the summation over $\omega$ equals $1$ at each expectile level. The normalization makes the estimation of the scaling factor defined in (\ref{eta}) unnecessary.  

\subsection{Expectile Spectrum}
Similar to the relationship between the PG and the power spectrum, we define the expectile spectrum via the autocovariance structure.
Suppose that the sequence $\{\dot{\rho}_\alpha(y_t - \mu(\alpha))\}$ is stationary,
and define the autocovariance function (ACF) of $\left\{ \dot{\rho}_\alpha(y_t - \mu(\alpha)) \right\}$ as
$$
\gamma(t-s, \alpha):={\rm Cov}\{\dot{\rho}_\alpha(y_t - \mu(\alpha)),\dot{\rho}_\alpha(y_s - \mu(\alpha))\} . 
$$
Assume that $\sum_{\tau}|\gamma(\tau,\alpha)| < \infty$ and $\omega \in [0, \pi]$. Then, the expectile spectrum is defined as
\begin{equation}\label{es}
	g(\omega , \alpha): = \eta^2(\alpha)h(\omega, \alpha),
\end{equation}	
where 
\begin{equation}\label{homega}
	h(\omega, \alpha) := \sum_{\tau=-\infty}^{\infty} \gamma(\tau,\alpha)\exp(-i \omega \tau) = \sum_{\tau=-\infty}^{\infty}\gamma(\tau,\alpha)\cos( \omega \tau),
\end{equation}
and the second equality holds because \(\gamma(\tau,\alpha) = \gamma(-\tau,\alpha)\). 
The scaling factor $\eta(\alpha)$ is defined as
\begin{equation}\label{eta}
	\eta(\alpha) := \frac{1}{2} \left\{ \alpha(1 - F(\mu(\alpha))) + (1 - \alpha) F(\mu(\alpha)) \right\}^{-1},
\end{equation}
which depends solely on the marginal distribution \(F(\cdot)\) and the expectile level $\alpha$.
The quantity
$$
\alpha(1 - F(\mu(\alpha))) + (1 - \alpha) F(\mu(\alpha)) = \alpha + (1 - 2\alpha) F(\mu(\alpha))
$$
always lies between $\alpha$ and $1 - \alpha$, and therefore, $\eta(\alpha)$ is a finite positive number for any given $\alpha \in (0,1)$. The value $2 \left\{ \alpha(1 - F(\mu(\alpha))) + (1 - \alpha) F(\mu(\alpha)) \right\} = {\rm E}\{\ddot \rho_{\alpha}(u)\}$ since 
$$
\ddot \rho_{\alpha}(u) = 
\begin{cases}
	2\alpha , & \text{if } u \geq 0, \\
	2(1 - \alpha) & \text{otherwise}.
\end{cases}
$$


\section{Asymptotic Analysis} \label{aa}
\subsection{Large Sample Theory}
It is well-established that a smoothed PG constitutes an estimate of the power spectrum of the underlying random process that generates the time series. In this section, we conduct a parallel investigation of the EP through large sample asymptotic analysis. Our results reveal that the EP is fundamentally connected to the expectile spectrum.

Let $\{y_t\}$ be a stationary process with cumulative distribution function $F(\cdot)$ and finite second moments. Consider the case of multiple frequencies, for fixed $q>1$, let the $q$ regressors defined in (\ref{xt}) be $\mathbf{x}_{jnt} =[ 1, \cos(\omega_{\nu_j} t) , \sin(\omega_{\nu_j} t) ]^{\top}, t = 1, \ldots, n,j=1,\ldots, q.$
Define 
\begin{equation}\label{zta}
	\boldsymbol{\zeta}_{jn}(\alpha):=n^{-1/2}\sum_{t=1}^{n}\dot{\rho}_{\alpha}(y_{t}-\mu(\alpha))\mathbf{x}_{jnt}.
\end{equation}
Assume that the following conditions are satisfied  for a fixed $\alpha$:
\begin{itemize}
	\item[C1.] The cumulative distribution function $F(\cdot)$ is Lipschitz continuous, i.e., there exists a constant $K>0$ such that $|F(y)-F(y^{\prime})|\leq K|y-y^{\prime}|$ for all $y,y^{\prime}\in\mathbb{R}$.

	\item[C2.] The process $\{y_{t}\}$ has the strong mixing property with mixing numbers $a_{\tau}$ ($\tau=1,2,\ldots$) satisfying $a_{\tau}\to 0$ as $\tau\to\infty$.

	\item[C3.] A central limit theorem is valid for \(\left[\boldsymbol{\zeta}_{1n}(\alpha),\ldots,\boldsymbol{\zeta}_{qn}(\alpha)\right]^{\top}\), i.e., \(\left[\boldsymbol{\zeta}_{1n}(\alpha),\ldots,\boldsymbol{\zeta}_{qn}(\alpha)\right]^{\top}\xrightarrow{D}N(\mathbf{0},[\mathbf{V}_{jk}(\alpha)]_{j,k=1}^q)\) as \(n\to\infty\), where $\mathbf{V}_{jk}(\alpha)$ is the asymptotic covariance matrix of $\boldsymbol{\zeta}_{jn}(\alpha)$ and $\boldsymbol{\zeta}_{kn}(\alpha)$ that are defined in (\ref{zta}). \footnote{C3 is a relaxation condition, see \textbf{Appendix A.4} for details.} 
\end{itemize}
Equipped with these concepts, we establish the following theorem regarding the EP.

\noindent
\textbf{Theorem 1} (Expectile Periodogram). {\it For fixed \(q>1\) and \(0<\lambda_{1}<\cdots<\lambda_{q}<\pi\), let \(\omega_{\nu_{1}},\ldots,\omega_{\nu_{q}}\) be Fourier frequencies satisfying \(\omega_{\nu_{j}}\to\lambda_{j}\) as \(n\to\infty\) for \(j=1,\ldots,q\). If the process $\{\dot{\rho}_{\alpha}(y_{t}-\mu(\alpha))\}$ is stationary  and its ACF \(\gamma(\tau,\alpha)\) is absolutely summable with \(h(\lambda_{j},\alpha)>0\) (\(j=1,\ldots,q\)), and if (C1), (C2), and (C3) hold, then
\begin{equation}\label{the2}
	\{{\rm EP}_{n}(\omega_{\nu_{1}},\alpha),\ldots,{\rm EP}_{n}(\omega_{\nu_{q}},\alpha)\}\xrightarrow{D}\{g(\lambda_{1},\alpha)(1/2)\chi^{2}_{2,1},\ldots,g(\lambda_{q},\alpha)(1/2)\chi^{2}_{2,q}\},
\end{equation} 
where \(\chi^{2}_{2,1},\ldots,\chi^{2}_{2,q}\) are independent standard chi-square random variables with two degrees of freedom. }

Theorem 1 shows that the EP exhibits scaled chi-square distributions with two degrees of freedom, which is similar to the behavior of the PG \citep{brockwell1991time}. The proof is provided in the appendix. 

Theorem 1 can be directly applied to the normalized EP. Define the normalized raw expectile periodogram
\(\widetilde{\mathrm{EP}}_n(\omega_{\nu_j},\alpha):=\mathrm{EP}_n(\omega_{\nu_j},\alpha)\big/\sum_{k=1}^q \mathrm{EP}_n(\omega_{\nu_k},\alpha)\)
and the normalized expectile spectrum
\(\widetilde{g}(\lambda_j,\alpha):=g(\lambda_j,\alpha)\big/\sum_{k=1}^q g(\lambda_k,\alpha)\), \(j=1,\ldots,q\).
Then under (C1) - (C3) and \(\omega_{\nu_{j}}\to\lambda_{j}\) as \(n\to\infty\) for \(j=1,\ldots,q\), as $n \to \infty$,
\[
\{\widetilde{\mathrm{EP}}_n(\omega_{\nu_1},\alpha),\ldots,\widetilde{\mathrm{EP}}_n(\omega_{\nu_q},\alpha)\}\rightarrow{D}
\{\widetilde{g}(\lambda_1,\alpha)(1/2)\chi^{2}_{2,1},\ldots,\widetilde{g}(\lambda_q,\alpha)(1/2)\chi^{2}_{2,q}\big\}.
\]
This follows by the continuous mapping theorem applied to the vector of raw EP ordinates because the normalization map \(x\mapsto x/\sum_{k=1}^q x_k\) is continuous on \((0,\infty)^q\).

\subsection{Properties of the Expectile Periodogram}
According to  Theorem 1, as $n \to \infty$ and $\omega_{\nu} \to \lambda$,
 \begin{equation}\label{p1}
 \operatorname{E}\{{\rm EP}_{n}(\omega_{\nu},\alpha)\} \rightarrow g(\lambda, \alpha),
 \end{equation}
 and
 \begin{equation}\label{p2}
 	{\rm Var}\{{\rm EP}_{n}(\omega_{\nu},\alpha)\} \rightarrow g^2(\lambda, \alpha).
 \end{equation}	
This means that the EP is an asymptotically unbiased but inconsistent estimator of the expectile spectrum. Let the smoothed estimator be
$$
	\hat{g}(\omega_{\nu}, \alpha) 
	= \sum_{s=-M_n}^{M_n} W_{n,s}\,
	{\rm EP}_{n}(\omega_{\nu+s},\alpha),
$$
where $\omega_{\nu+M_{n}}, \omega_{\nu-M_{n}} \in [0, \pi]$  and $W_{n,s}\geq 0$ are weights satisfying $\sum_{s=-M_n}^{M_n} W_{n,s} = 1$. In addition to (C1)--(C3), assume
\begin{itemize}
	\item[C4.]  $g(\omega, \alpha)$ is continuous in $\omega$.
	\item[C5.] The smoothing span $M_n \to \infty$ with $M_n / n \to 0$ as $n\to\infty$.
	\item[C6.] The weights $\{W_{n,s}\}$ form a bounded kernel, i.e.
	$$
	\sup_n \max_{|s|\leq M_n} W_{n,s} < \infty,
	\qquad 
	\sum_{s=-M_n}^{M_n} W_{n,s}^2 \to \|K^*\|_2^2 < \infty,
	$$
	where $K^*$ is the limiting kernel function.
\end{itemize}
Then, we establish the following theorem regarding the smoothed EP.

\noindent
{\bf Theorem 2} (Smoothed Expectile Periodogram). {\it 
	Suppose (C4)--(C6) hold, 
	\begin{equation}\label{p3}
			\hat{g}(\omega_{\nu}, \alpha)  \;\overset{p}{\longrightarrow}\; g(\lambda, \alpha).
	\end{equation}}

Theorem 2 indicates that the smoothed EP provides a consistent estimator of the expectile spectrum.  This implies that the expectile spectrum can be estimated directly by smoothing the  raw EP, thereby removing the need to estimate the scaling factor $\eta^2(\alpha)$. The proof is provided in the appendix. The three properties (\ref{p1}), (\ref{p2}), and (\ref{p3}) are similar to the corresponding properties of the PG.

\subsection{Asymmetrically-Scaled Expectile Crossing Process}
\(h(\omega,\alpha)\) in (\ref{homega}) is the ordinary spectrum of the stationary process $\{\dot{\rho}_{\alpha}(y_{t}-\mu(\alpha))\}$. This process is an asymmetrically-scaled version of the process \(\{y_{t}-\mu(\alpha)\}\) around \(\mu(\alpha)\). We define $\{\dot{\rho}_{\alpha}(y_{t}-\mu(\alpha))\}$ as the asymmetrically-scaled expectile crossing process (ASECP) of $\{y_t\}$, which is analogous to the level-crossing process (LCP) $\{I(y_t > \theta)\}$  used in defining the quantile spectrum in \cite{li2012quantile}.

Specifically, the check loss function in quantile regression is defined as
\[
\rho^*_{\theta}(u) = u\{\theta - I(u<0)\} =
\begin{cases}
	\theta\,u, & \text{if } u \ge 0,\\[4pt]
	-(1-\theta)\,u, & \text{if } u < 0.
\end{cases}
\]
Since $\rho^*_{\theta}(u)$ is not differentiable at $u=0$, we consider its weak derivative,
\begin{equation}\label{qdiff}
	\dot\rho^*_{\theta}(u)=
	\begin{cases}
		\theta, & u>0,\\[4pt]
		-(1-\theta), & u<0,
	\end{cases}
\end{equation}
and any value in the interval $[\theta-1,\theta]$ may be assigned at $u=0$.
Under the assumption that $\Pr(u=0)=0$, the exact value at $u=0$ is asymptotically irrelevant. Then, $\dot\rho^*_{\theta}(u)$ is linearly related to the level-crossing process (LCP) through
\[
I\{y_t - \mu^*(\theta) > 0\} = (1-\theta) + \dot\rho^*_{\theta}(u),
\]
or equivalently, $I\{y_t - \mu^*(\theta) \leq 0\} = \theta - \dot\rho^*_{\theta}(u)$.
Hence, the quantile spectrum defined by the LCP is equivalent (up to a linear transformation, which only affects the scaling factor)
to that defined via the derivative of the loss function in (\ref{qdiff}).
Thus, both the quantile spectrum and the expectile spectrum characterize serial dependence through the covariance of transformed residuals defined by their respective loss derivatives.

The function $\gamma(\tau,\alpha)$ measures serial dependence in asymmetric deviations from the expectile $\alpha$. For $\alpha>0.5$, larger weights are assigned to positive deviations, so $\gamma(\tau,\alpha)$ measures persistence in upper-tail dynamics; for $\alpha<0.5$, negative deviations receive greater weight, emphasizing lower-tail dependence. When $\alpha=0.5$, $\gamma(\tau,0.5)$ coincides with the ordinary ACF. Hence, varying $\alpha$ allows the expectile spectrum to reveal frequency-domain characteristics specific to different parts of the distribution, providing an asymmetric generalization of the ordinary spectrum.

\section{Numerical Results}\label{NR}
In this section, we present simulations to demonstrate the efficiency of the EP in detecting hidden periodicities in time series.

\subsection{Hidden Periodicities Detection}
We consider the following model: 
\begin{equation}\label{hidden}
	y_t = a_t x_t,
\end{equation} 
where
$$	a_t =b_0 + b_1\cos(\omega_{\nu_0} t) + b_2\sin(\omega_{\nu_1} t),$$
with $b_0=1, b_1=0.9, b_2=1$. $\{x_t\}$ is an AR(2) process satisfying
\begin{equation}\label{ar2}
	x_t = \phi_1 x_{t-1} + \phi_2 x_{t-2} + \epsilon_t, 
\end{equation}
where $\phi_1 = 2r \cos(\omega_{\nu_c}), \phi_2 = -r^2$ ($r=0.6)$ and $\{\epsilon_t\}$ is the standard Gaussian white noise. Additionally, we set $\omega_{\nu_0} = 0.1 \times 2 \pi, \omega_{\nu_1} = 0.12 \times 2\pi$. This setup aims to evaluate the effectiveness of different types of periodograms in detecting multiple closely spaced periodicities.

\begin{figure}[]
	\centering
	\subfigcapskip = -0.3cm
	\subfigure[]{\includegraphics[width=0.38\textwidth]{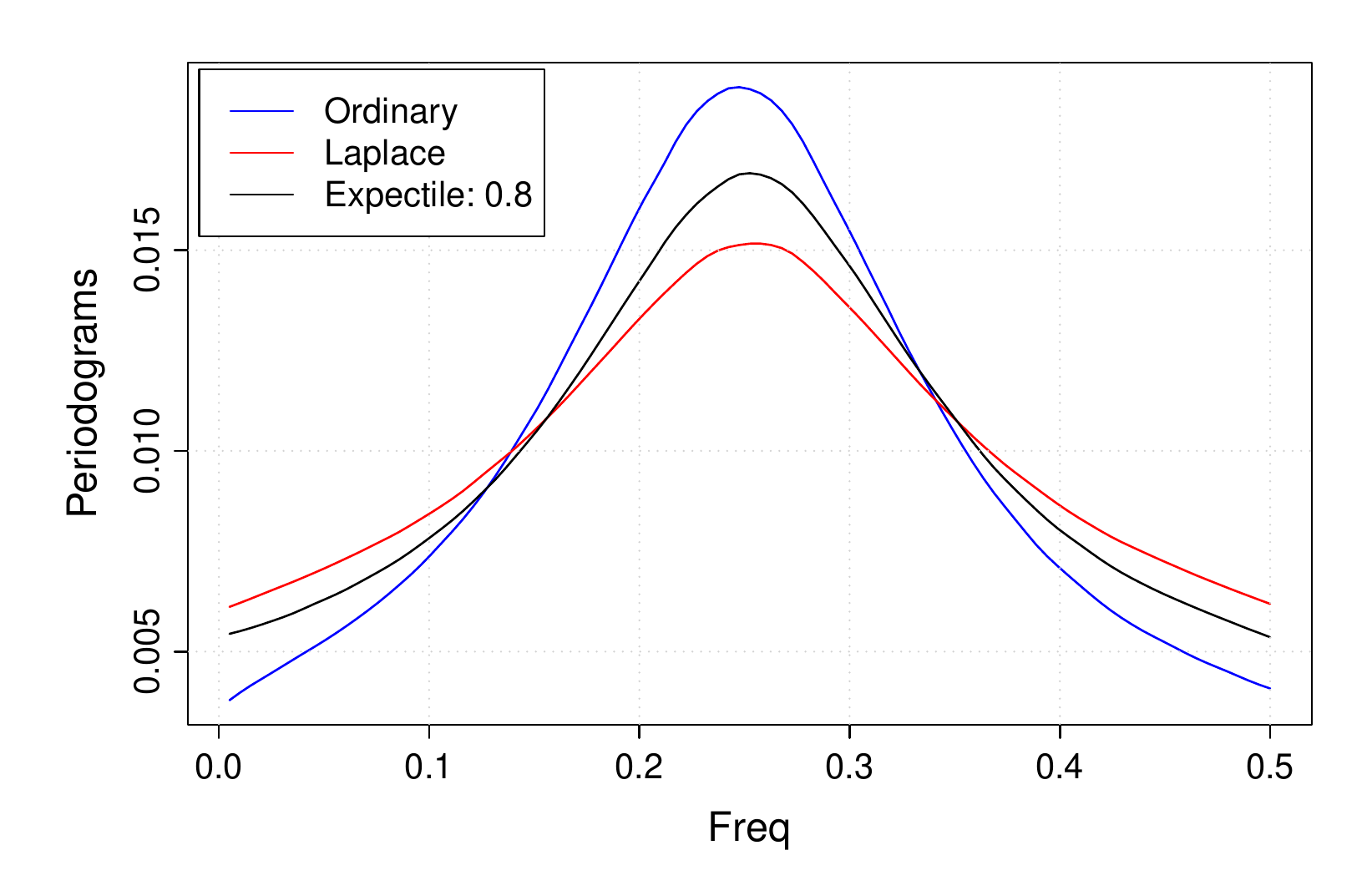}}\vspace{-3mm}
	\subfigure[]{\includegraphics[width=0.38\textwidth]{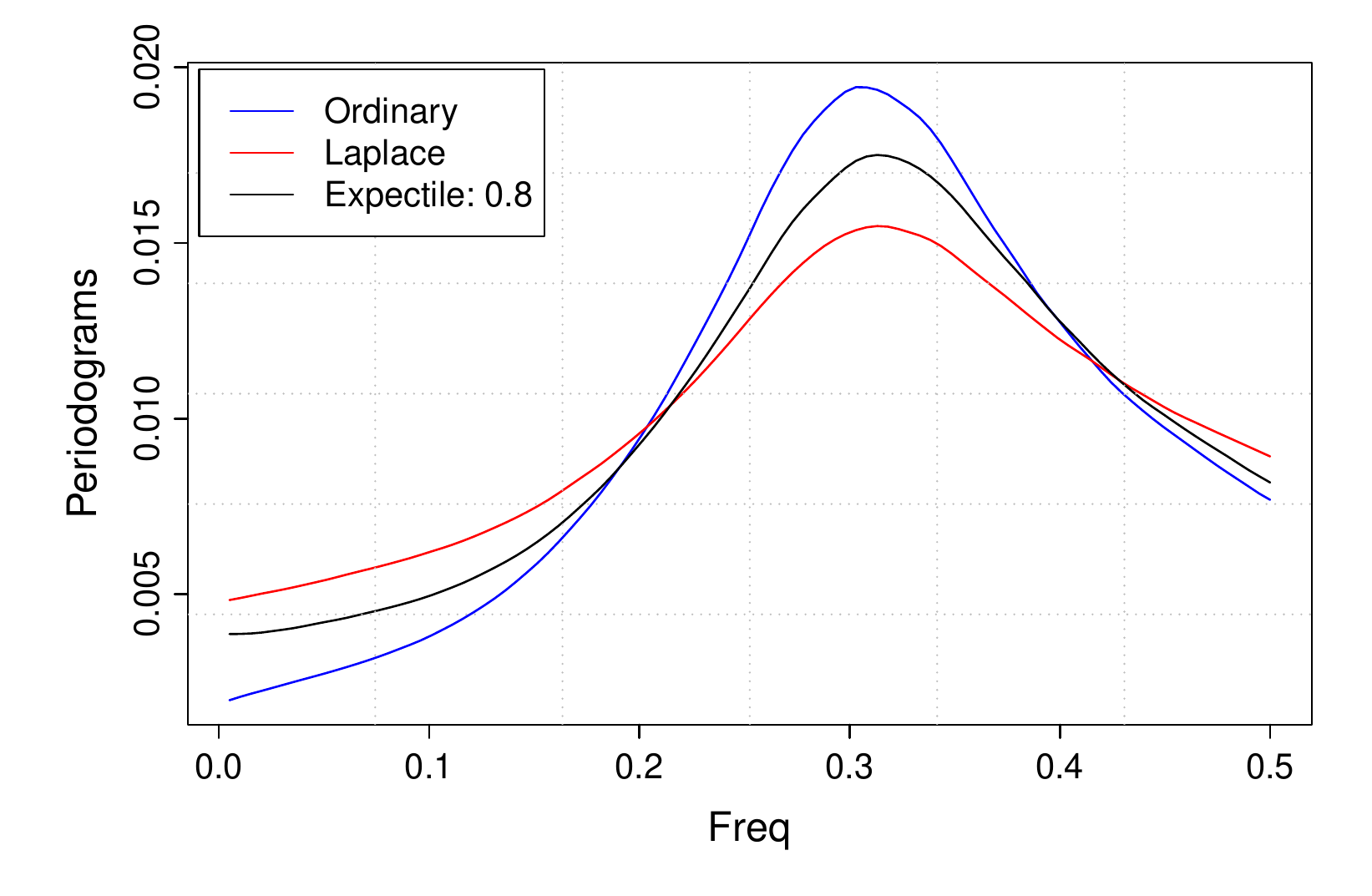}}
	\subfigure[]{\includegraphics[width=0.38\textwidth]{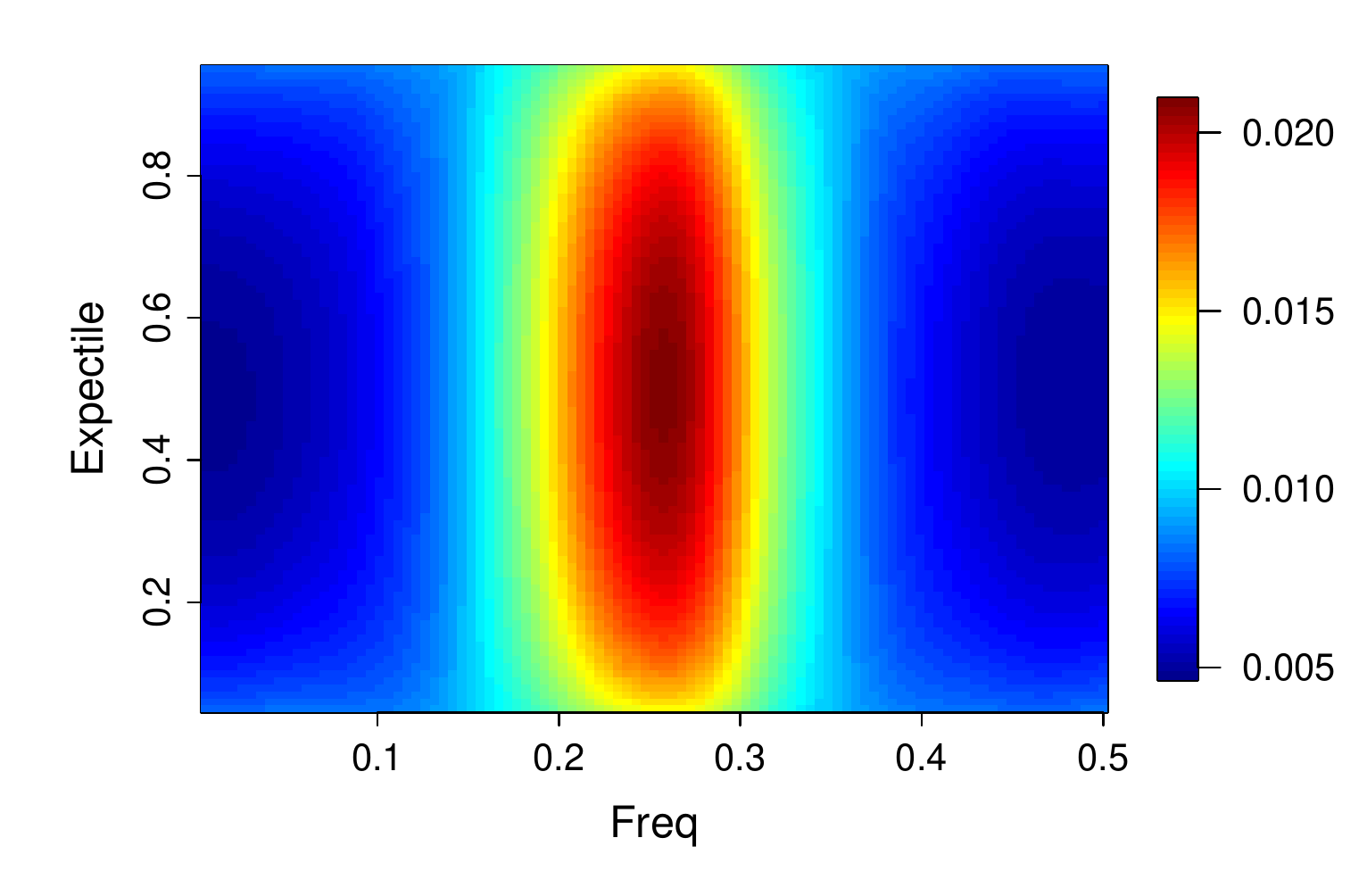}}
	\subfigure[]{\includegraphics[width=0.38\textwidth]{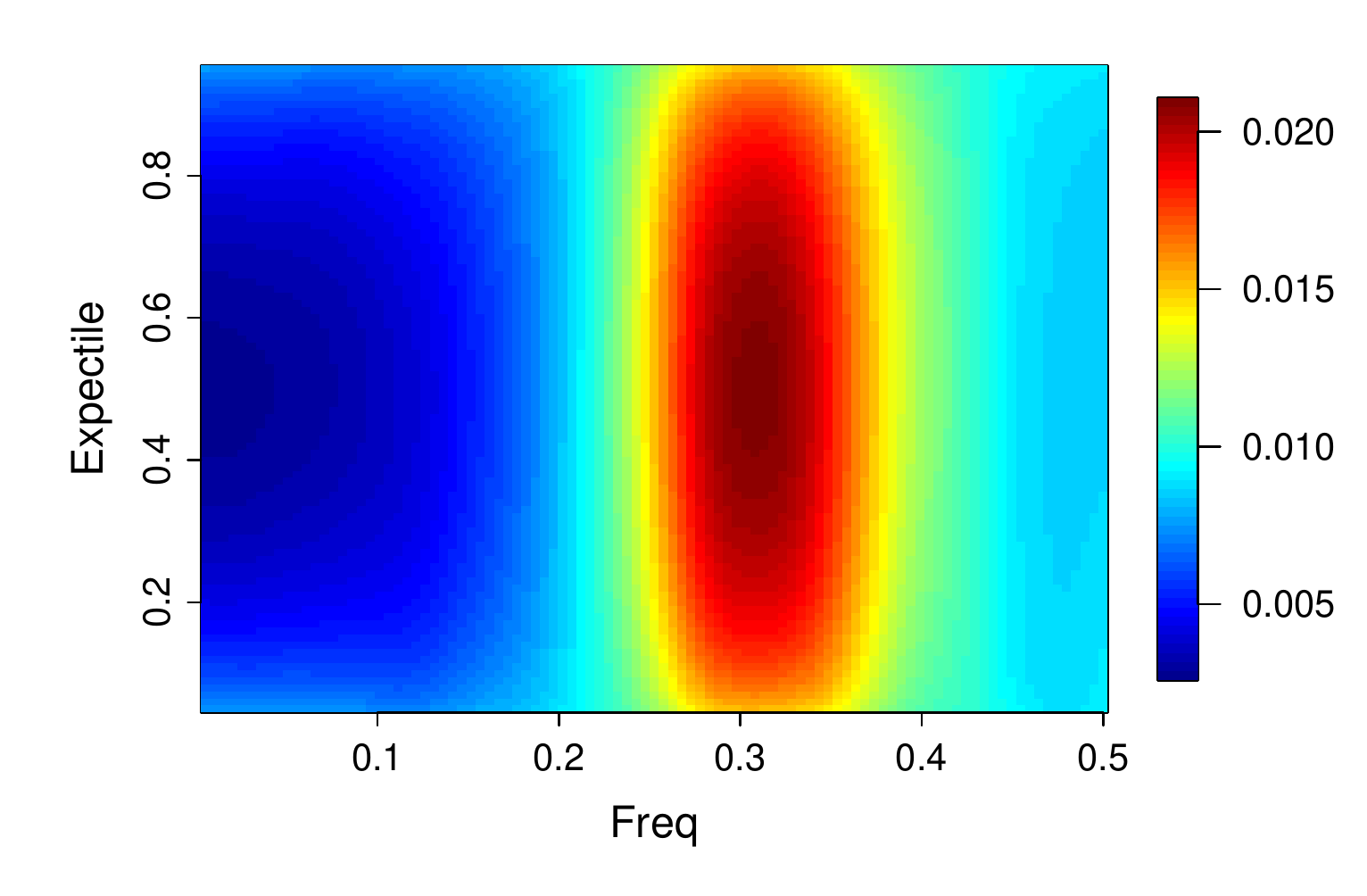}}\vspace{-3mm}
		\subfigure[]{\includegraphics[width=0.38\textwidth]{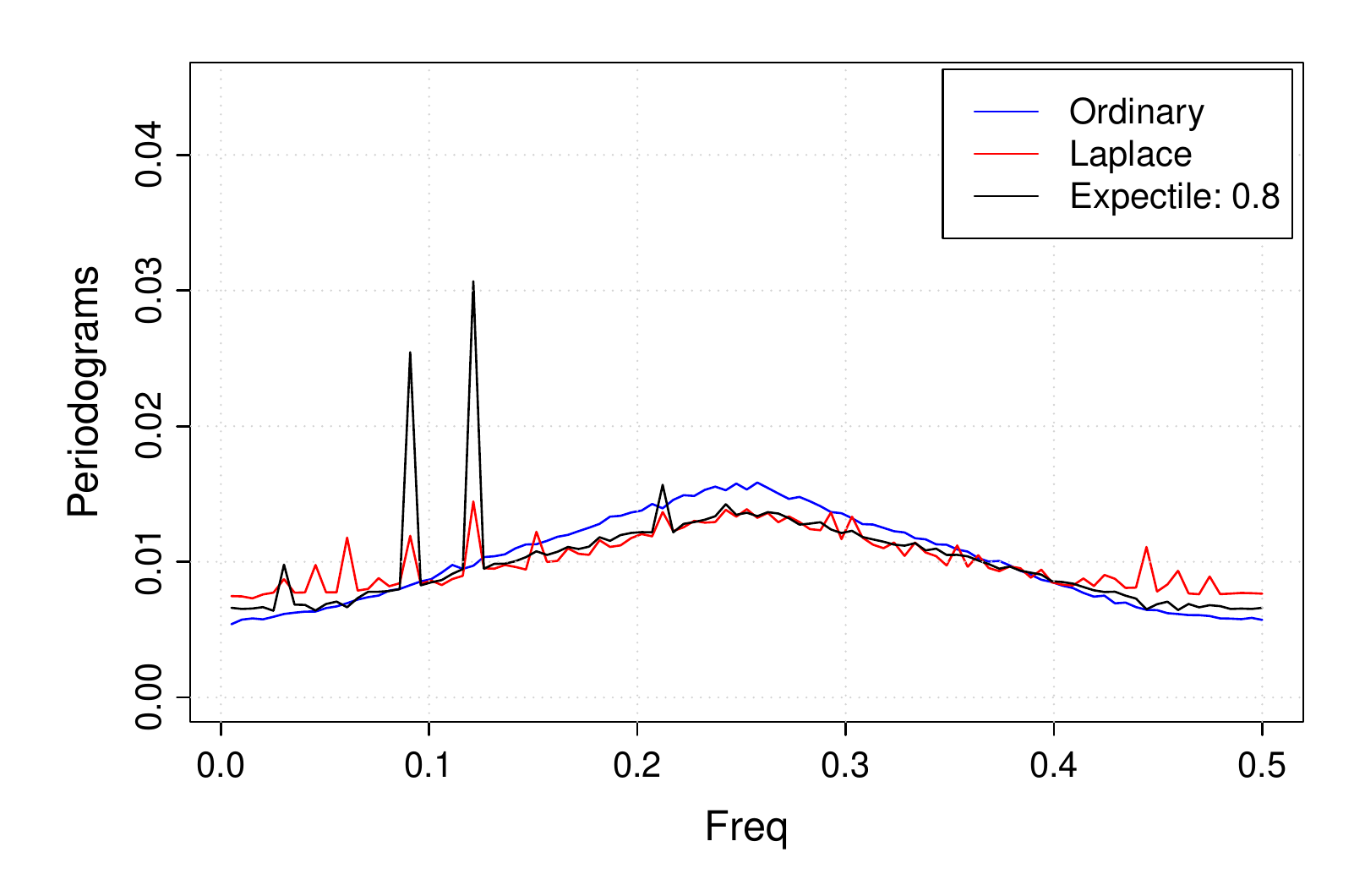}}
	\subfigure[]{\includegraphics[width=0.38\textwidth]{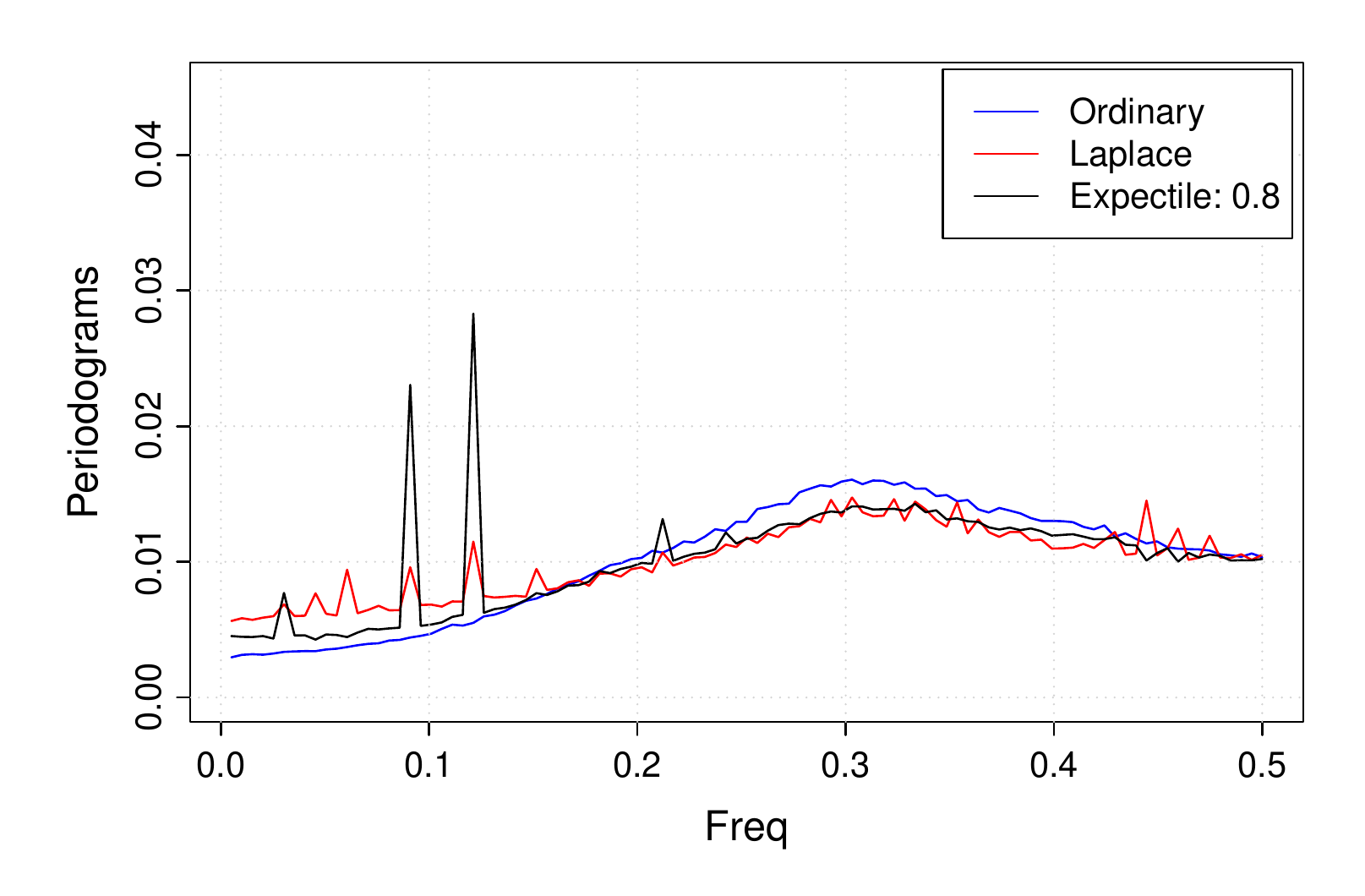}} 
	\caption{(a) and (b): Ensemble means of smoothed periodograms of model (\ref{ar2}) with $\omega_{\nu_c} = 2 \pi \times 0.25$ and $\omega_{\nu_c} = 2 \pi \times 0.3$, respectively; (c) and (d): the EPs of model (\ref{ar2}) at  $\alpha = \{0.05, 0.06,...,0.95\}$ for $\omega_{\nu_c} = 2 \pi \times 0.25$ and $ 2 \pi \times 0.3$, respectively; (e) and (f): ensemble means of periodograms of model (\ref{hidden}) with $\omega_{\nu_c} = 2 \pi \times 0.25$ and $\omega_{\nu_c} = 2 \pi \times 0.3$, respectively.}
	\label{fig-ar}
\end{figure}

We first present the periodograms of model (\ref{ar2}) with $\omega_{\nu_c} = 2\pi \times 0.25$ and $2 \pi \times 0.3$, representing cases without hidden periodicities. As shown in Figure \ref{fig-ar} (a) and \ref{fig-ar} (b), the PGs, the EPs, and the Laplace periodograms exhibit similar bell-shaped patterns, with spectral peaks around $\omega_{\nu_c}$. These results are consistent with the spectral properties of the AR(2) process. Specifically, the characteristic polynomial of model (\ref{ar2}) is $\phi(z)=1-\phi_1 z-\phi_2 z^2$, where the roots $z_1$ and $z_2$ are complex conjugates. The magnitude $|z_1|=|z_2|=1/r > 1$ ensures causality. The AR coefficients $\phi_1$ and $\phi_2$ determine the location and narrowness of the spectral peak. Specifically, the peak frequency is located at $\omega_{\nu_c}/ 2 \pi$, and as $r \rightarrow 1^-$, the peak becomes narrower \citep{shumway2016time}. Figure \ref{fig-ar} (c) and (d) show the EPs at expectiles $\{0.05, 0.06,...,0.95\}$, which share similar features to the quantile spectrum in \cite{chen2021semi}. The number of realizations is 5,000 and the sample size is $n=200$. 

Figure \ref{fig-ar} (e) and (f) illustrate the ability of the EP in detecting hidden periodicities. We present the ensemble mean of 5,000 realizations of the PGs and EPs of model (\ref{hidden}). The EPs detect the hidden periodicities of the non-stationary process $\{y_t\}$, manifesting as large spikes at $\omega_{\nu_0}$ and $\omega_{\nu_1}$, whereas the PGs do not exhibit this capability.  The process $\{y_t\}$ is a product of a slowly varying amplitude modulator $\{a_t\}$ (which contains the hidden frequencies $\omega_{\nu_0}$, $\omega_{\nu_1}$) and a carrier process $\{x_t\}$ (an AR(2) with its own spectral peak near $\omega_{\nu_c}$). The expectation $\mathrm{E}\{y_t\}=0$, indicating that the hidden frequencies only affect the variance of $\{y_t\}$ rather than its mean.  The high/low expectiles act like upper/lower envelopes of $\{y_t\}$, making the EP at these expectile levels sensitive to the variance modulation induced by $\{a_t\}$.  Thus, the EP reveals the hidden frequencies $\omega_{\nu_0}$ and $\omega_{\nu_1}$ at high/low expectiles. 

Stationarity is not required for computing the periodogram,
which is obtained via trigonometric ER or EDFT. However, stationarity is required when defining the theoretical expectile spectrum and for establishing the asymptotic properties. The EP can still capture dominant periodicities and distributional asymmetries in mildly non-stationary settings.
It is also important to note that spectral leakages, which have been observed for the Laplace periodogram, can also occur in the EP. Therefore, small spikes may take place at some other frequencies, as indicated by Theorem 2 in \cite{li2012quantile}. To address the issue of spectral leakage, incorporating an $\ell_1$ regularization into the loss function is beneficial \citep{meziani2020penalised}.

\begin{figure}[ht]
	\centering
	\subfigcapskip = -0.3cm
	\subfigure[]{\includegraphics[width=0.4\textwidth]{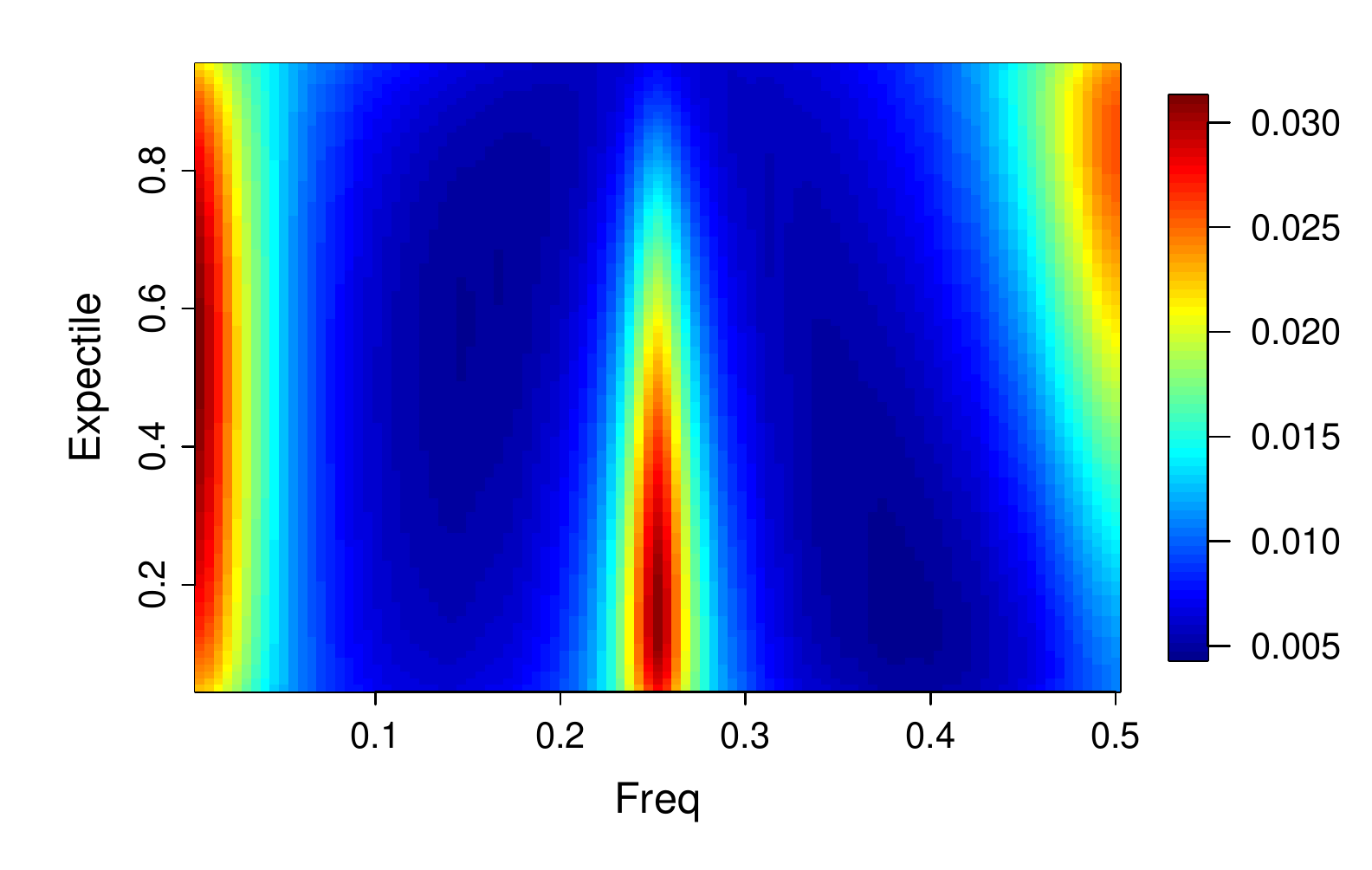}}
	\subfigure[]{\includegraphics[width=0.4\textwidth]{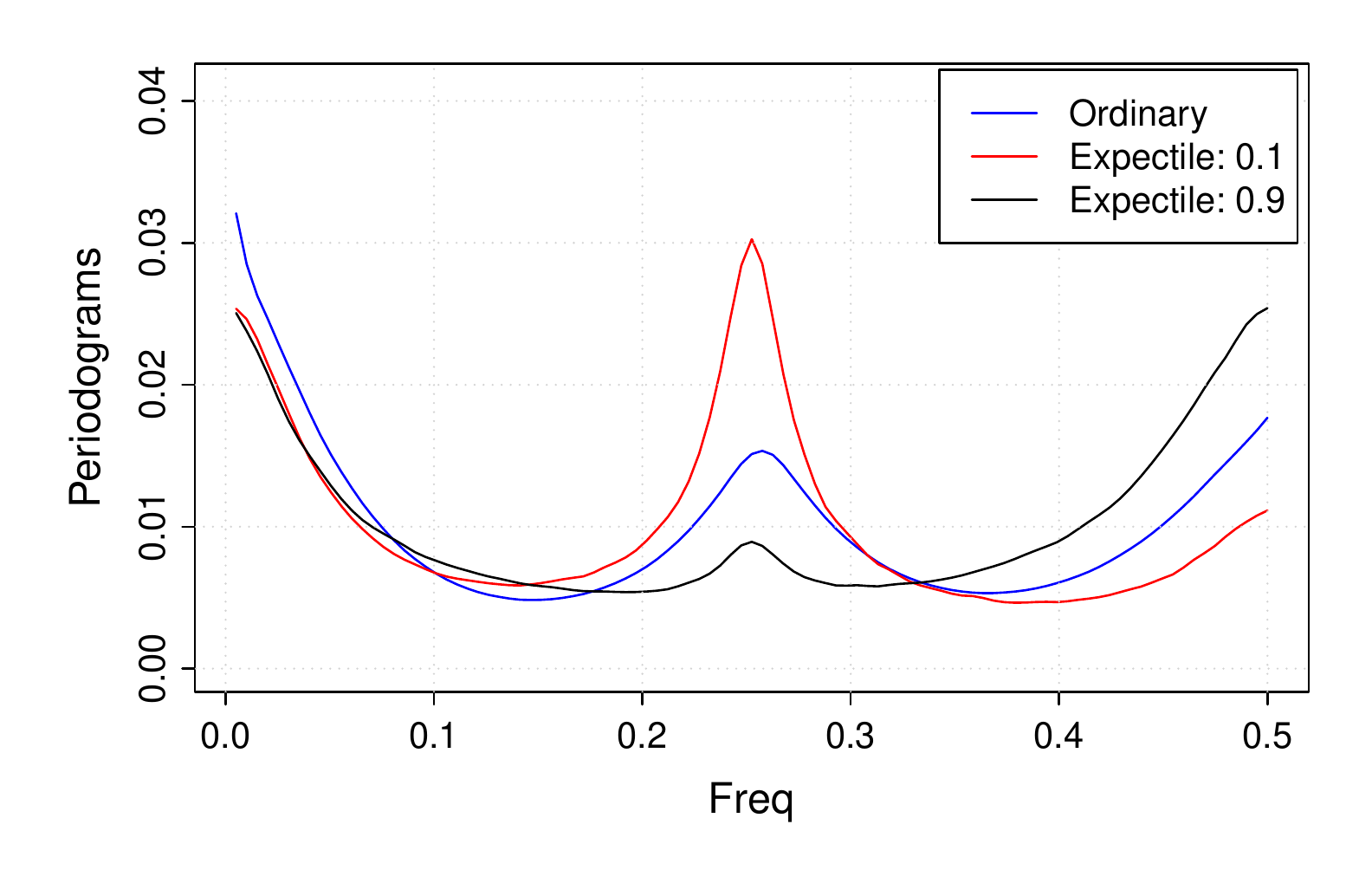}} 
	
	\caption{The periodograms of the mixture model (\ref{eq-mix}). (a) The EP with asymmetric pattern across the expectile levels, and (b) the EPs at $\alpha =0.1$ and $0.9$, along with the PG. The number of realizations is 5,000 and the sample size is $n=200$.}
	\label{fig-asy}
\end{figure}

The EPs in Figure \ref{fig-ar} (c), (d) are symmetric across the expectile levels. To demonstrate the capability of the EP in handling more complex spectral features, we construct $\{y_t\}$, whose EP is intentionally made asymmetric across the expectile levels. $\{y_t\}$ is defined as a nonlinear mixture of three components:
$$z_t  =   W_1(x_{t,1}) \, x_{t,1} + \{1-W_1(x_{t,1})\} \, x_{t,2},$$
\begin{equation}\label{eq-mix}
	y_t  =   W_2(z_t) \, z_t + \{1-W_2(z_t)\} \, x_{t,3}.
\end{equation}
The components $\{x_{t,1}\}$, $\{x_{t,2}\}$, and $\{x_{t,3}\}$ are independent Gaussian AR processes satisfying
$$x_{t,1}  =  0.8 x_{t-1,1} + w_{t,1},$$
$$x_{t,2}  =  -0.75 x_{t-1,2} + w_{t,2}, $$
$$x_{t,3} =  -0.81 x_{t-2,3} + w_{t,3},$$
where $w_{t,1}, w_{t,2}, w_{t,3}$ are standard Gaussian white noise. From the perspective of traditional spectral analysis, the series $\{x_{t,1}\}$ has a lowpass spectrum, $\{x_{t,2}\}$ has a highpass spectrum, and $\{x_{t,3}\}$ has a bandpass spectrum around frequency $0.25$. The mixing function $W_1(x)$ is equal to 0.9 for $x < -0.8$, 0.2 for $x > 0.8$, and a linear transition for $x$ in between. The mixing function $W_2(x)$ is similarly defined except that it equals 0.5 for $x < -0.4$ and 1 for $x > 0$. Figure \ref{fig-asy} (a) shows the EPs of model (\ref{eq-mix}), where the EP is asymmetric across the expectile levels. Figure \ref{fig-asy} (b) compares the PGs and the EPs at $\alpha=0.1$ and $0.9$, illustrating that the PG is restricted to capturing spectral features near the central expectiles. In contrast, the EP offers a broader perspective, effectively analyzing the time series across the entire range of  $\alpha \in (0, 1)$.

\subsection{Fisher's Test}
One commonly used hypothesis test to detect periodicities for the PG is Fisher's test \citep{brockwell1991time}. The null hypothesis is that the time series is Gaussian white noise and the alternative hypothesis is that the time series contains at least one deterministic periodic component.   For frequencies $\{\omega_{\nu_1},\omega_{\nu_2},...,\omega_{\nu_q}\}$, the test statistic is defined as
$$
\text{g}_{{Fisher}} = \frac{\max_{1\leq k \leq q} \{I_n(\omega_{\nu_k})\} }{\sum_{k=1}^qI_n(\omega_{\nu_k})}.
$$  
A sufficiently large value of 
${\rm g}_{{Fisher}}$ indicates the presence of a hidden periodicity. Under the null hypothesis, each periodogram ordinate $ I_n(\omega_{\nu_k})$ is asymptotically distributed as a $\chi^2$ random variable (equivalently, a scaled exponential random variable). With $S := \sum_{k=1}^{q} I_n(\omega_{\nu_k})$, the vector of normalized periodogram ordinates
\[
\left[ \frac{I_n(\omega_{\nu_1})}{S}, \frac{I_n(\omega_{\nu_2})}{S}, \dots, \frac{I_n(\omega_{\nu_q})}{S} \right]
\]
follows a symmetric Dirichlet(1,...,1) distribution. Then,
${\rm g}_{{Fisher}} = \max_{1\leq k \leq q} \frac{I_n(\omega_{\nu_k})}{S},$
and the tail probability derived by \cite{fisher1929tests} is 
\[
P({\rm g}_{{Fisher}} > x) = \sum_{k=1}^{\lfloor 1/x \rfloor} (-1)^{k-1} \binom{q}{k} (1 - kx)^{q-1},
\]
where $\lfloor \cdot \rfloor$ denotes the floor. We apply Fisher's test to the EP by replacing $I_n(\omega_{\nu_k})$ with ${\rm EP}_{n}(\omega_{\nu_k},\alpha)$ in the test statistic. The probabilities of detection are obtained by 5,000 Monte Carlo simulations for time series ($n=200$) defined by model (\ref{hidden}), with $\omega_{\nu_0} = 0.1 \times 2 \pi, \omega_{\nu_c}= 0.3 \times 2\pi,$ and $b_2 = 0.$ As can be seen in Table \ref{tb1}, both the EP and QP outperform the PG. At a significance level of 0.05, the EP ($\alpha = 0.9$) achieves a detection rate of 84.26\%, whereas the PG achieves only 29.78\%. Detection rates are sensitive to the expectile level, showing a clear trend in this experiment.  As the expectile or quantile approaches 1, the detection rate of the EP increases and eventually surpasses that of the QP. 
\begin{table}[ht]
	\begin{center}
		
		\caption{Fisher's test of different types of periodograms.}
		\label{tb1}
		\begin{tabular}{cccccccc}
			\toprule[1.2pt]
			\multirow{2}{*}{\begin{tabular}[c]{@{}c@{}}Significa- \\nce level\end{tabular}} & \multicolumn{3}{c}{EP} & \multicolumn{3}{c}{QP}& \multirow{2}{*}{\begin{tabular}[c]{@{}c@{}} PG \end{tabular}}  \\
			
			& $\alpha$=0.85    & $\alpha$=0.9    & $\alpha$=0.95 & $\alpha$=0.85    & $\alpha$=0.9    & $\alpha$=0.95  &         \\\hline
			0.01   & 0.4048    & 0.5608    & 0.5850   &0.6898 &0.6328 & 0.4428  & 0.1224   \\
			0.05   & 0.7158    & 0.8426    & 0.8510   &0.8720 &0.8260 & 0.6646  & 0.2978   \\
			0.10   & 0.8308    & 0.9262    & 0.9306   &0.9278 &0.8952 & 0.7678  & 0.4258   \\\toprule[1.2pt]

		\end{tabular}
	\end{center}
	
\end{table}

\subsection{Smoothed EP}
\begin{figure}[ht]
	\centering
	\subfigure[]{\includegraphics[width=0.4\textwidth]{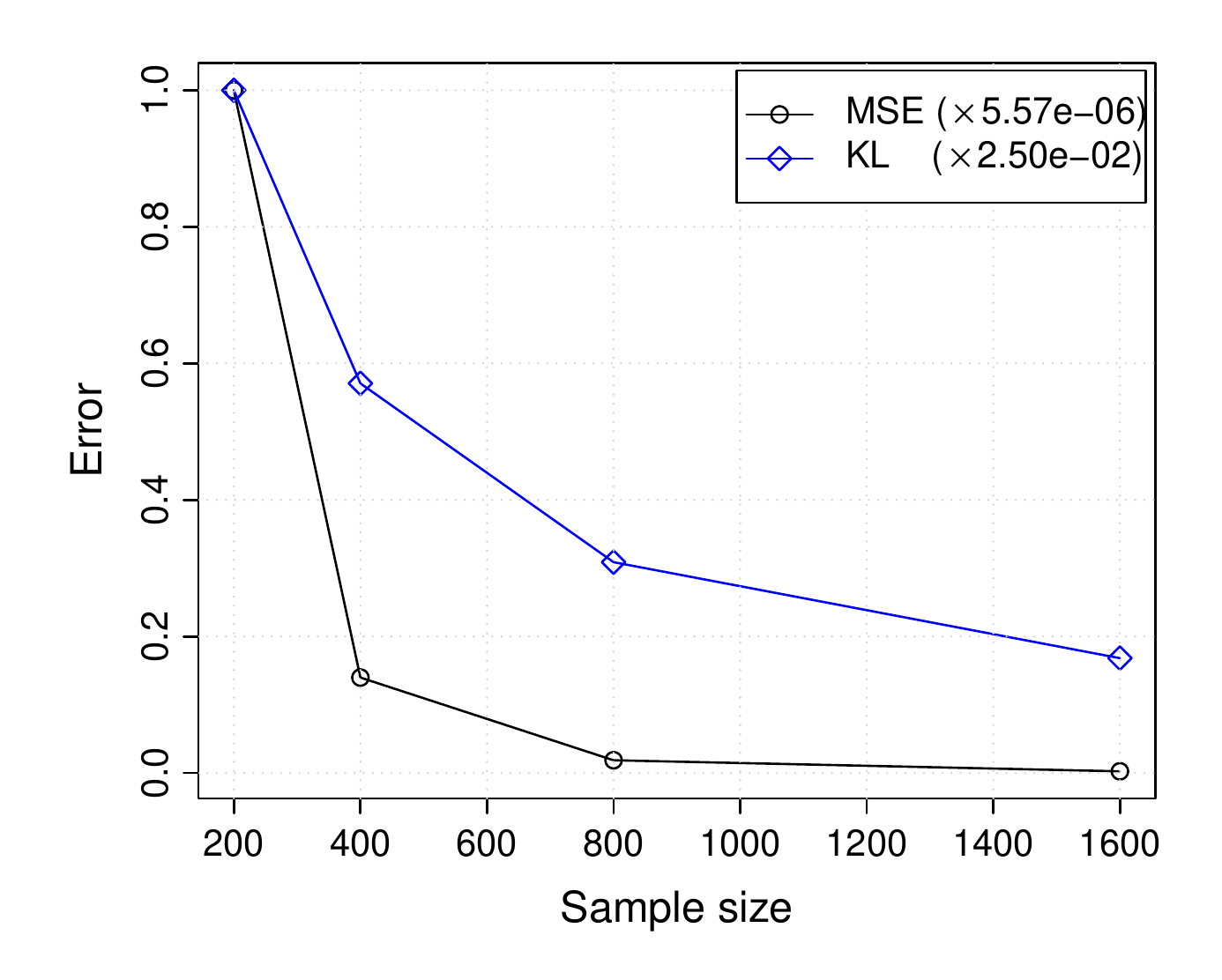}}
	\subfigure[]{\includegraphics[width=0.4\textwidth]{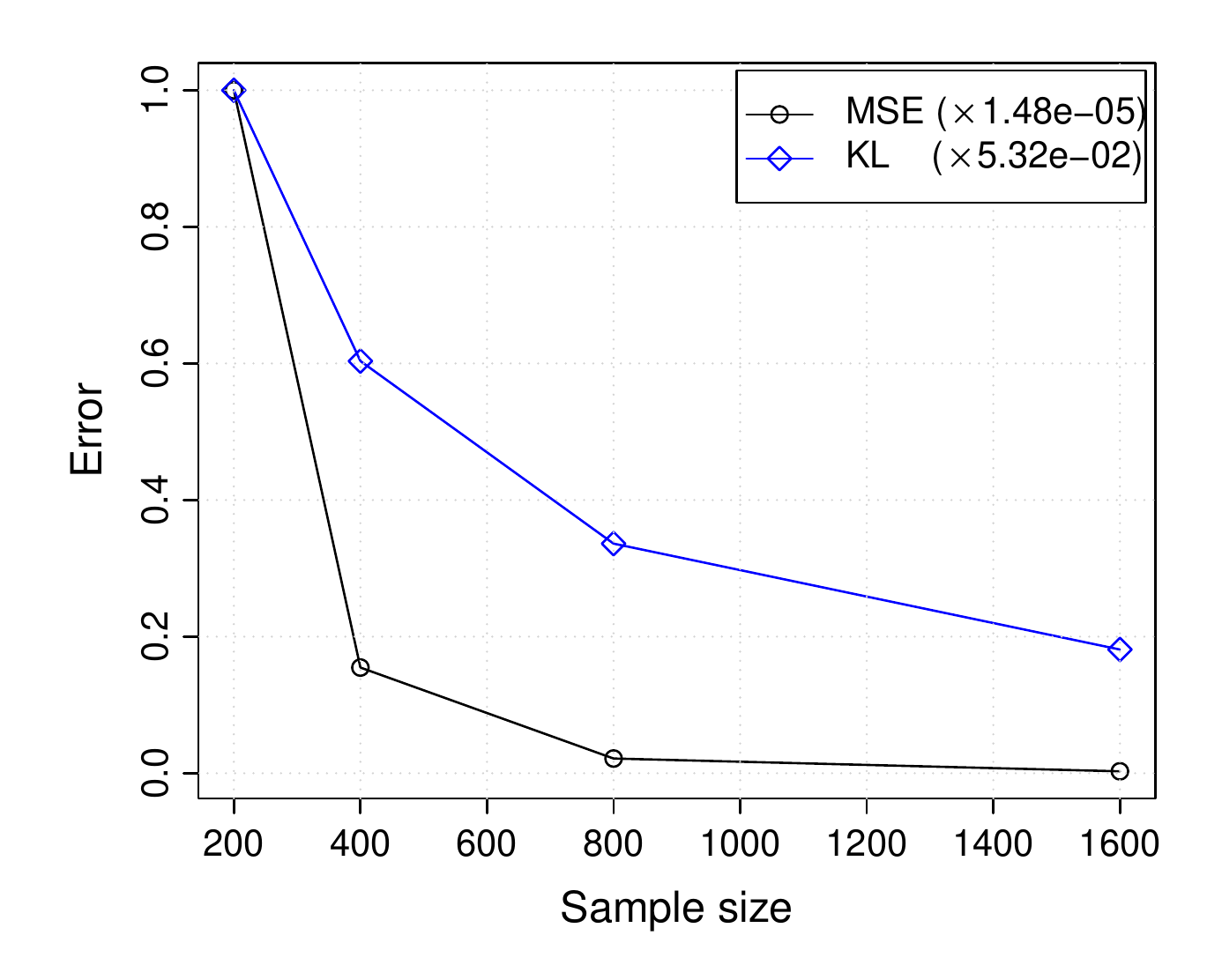}} 
	
	\caption{MSE and KL divergence of smoothed periodogram. (a) AR(2) model (\ref{ar2}); (b) mixture model (\ref{eq-mix}). The results are based on 5,000 simulation runs.} 
	\label{mse}
\end{figure}

It is well known that the power spectrum can be estimated consistently by a properly smoothed PG \citep{brockwell1991time}. Theorem 2 suggests that the expectile spectrum, defined by (\ref{es}), can be estimated non-parametrically by smoothing the raw EP ordinates at the Fourier frequencies in the same way. 
The simulation studies confirm that the estimation accuracy increases as $n$ increases. We measure the distance between the smoothed EP and the expectile spectrum (the average of 5,000 smoothed EPs) using the mean squared error (MSE) and Kullback-Leibler (KL) divergence \citep{kullback1951information}. Models (\ref{ar2}) and (\ref{eq-mix}) at $\alpha=0.9$ are used for illustration. We smooth the raw EP for a given expectile level.
Figure \ref{mse} depicts the MSE and KL divergence as a function of $n$, with $n$ taking values of 200, 400, 800, and 1,600. To highlight the decreasing trend as $n$ increases, the values are scaled by setting the maximum to 1.

\subsection{Comparison with the Quantile Periodogram}
\begin{figure}[]
	\centering
	\subfigcapskip = -0.3cm
	\subfigure[]{\includegraphics[width=0.38\textwidth]{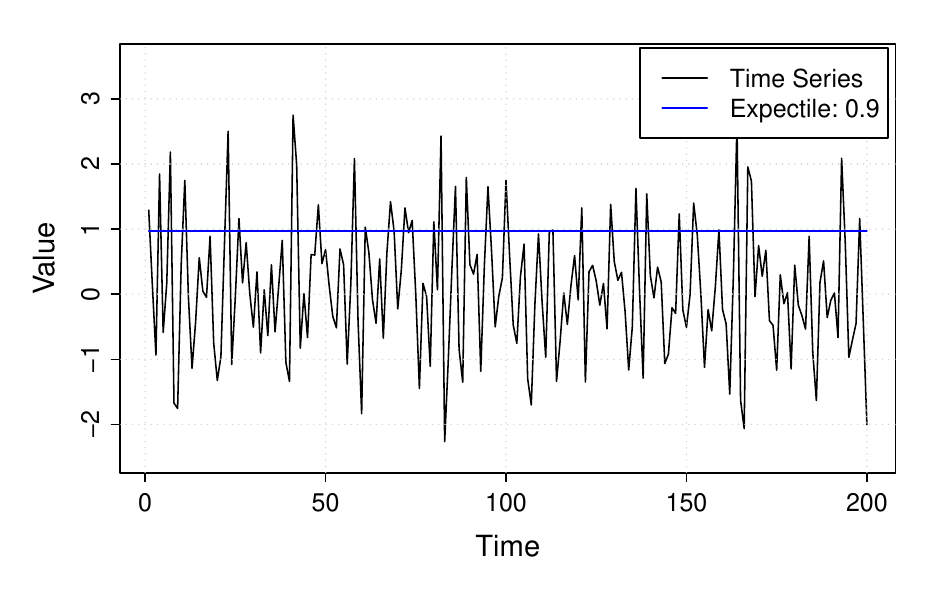}}\vspace{-3.1mm}
	\subfigure[]{\includegraphics[width=0.39\textwidth]{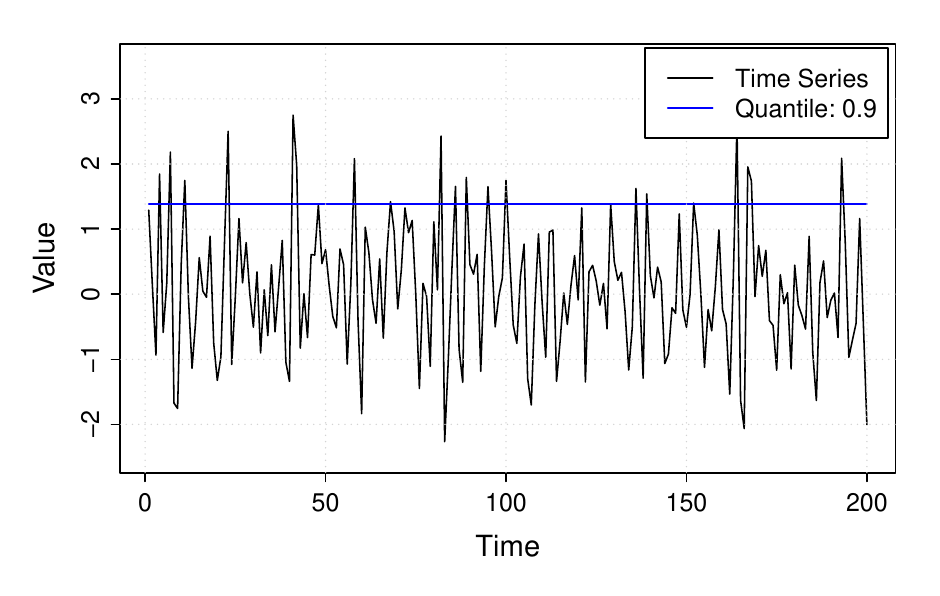}}
	\subfigure[]{\includegraphics[width=0.39\textwidth]{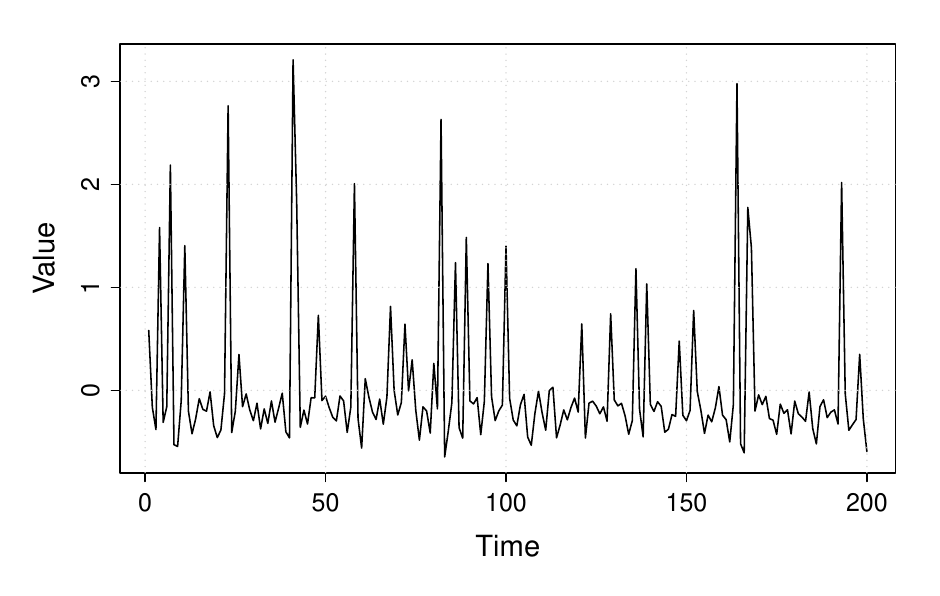}}\vspace{-3.1mm}
	\subfigure[]{\includegraphics[width=0.39\textwidth]{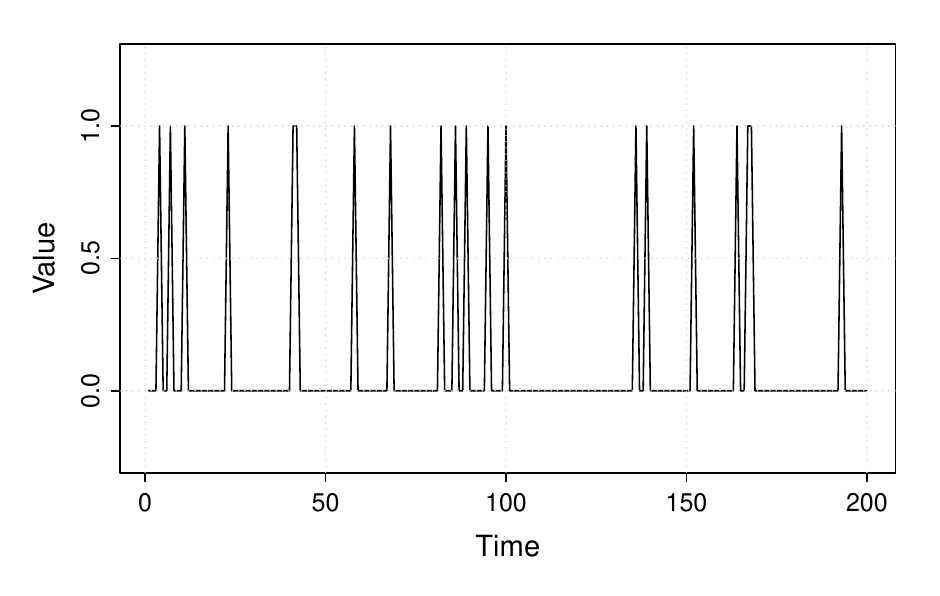}}
	\subfigure[]{\includegraphics[width=0.39\textwidth]{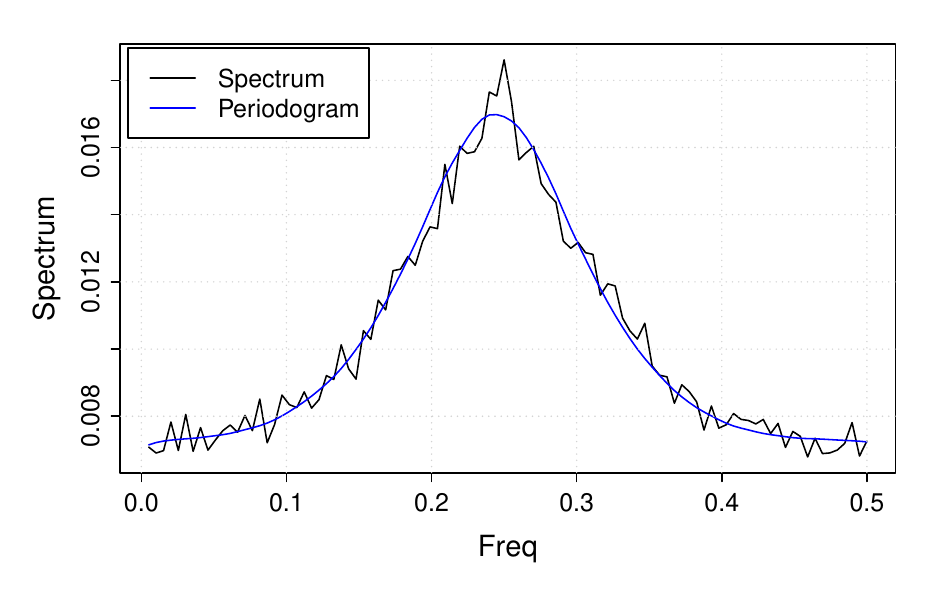}}\vspace{-3.1mm}
	\subfigure[]{\includegraphics[width=0.39\textwidth]{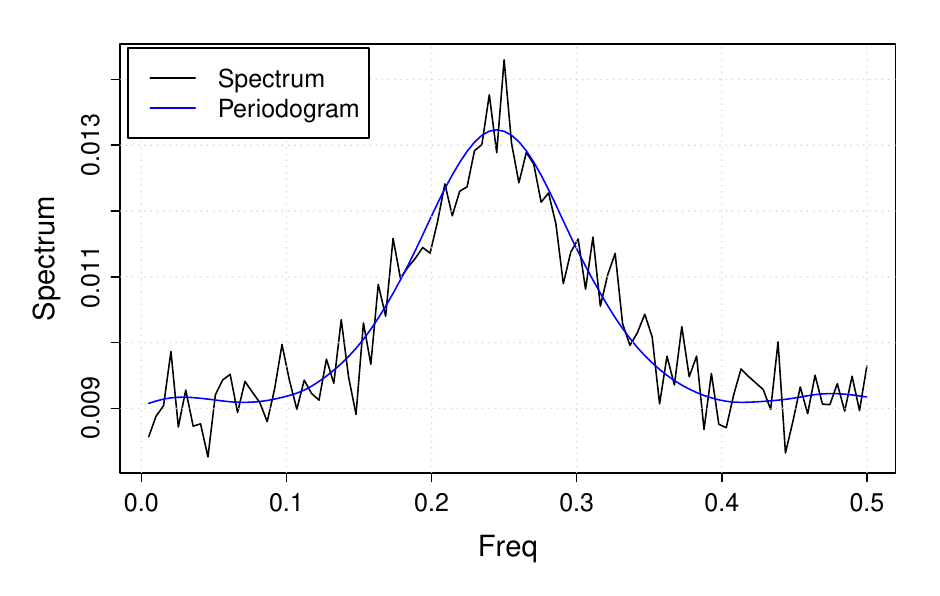}}	
	\subfigure[]{\includegraphics[width=0.39\textwidth]{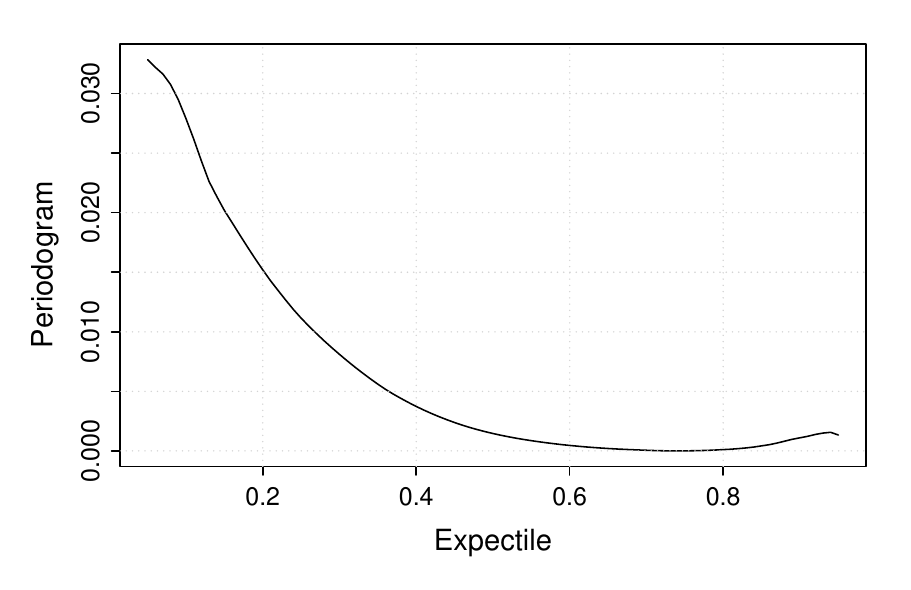}}\vspace{-3.1mm}
	\subfigure[]{\includegraphics[width=0.39\textwidth]{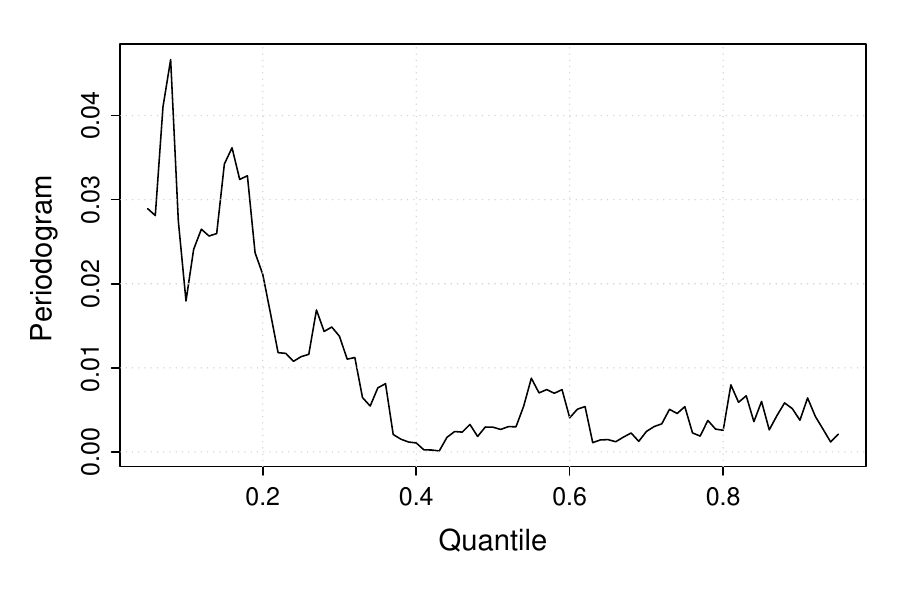}}			
	\caption{Left panel: expectile; right panel: quantile. First row: time series and their corresponding 0.9 expectile and quantile; second row: the ASECP and LCP of the time series; third row: the spectra and the averaged periodograms of 500 realizations; fourth row: the periodograms across the expectile and quantile levels at $\omega = 0.1 \times 2\pi$ of one realization.  The spectra are computed by averaging 5,000 smoothed periodograms.}
	\label{asec}
\end{figure}

Besides computational efficiency, another advantage compared with the QP can be seen in Figure \ref{asec}, where the top row illustrates a realization of model (\ref{ar2}) with $\omega_{\nu_c} = 0.25 \times 2 \pi$, along with the corresponding 0.9 expectile and quantile in the second row, respectively. As shown in Figure \ref{asec} (c) and (d), the ASECP contains more information, whereas the LCP loses information when transforming a real-valued time series into a binary-valued process.  We also observe that the averaged EP exhibits smaller divergence from the true spectrum and lower sampling variability than the averaged QP, as illustrated in Figure \ref{asec}(e) and (f), where we average 500 periodograms and compare the results with the true spectrum.
	
\begin{figure}[ht]
	\centering
	\subfigcapskip = -0.3cm
	\subfigure[]{\includegraphics[width=0.38\textwidth]{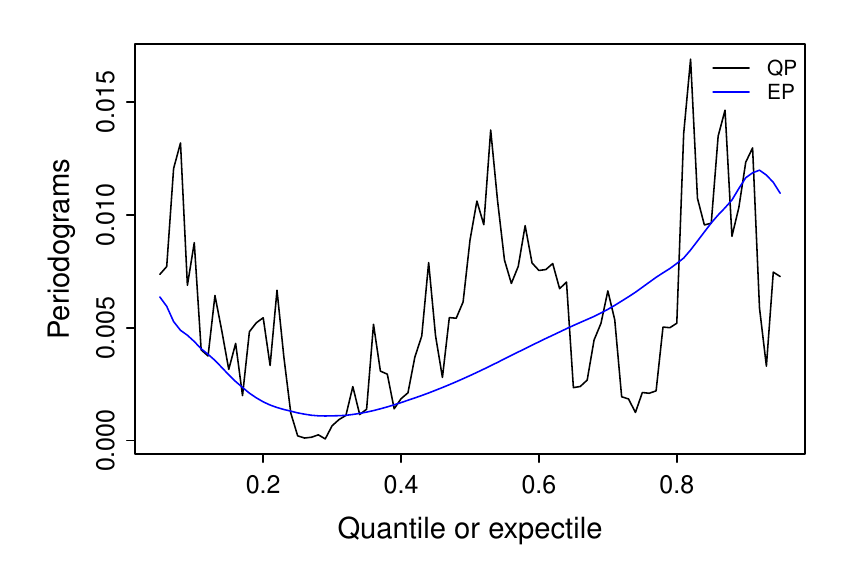}}\vspace{-3.1mm}
	\subfigure[]{\includegraphics[width=0.39\textwidth]{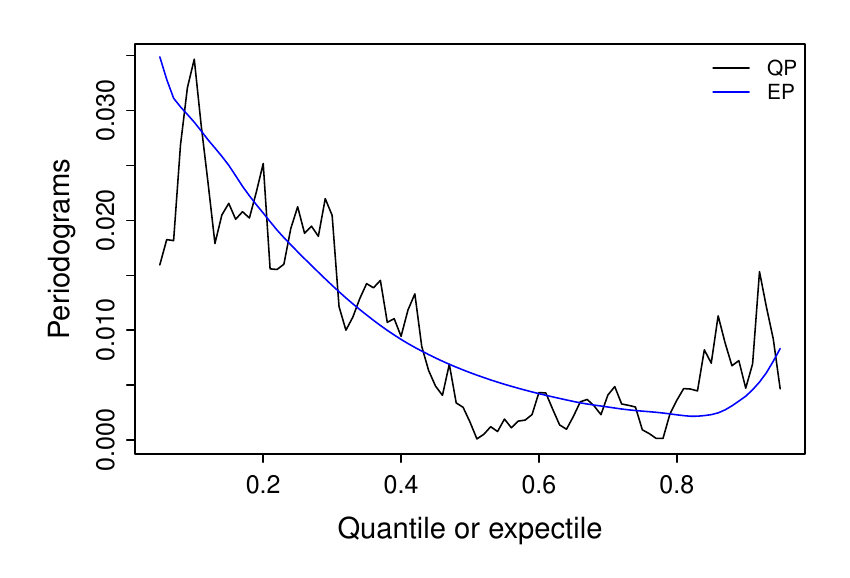}}
	\subfigure[]{\includegraphics[width=0.39\textwidth]{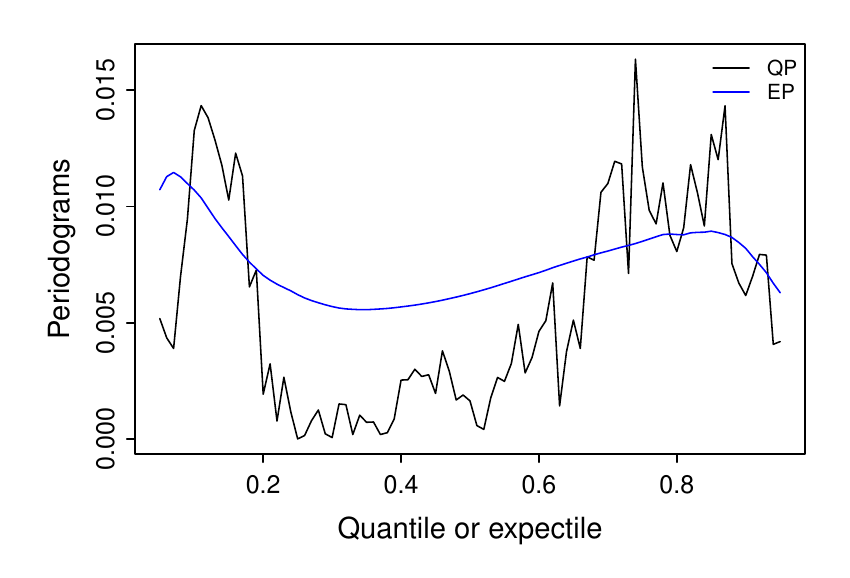}}\vspace{-3.1mm}
	\subfigure[]{\includegraphics[width=0.39\textwidth]{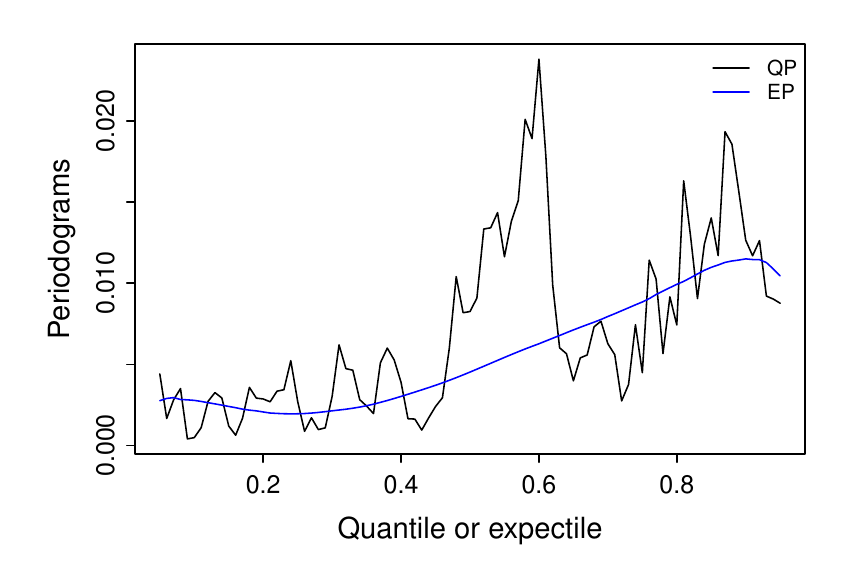}}
	\caption{The EP across the expectile levels and the QP across the quantile levels. Left panel: model (\ref{gar}); right panel: model (\ref{eq-mix}). Top panel: $\omega = 0.05\times 2\pi$; bottom panel: $\omega = 0.45 \times 2\pi$. The expectile or quantile levels we use are $0.05,0.06,\dots, 0.95$.}
	\label{smooth}
\end{figure}

Additionally, the EP is smoother across the expectile levels than the QP is across the quantile levels, as shown in Figure \ref{asec} (g) and (h). We quantified this smoothness in two ways. First, we computed the variances $\mathrm{Var}_{\alpha}\{\mathrm{EP}(0.1\times 2\pi, \alpha)\}$ and $\mathrm{Var}_{\tau}\{\mathrm{QP}(0.1\times 2\pi, \tau)\}$ for each realization. The average variance of the 500 realizations was $1.492 \times 10^{-5}$ for the EP and $3.932 \times 10^{-5}$ for the QP. Second, we computed the roughness based on the squared second-order differences of the periodograms at $\omega =0.1\times 2\pi$ for each realization. The average roughness across the 500 realizations was $0.028$ for the EP and $0.159$ for the QP.
The reason follows from the fact that expectile regression is based on a smooth, differentiable quadratic loss, which leads to estimators that vary smoothly with respect to $\alpha$. In contrast, quantile regression relies on the non-differentiable check loss, resulting in piecewise-linear and potentially kinked trajectories across quantile levels. Figure \ref{smooth} further validates the conclusion using models (\ref{gar}) and (\ref{eq-mix}).
\subsection{Computation details}
All computations were performed in R, except for training the deep learning model in Section 5. ER were solved using an iterative weighted least squares algorithm using the \texttt{expectreg.ls} function in package \texttt{expectreg}. The smoothing of periodograms was conducted via the \texttt{smooth.spline} function with the smoothing parameter \texttt{spar} automatically selected by cross-validation (\texttt{cv = TRUE}). Parallel computation was employed through the \texttt{foreach} and \texttt{doParallel} packages 
to speed up the computation; however, only a single thread was used when recording computation time. The computations were performed  on a workstation equipped with an Intel Core i9-13900KF CPU @5.2GHz, 64 GB of RAM, and an Nvidia RTX 4090 GPU.                  

\section{Real Data Application}\label{EC}
In this section, we first present two short examples in different scientific fields, in which the PG fails to capture full spectral features of the time series in Section \ref{example}. Then, we apply the EP to an earthquake classification problem in Section \ref{eq}.
\subsection{Two Examples}\label{example}
\begin{figure}[!p]
	\centering
	\vspace{-0.2cm}
	\subfigcapskip = -0.3cm
	\subfigure[]{\includegraphics[width=0.6\textwidth]{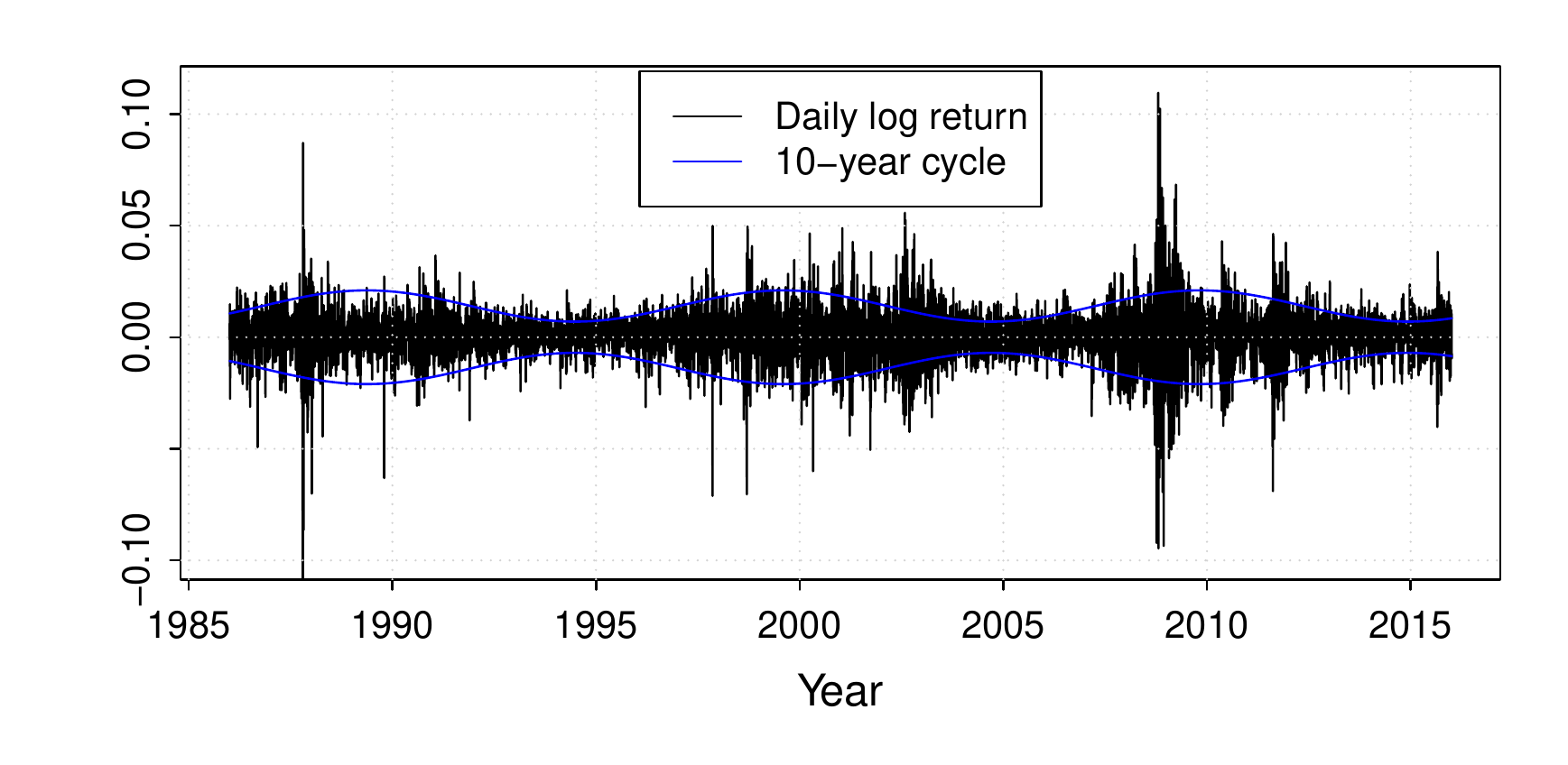}}\vspace{-3mm}
	\subfigure[]{\includegraphics[width=0.4\textwidth]{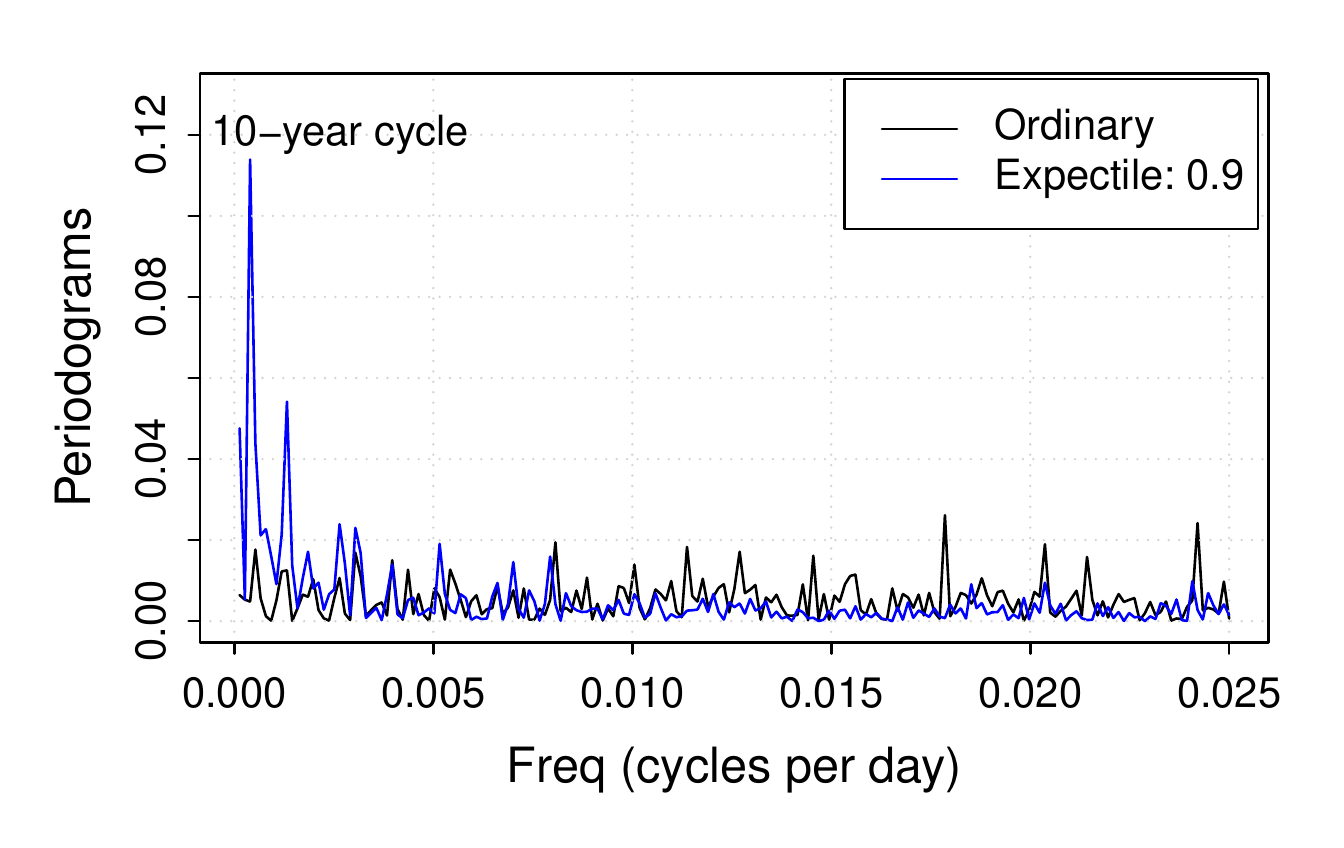}}\vspace{-2mm}
	\subfigure[]{\includegraphics[width=0.4\textwidth]{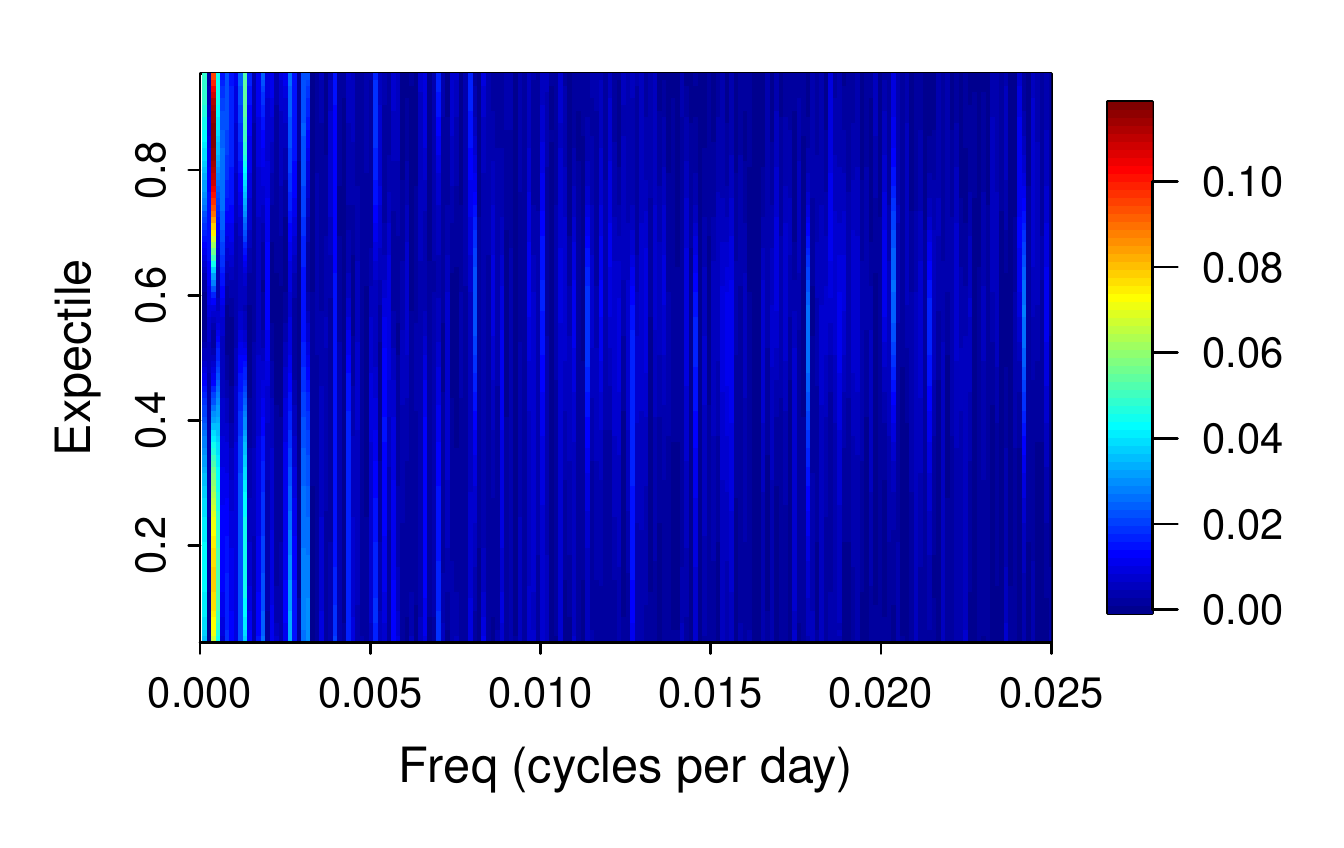}}\vspace{-2mm}
	\subfigure[]{\includegraphics[width=0.4\textwidth]{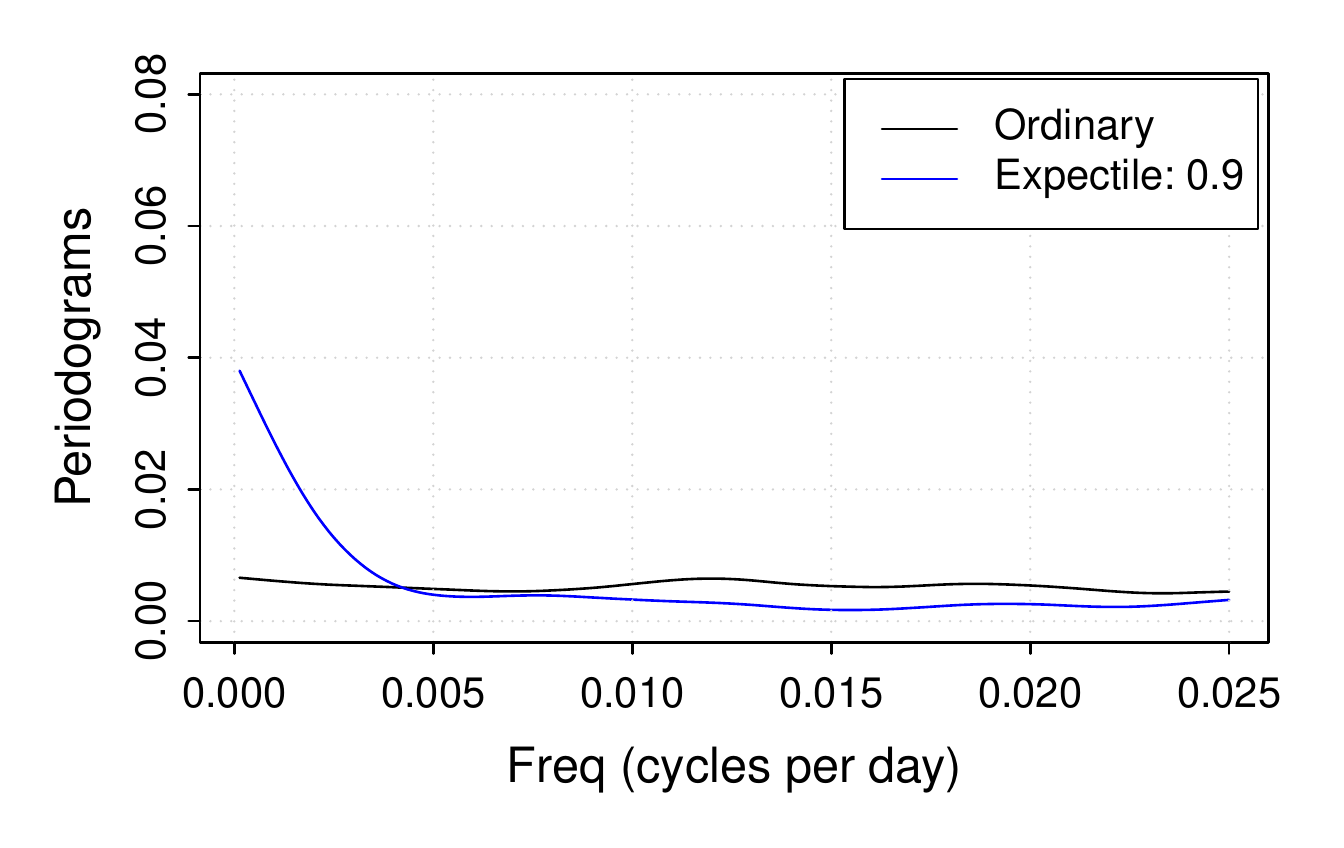}}\vspace{-2mm}
	\subfigure[]{\includegraphics[width=0.4\textwidth]{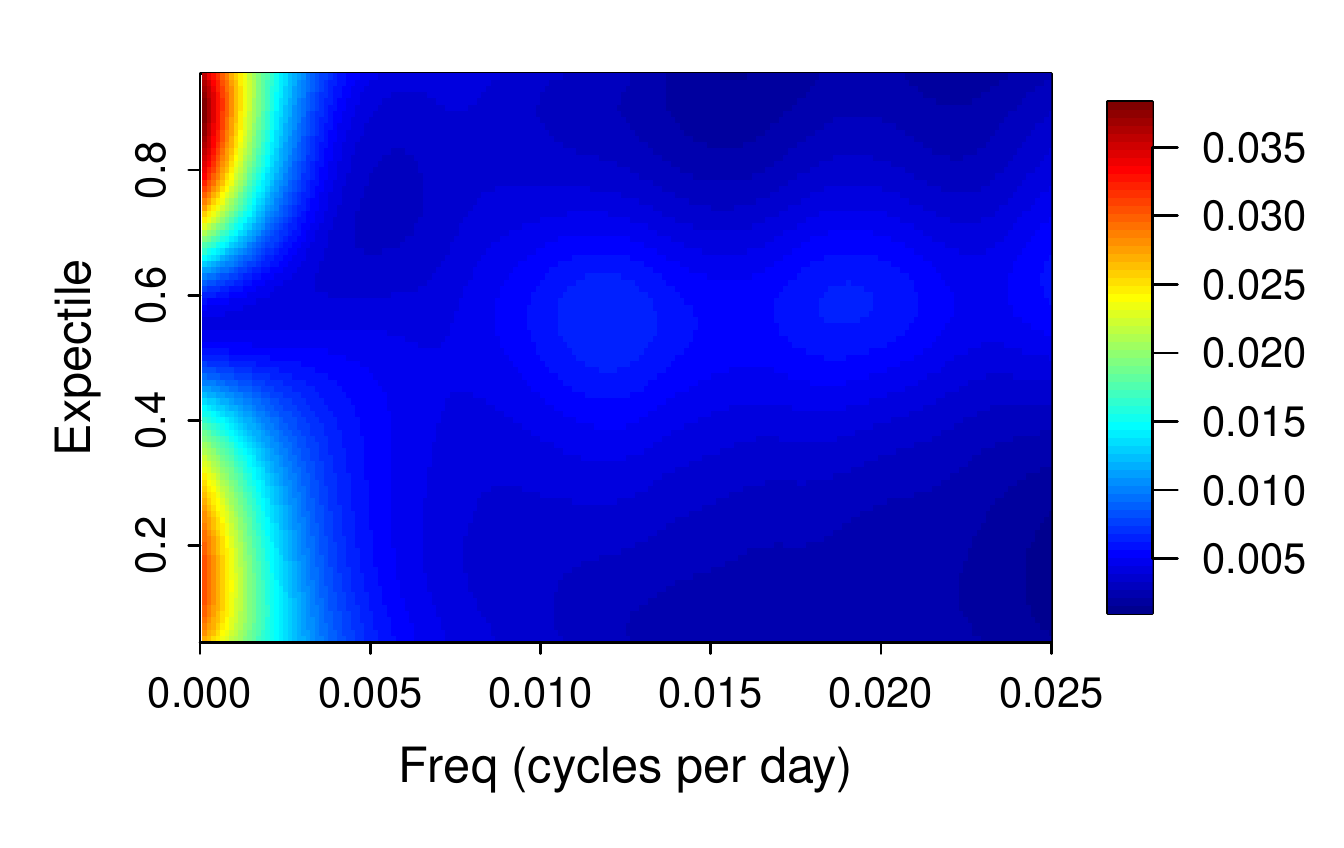}}\vspace{-2mm}
	\subfigure[]{\includegraphics[width=0.4\textwidth]{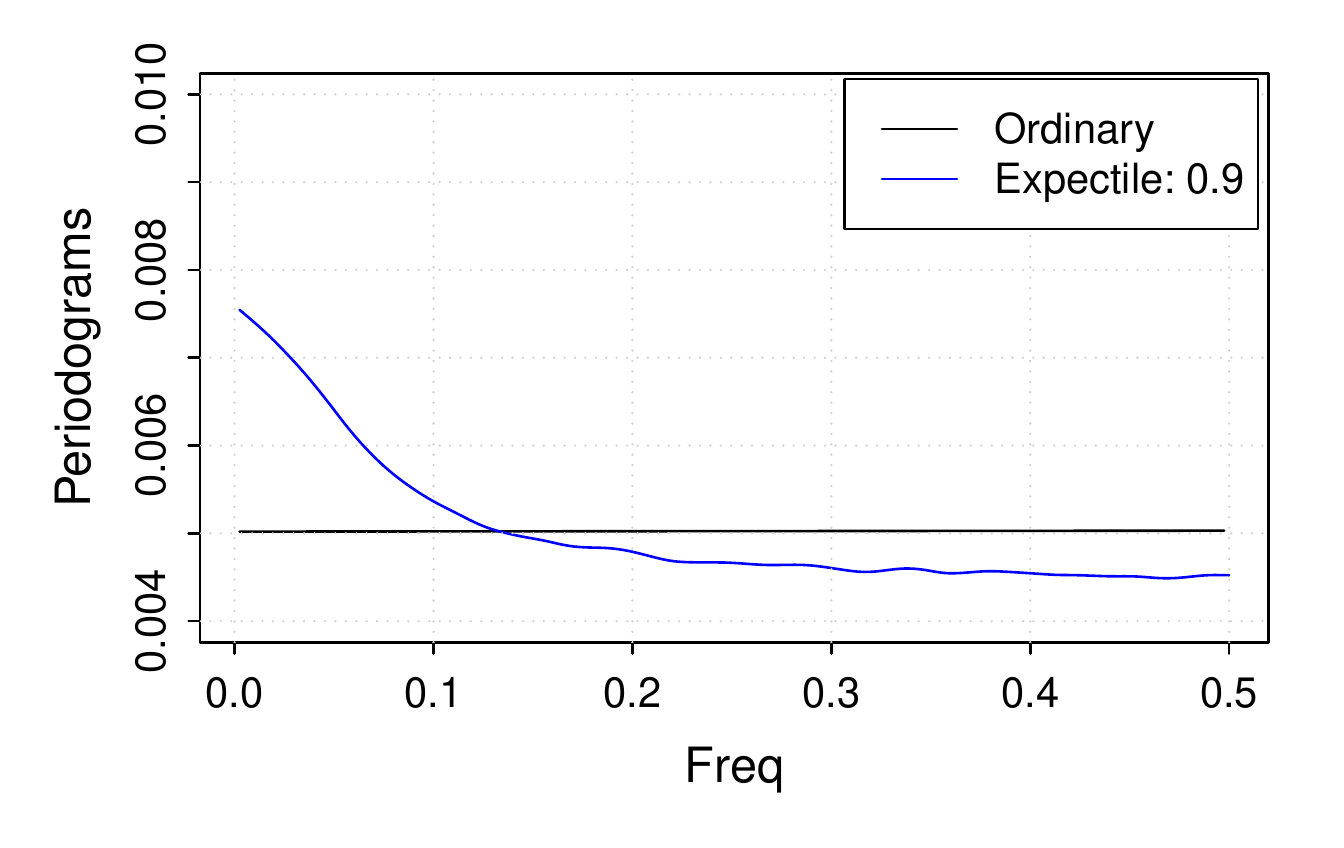}}\vspace{-2mm}
	\subfigure[]{\includegraphics[width=0.4\textwidth]{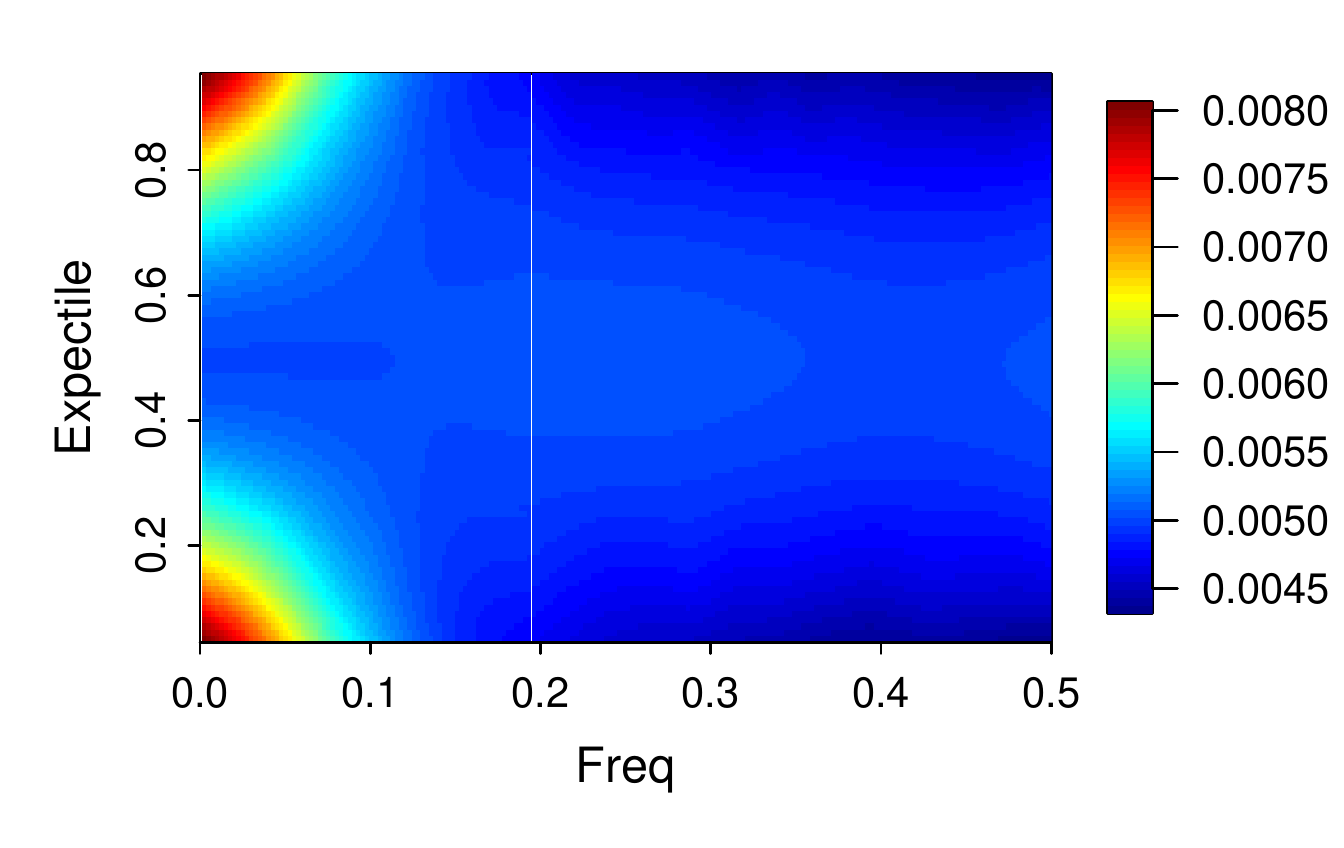}}\vspace{-2mm}
	
	\caption{(a) Daily log returns of the S$\&$P 500 Index data; (b) the PG and the EP at expectile $\alpha=0.9$; (c) the EP at expectiles $\{0.05, 0.06,...,0.95\}$; (d) and (e) the smoothed versions of (b) and (c), respectively; (f) the averaged PG and EP ($\alpha=0.9$) of model (\ref{gar}); and (g) the averaged EP of model (\ref{gar}) at expectiles $\{0.05, 0.06,...,0.95\}$. (f) and (g) are ensemble means of 5,000 realizations.}
	\label{sp}
\end{figure}

In the first example, we analyze the daily log returns of the S$\&$P 500 Index data from 1986 to 2015, as shown in Figure \ref{sp}(a). The data are publicly available at \url{finance.yahoo.com} or \url{fred.stlouisfed.org}. The EP in Figure \ref{sp}(b) successfully identifies the approximately 10-year cycle of market volatility (highlighted by the blue lines in Figure \ref{sp}(a)) at both lower and upper expectiles. In contrast, the PG yields a relatively featureless flat line. The smoothed periodograms in Figure \ref{sp}(d) and (e) exhibit a spectral feature similar to a GARCH(1,1) \citep{engle1982autoregressive,bollerslev1986generalized} model:
\begin{equation}\label{gar}
	y_t \sim N(0,\sigma_t^2),
\end{equation}
where $\sigma_t^2=10^{-6}+0.49y^2_{t-1}+0.49\sigma_{t-1}^2$.
We generate 5,000 realizations of model (\ref{gar}), and average the smoothed periodograms to represent the ground truth in Figure \ref{sp}(f) and (g). Notably, while the PG of model  (\ref{gar}) is a constant, the EPs detect the low-frequency feature at both the lower and higher expectiles.  This example demonstrates that the EP can reveal long-term periodicities that are invisible to the PG. 

\begin{figure}[ht]
	\centering
	\vspace{-0.5cm}
	\subfigcapskip = -0.3cm
	\subfigure[]{\includegraphics[width=0.4\textwidth]{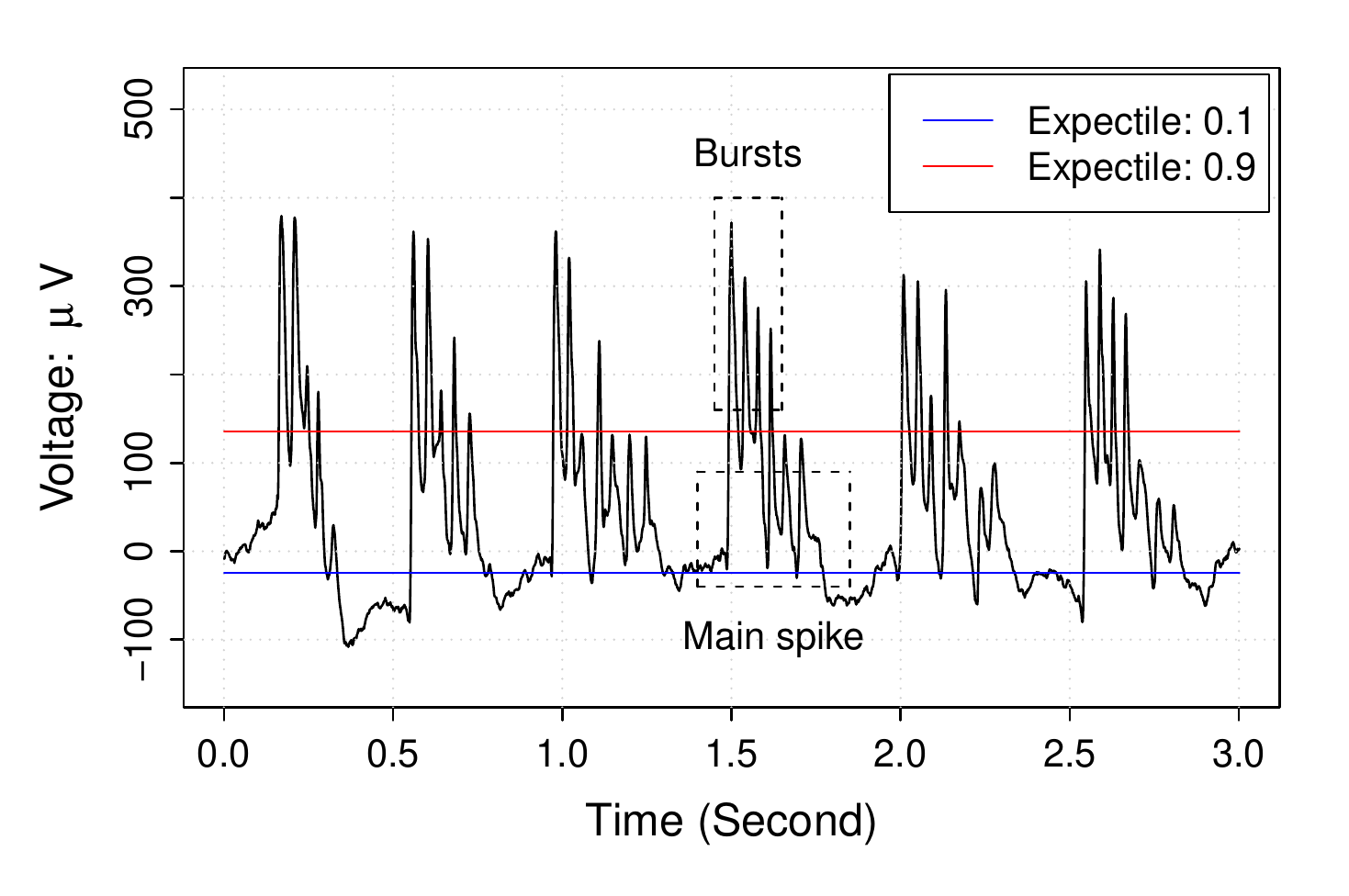}}\vspace{-3mm}
	\subfigure[]{\includegraphics[width=0.4\textwidth]{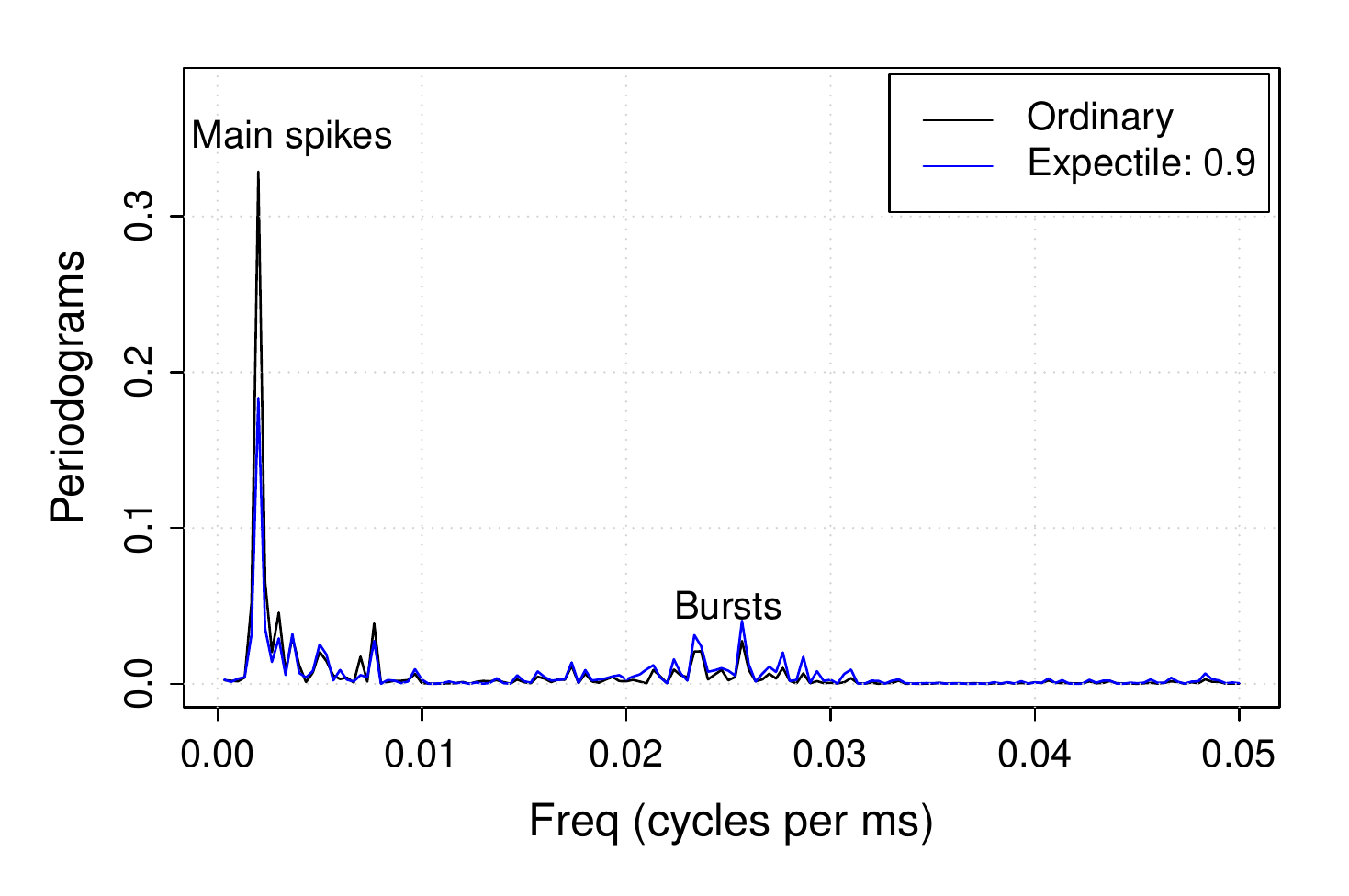}}
	\subfigure[]{\includegraphics[width=0.4\textwidth]{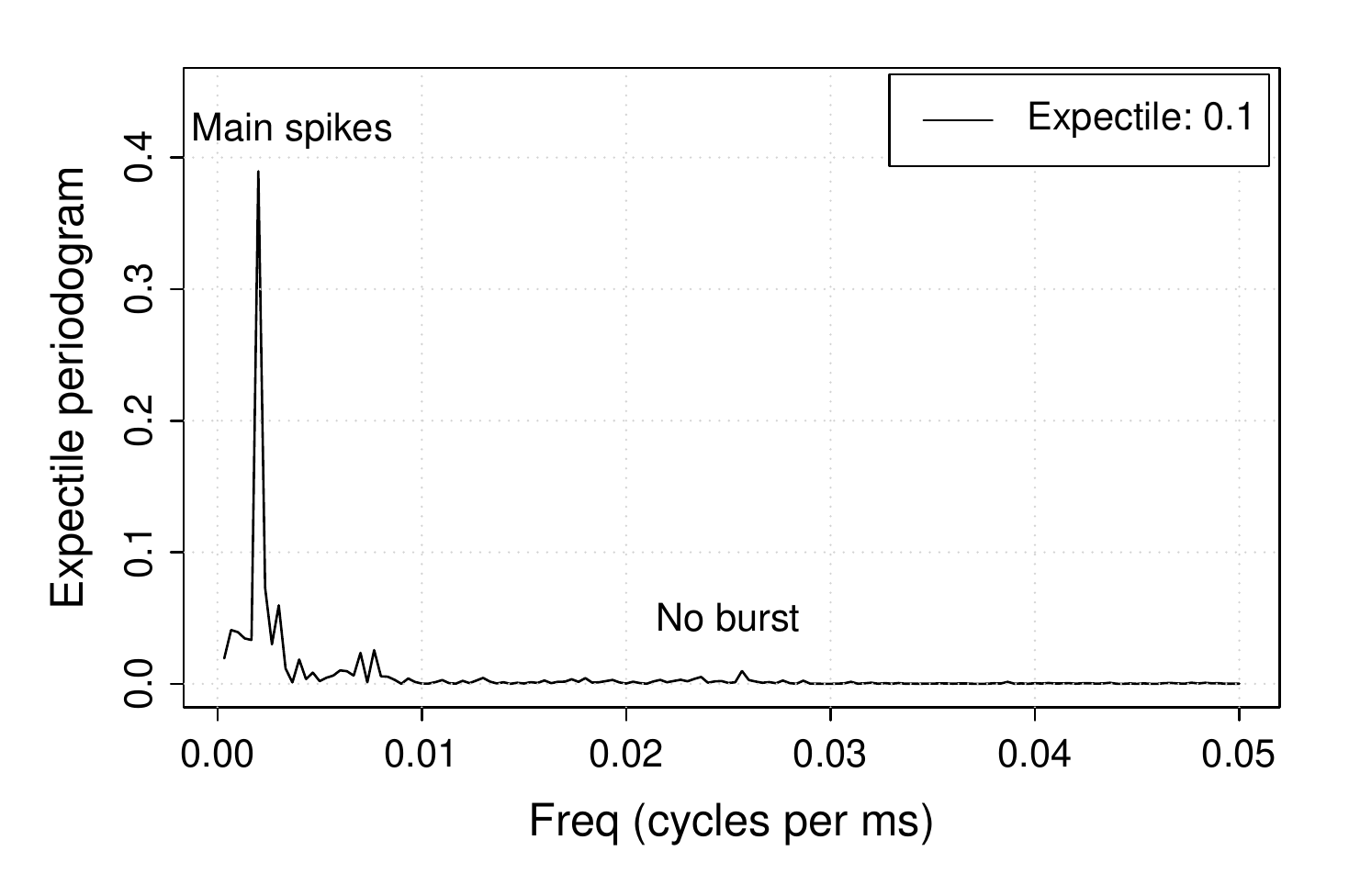}}
	\subfigure[]{\includegraphics[width=0.4\textwidth]{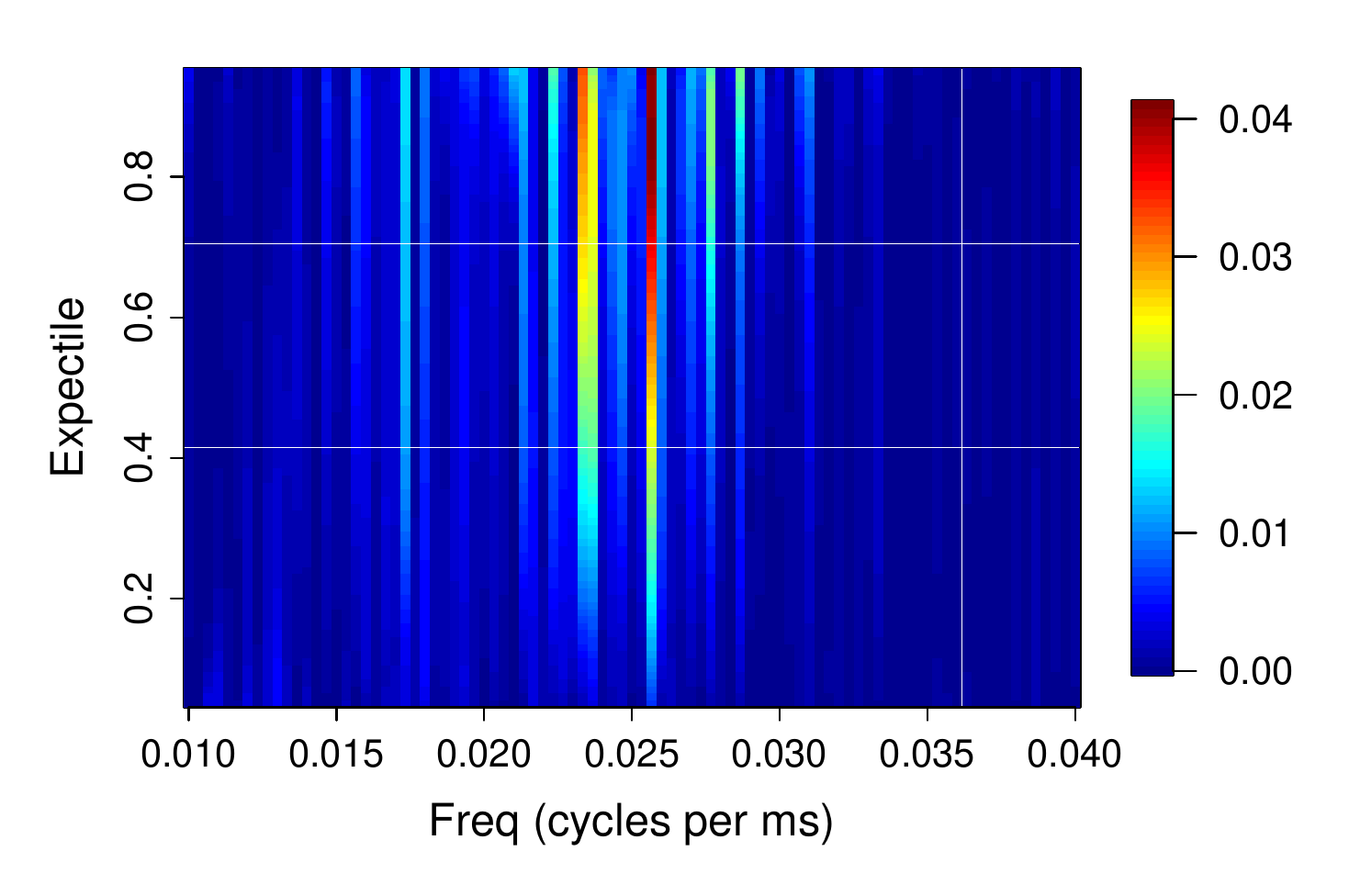}}
	\caption{(a) The EEG data over a 3-second interval, along with its sample expectiles 0.1 and 0.9; (b) the PG and EP ($\alpha = 0.9$); (c) the EP at $\alpha = 0.1$; and (d) the EP at expectiles $\{0.05, 0.06,...,0.95\}$. The time resolution of the EEG data is 1000 Hz.}
	\label{eeg}
\end{figure}

The second example is based on electroencephalogram (EEG) data  containing delta-burst activity, which were collected from an epilepsy patient during a seizure interval by South China Normal University School of Psychology and is authorized for use in this paper. Delta-burst activity in EEG signals is often used in the diagnosis and characterization of epileptic activity \citep{schmitt2012extreme,whitehead2017characteristics,li2023epileptogenicity}.
Figure \ref{eeg}(a) presents the EEG segments along with their corresponding sample expectiles. The PG in Figure \ref{eeg}(b) successfully detects both low-frequency and high-frequency components, corresponding to the six main spikes and the associated bursts, respectively. The EPs in Figure \ref{eeg}(b) and (c) provide additional information, demonstrating that the high-frequency patterns appear predominantly at higher expectile levels. This observation is consistent with the EEG data, where the bursts primarily occur at the top of the main spikes. Furthermore, the intensity of the bursts increases with higher expectiles, particularly at frequencies $\omega \in [0.02, 0.03]\times 2\pi$, as shown in Figure \ref{eeg}(d).  This example demonstrates that the EP is able to distinguish asymmetric burst activities across expectile levels, revealing structure that may be obscured in the PG.
\subsection{Earthquake Data Classification}\label{eq}
Section \ref{data} introduces the earthquake data and Section \ref{cnn} describes the deep learning model and presents the classification results.

\subsubsection{Data Description}\label{data}
The earthquake waveform data with a sampling rate of 100 Hz, were collected in February 2014 in Oklahoma State. These data are available at \url{https://www.iris.edu/hq/} and \url{http://www.ou.edu/ogs.html}. Details about the catalog data are provided in \cite{benz2015hundreds}, including labels for earthquake magnitudes and occurrence times.  We extract 2,000 non-overlapping segments, each being a time series with $n=2,000$, corresponding to 20 seconds of data. The choice of longer time series ensures that the segments contain complete earthquake events. Among these segments, 1,000 contain an earthquake with magnitudes greater than 0.25, while the remaining 1,000 segments contain no earthquakes. We smooth the raw EPs of the 2,000 segments using the semi-parametric method proposed in \cite{chen2021semi}, ensuring smoothness across both the expectile and frequency dimensions. Specifically, we apply an AR spectrum approximation at each expectile level to smooth the raw EP across $\omega$. Then, a non-parametric method is used to smooth the raw EP across $\alpha$.  In this experiment, we use the lower half of the frequencies ($\omega \leq 0.25\times 2\pi$) and 46 expectiles \{$0.05,0.07,...,0.93,0.95$\}. Since we focus on serial dependence and normalize the EPs, amplitude considerations are excluded, making the classification more challenging. Additionally, we incorporate two competitive periodograms: the smoothed PG and the QP.

\begin{figure}[]
	\centering
	\subfigcapskip = -0.3cm
	\subfigure[]{\includegraphics[width=0.8\textwidth]{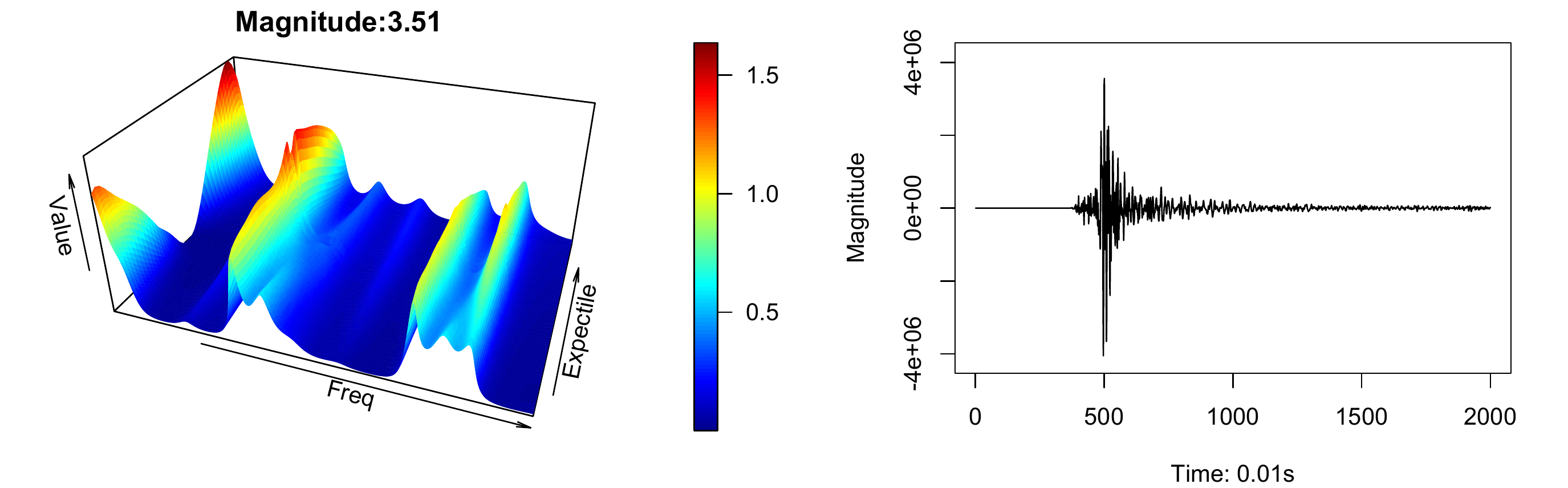}}\vspace{-3mm}
	\subfigure[]{\includegraphics[width=0.8\textwidth]{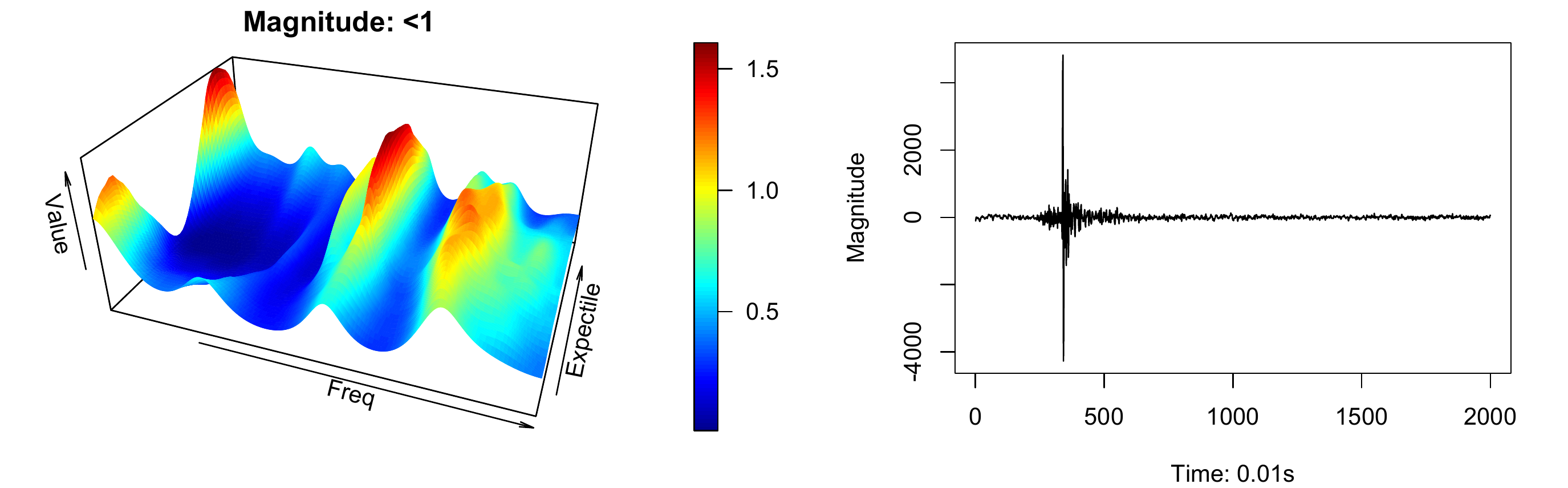}} \vspace{-3mm}
	\subfigure[]{\includegraphics[width=0.8\textwidth]{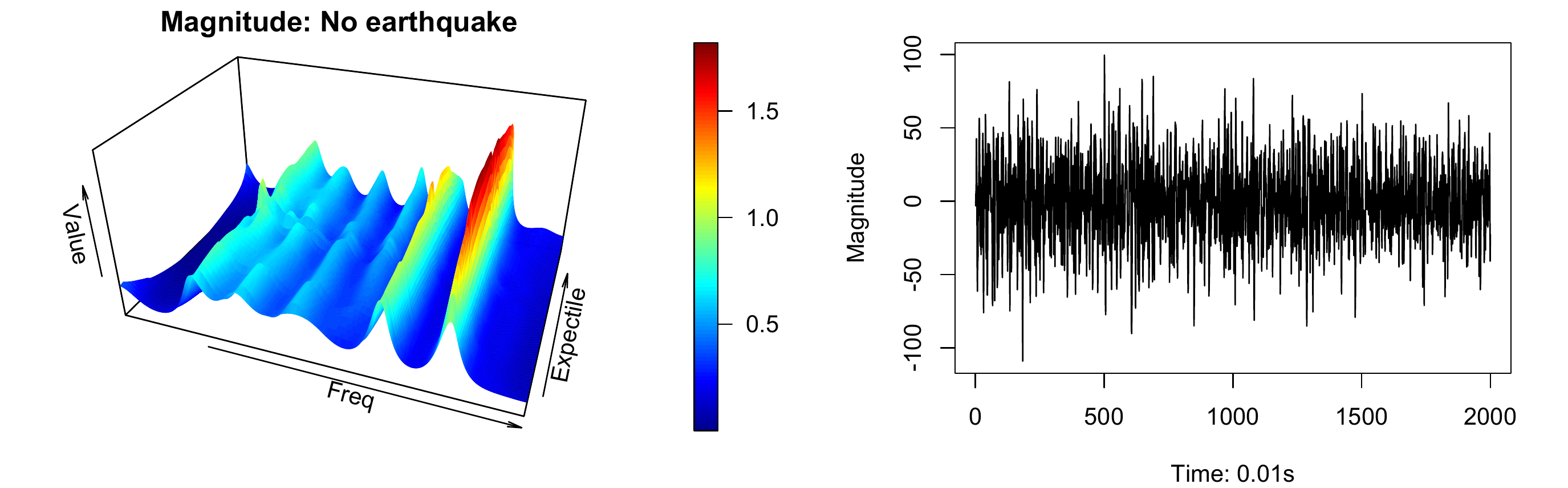}} 
	
	\caption{Three segments and the corresponding smoothed EPs. (a) The segment with an earthquake with a magnitude $>3$, (b) the segment with an earthquake with a magnitude $<$1, and (c) the segment with no earthquake. $n = 2000$, $\omega \leq 0.25 \times 2 \pi$, and $\alpha = 0.05, 0.07, ...,0.95$.  ``E-PER" denotes the EP.}
	\label{fig-3epo}
\end{figure}

We show three representative segments along with their corresponding smoothed EPs in Figure \ref{fig-3epo}. Specifically, Figure \ref{fig-3epo}(a) contains a large earthquake with a magnitude larger than 3, Figure \ref{fig-3epo}(b) contains a relatively small earthquake with a magnitude less than 1, and Figure \ref{fig-3epo} (c) contains no earthquake. Based on the three segments, we observe the following features: 
\begin{itemize}
	\item The smoothed EP for the segment with a large earthquake exhibits spectral peaks at the low-frequency band at both  higher and lower expectiles.
	\item The smoothed EP for the segment with a small earthquake exhibits spectral peaks at both the low-frequency band (at lower and higher expectiles) and the high-frequency band (at middle expectiles).
	\item The smoothed EP for the segment with no earthquake exhibits peaks only in the high-frequency bands.
	
\end{itemize}

\subsubsection{Classification using Deep Learning Model}\label{cnn}
\begin{figure}
	\centering
	\includegraphics[width=0.9\textwidth]{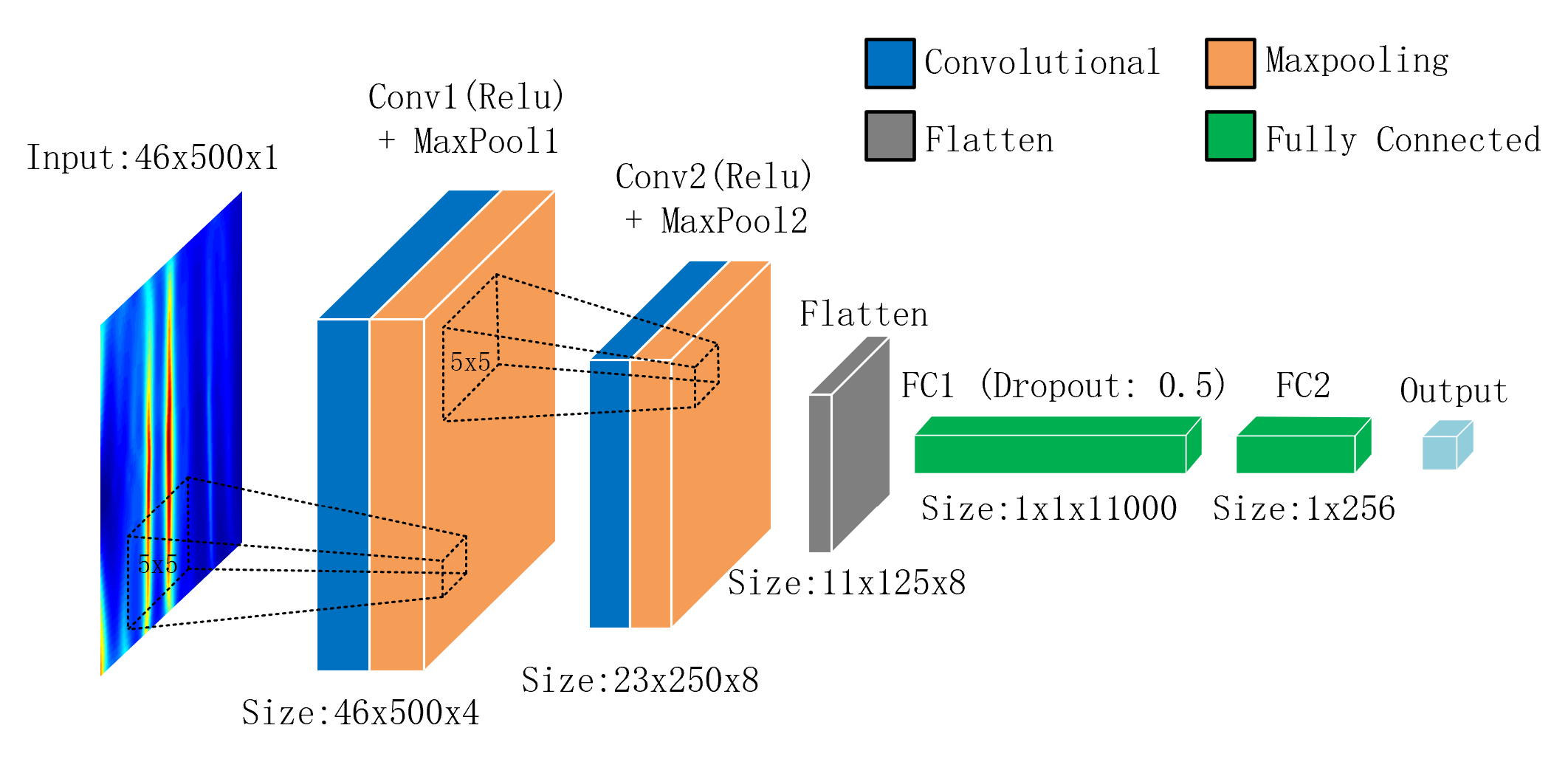}
	\caption{Structure of the deep learning model.}
	\label{fig-model}
\end{figure}

In this section, we use the three types of smoothed periodograms as features to classify the segments into those that contain earthquakes and those that do not. We randomly split the segments into training and testing sets, comprising 1,600 and 400 samples, respectively. To classify the EPs and QPs, we employ a model with two convolutional layers for feature extraction, each followed by a max-pooling layer. The second pooling layer connects to a fully connected (FC) layer after flattening the output. A dropout layer with a rate of 50\% is applied to the FC layer, leading to the output layer.  The total number of trainable parameters is 2,817,682, and the learning rate is set to $10^{-4}$ with a reduction rate of 0.5 every 20 epochs. Additional details about the model are presented at \url{https://github.com/tianbochen1/Expectile-Periodogram}. The model structure is illustrated in Figure \ref{fig-model}. To classify the PGs, which are one-dimensional with respect to $\omega$, we adapt the model to use different input dimensions ($1\times 500$ instead of $46\times 500$) and kernel size ($1\times 5$ instead of $5\times 5$). 

We conduct the training ten times, each with a randomly constructed training–testing split and randomly initialized weights. Over 80\% of the training runs converged within 30 epochs. The testing accuracies for the three types of periodograms are as follows: 
\begin{itemize}
	\item EP: \{0.9900, 1.0000, 0.9925, 1.0000, 0.9975, 0.9950, 0.9925, 0.9925, 0.9950, 0.9925\}.
	\item QP: \{0.9825, 0.9900, 0.9925, 0.9925, 0.9825, 0.9950, 0.9925, 0.9900, 0.9875, 0.9825\}.
	\item PG: \{0.9875, 0.9725, 0.9825, 0.9825, 0.9975, 0.9850, 0.9850, 0.9825, 0.9775, 0.9900\}. 
\end{itemize}

The averaged confusion matrices are shown in Table \ref{tb2}, in which true positive (TP) indicates that a segment with an earthquake is correctly classified as containing an earthquake; true negative (TN) indicates that the segment without an earthquake is correctly classified as not containing an earthquake; false positive (FP) indicates that a segment without an earthquake is incorrectly classified as containing one; and false negative (FN) indicates that a segment with an earthquake is incorrectly classified as not containing one (where P denotes positive, N denotes negative, T denotes true, and F denotes false).

Table \ref{tb3} shows three classification metrics: accuracy, precision ($\frac{TP}{TP+FP}$), and recall ($\frac{TP}{TP+FN}$). The optimal value in each row for the three types of periodograms is highlighted in bold. Additionally, we present the time required for estimation and training (per epoch) to compare the computational complexity of the EP and QP.                 
\begin{table}[!h]
	\centering
	\caption{The averaged confusion matrices of the classification.}
	\label{tb2}
	\subtable[EP]{
		\begin{tabular}{ccc}
			\toprule[1.2pt]
			&  P   & N  \\\hline
			T  & 199.3   & 198.6  \\
			F  &  0.7   & 1.4  \\\toprule[1.2pt]
		\end{tabular}
		
	}
	\qquad
	\subtable[QP]{
		\begin{tabular}{ccc}
			\toprule[1.2pt]
			&  P   & N  \\\hline
			T& 196.2 & 199.3  \\
			F  &  3.8   & 0.7 \\\toprule[1.2pt]
			
		\end{tabular}
	}
	\qquad
	\subtable[PG]{
		\begin{tabular}{ccc}
			
			\toprule[1.2pt]
			&  P   & N  \\\hline
			T  & 196.2   & 197.5  \\
			F  & 3.8   & 2.5  \\\toprule[1.2pt]
			
		\end{tabular}
	}
	
\end{table}

\begin{table}[ht]
	\begin{center}
		\caption{The classification results.}
		\label{tb3}
		\vskip 0.3cm
		\begin{tabular}{ccccccc}
			\toprule[1.2pt]
			Metrics   &              &     EP          &  QP         & PG         \\\hline			
			Accuracy  &  Averaged    &  {\bf 0.9948}          & 0.9880           & 0.9858           \\
			&  [min, max]  &  [0.9925, 1.0000] & [0.9725, 0.9950] & [0.9750, 0.9925] \\\hline
			Precision &  Averaged    &  {\bf 0.9965}          & 0.9840           & 0.9843           \\
			&  [min, max]  &  [0.9896, 1.0000] & [0.9559, 0.9952] & [0.9609, 0.9902] \\\hline
			Recall    &  Averaged    &  {\bf0.9931}           & 0.9921           & 0.9877           \\
			&  [min, max]  &  [0.9858, 1.0000] & [0.9794, 1.0000] & [0.9653, 1.0000] \\\hline
			Time (s)  &  Estimation  &  12.6684               & 19.4528          &   -- \\
			&  Training    &  0.3739    & 0.3731    & --  \\\toprule[1.2pt]
		\end{tabular}
	\end{center}
\end{table}

\begin{figure}[ht]
	\centering
	
	\includegraphics[width=0.8\textwidth]{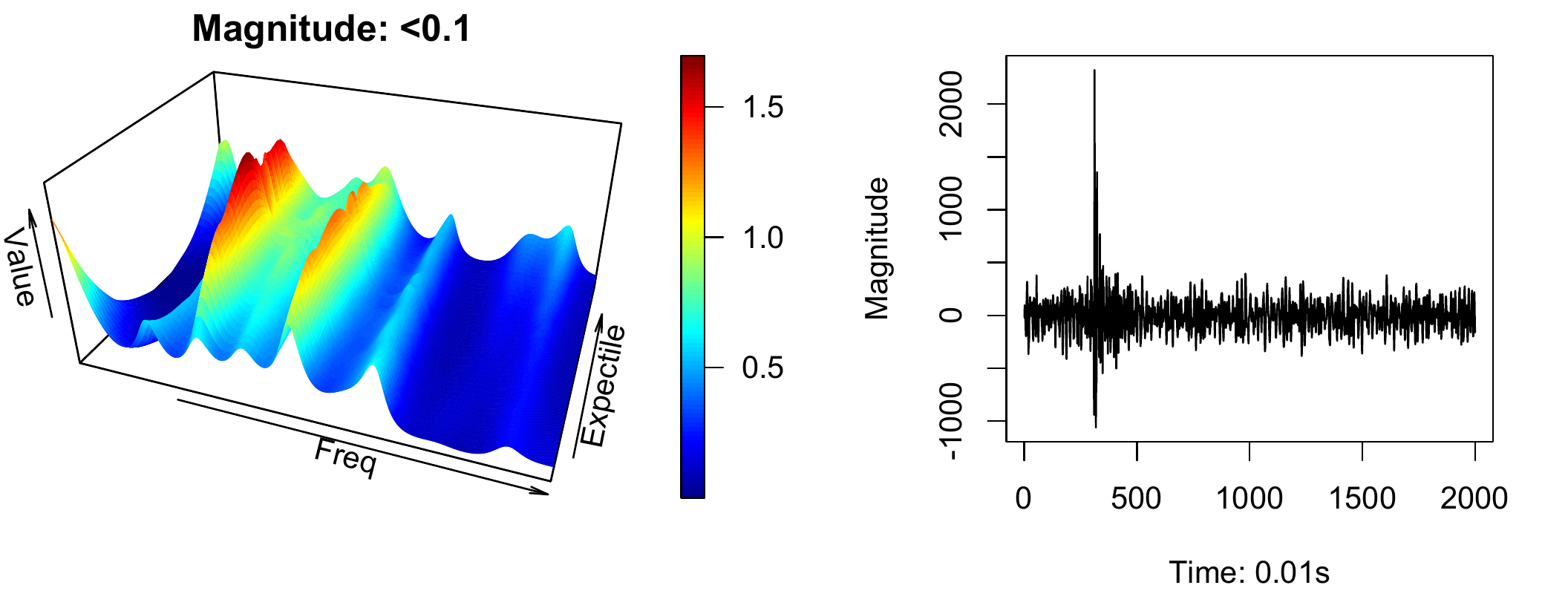}
	\caption{A misclassification case (FN). ``E-PER" denotes the EP.}
	\label{fig-mis}
\end{figure}

From the results, we can see that:
\begin{itemize}
	\item  The classification based on the EP has higher testing accuracy, precision, and recall rate than both the QP and PG. This indicates that the EPs are suitable features for time series classification. 
	\item  A false-negative example using the EPs is shown in Figure \ref{fig-mis}. Since the magnitude is too small ($<0.1$), the power at low frequencies is not as large as the power at high frequencies, which causes the misclassification.
	\item The EP incurs a lower estimation complexity than the QP. However, its computational cost is higher than that of the PG, as the dimension is multiplied by the number of expectile levels. 
\end{itemize}

\section{Conclusion}\label{CO}
We have proposed the EP as a counterpart to the PG and investigated its potential as a non-parametric tool for time series analysis. We also established the asymptotic theory and investigated the relationship between the EP and the expectile spectrum.  We conducted real-world examples and simulation studies to highlight its proficiency in detecting hidden periodicities within time series. In the earthquake data classification task, we leveraged the inherent two-dimensional characteristics of the EP using a deep learning model, which is a powerful technique in image classification. 

The EP provides richer information than the PG by examining the serial dependence at different expectile levels, while exhibiting higher computational efficiency than the QP. In principle, the conditional expectile function is monotone in the expectile level $\alpha$. However, in small-sample estimation, minor crossings may occur, especially at extreme expectiles (e.g., for $\alpha$ close to 0 or 1). In our simulations and applications, such crossings were not observed. The reason is  that the smoother nature of the expectile loss ensures greater stability than the non-differentiable check loss used in quantile regression. \cite{newey1987asymmetric,efron1991regression,waltrup2015expectile} discussed this issue and noted that the expectile serves as a smoothed alternative to the quantile, with crossing expectiles occurring less frequently.  If crossings do occur, they can be corrected by monotone rearrangement \citep{chernozhukov2010quantile}. This post-processing method can enforce monotonicity with acceptable cost.

Despite its computational efficiency relative to the QP, the EP demands higher computational resources than the PG. This computational burden arises from the fact that the dimensionality is multiplied by the number of expectiles used. One solution to reduce this computational cost is to use fewer expectiles. In this paper, we selected a large number of expectiles uniformly across $(0,1)$. Researchers may choose to focus exclusively on a subset of expectiles with sufficient discriminative power (e.g., high or low expectiles). Furthermore, we utilized multi-thread parallelization to speed up computation using the packages \texttt{foreach, doParallel} in \texttt{R}.

In closing, further developments are needed. First, it is worthwhile to derive the joint asymptotic distribution of the EP at multiple expectile levels to better understand the behavior of the EP for expectile-frequency analysis. Second, future work may focus on the properties of the EP at extremely high or low expectiles, where crossings may occur. A better understanding of its asymptotic and numerical behavior in this region could improve stability for tail inference. Such extensions would be particularly valuable for modeling rare or extreme events, including expectile-based Value-at-Risk and Expected Shortfall in finance, or detecting abnormal EEG activity.

\section*{Data Availability Statement}
 The data and code for estimating the EP and reproducing the results in Section \ref{NR} and Section \ref{EC} is accessible at: \url{https://github.com/tianbochen1/Expectile-Periodogram}. 
The data used in this paper have already been provided in the text.

\section*{Funding}
This work was supported by the National Natural Science Foundation of China under Grant No.12301326 and No.12201218; Anhui Provincial Natural Science Foundation under Grant No.2308085QA05; Natural Science Research Project of
Anhui Province under Grant No.2023AH050099 and No.2023AH033436.

\section*{Disclosure statement}\label{disclosure-statement}
The authors report there are no competing interests to declare.

\bibliographystyle{apalike}
\bibliography{reff}

\section*{Appendix A}
\subsection*{A.1 Lemma and Proof}
 Consider a more general ER problem. Given a time series $\{y_t : t = 1, \ldots, n\}$  with cumulative distribution function $F(\cdot)$ and finite second moments, let $\{\mathbf{x}_{jnt}\} $ be a sequence of deterministic vectors (the first element in the vector is set to be $1$) in $\mathbb{R}^p$ for each $j = 1,...,q$.
Let the following assumptions be satisfied:
\begin{itemize}
	\item[A1.] The cumulative distribution function $F(\cdot)$ is Lipschitz continuous, i.e., there exists a constant $K>0$ such that $|F(y)-F(y^{\prime})|\leq K|y-y^{\prime}|$ for all $y,y^{\prime}\in\mathbb{R}$.
	
	\item[A2.] The regressor sequence $\{\mathbf{x}_{jnt}\}$ is bounded (in the $\ell_{2}$-norm).

	\item[A3.] The process $\{y_{t}\}$ has the strong mixing property with mixing numbers $a_{\tau}$ ($\tau=1,2,\ldots$) satisfying $a_{\tau}\to 0$ as $\tau\to\infty$.

	\item[A4.] There exists a positive definite matrix \(\mathbf{D}_j\) for all $j$ such that \(\mathbf{D}_{jn}\to\mathbf{D}_j\) as \(n\to\infty\), where 
	$$
	\mathbf{D}_{jn}:=n^{-1}\sum_{t=1}^{n}\mathbf{x}_{jnt}\mathbf{x}_{jnt}^{\top}.
	$$

	\item[A5.] There exists a positive definite matrix \(\mathbf{V}_{jk}(\alpha)\) for all $j,k$  such that \(\mathbf{V}_{jkn}(\alpha)\to\mathbf{V}_{jk}(\alpha)\) as \(n\to\infty\), where 
	\begin{equation}\label{vnj}
		\mathbf{V}_{jkn}(\alpha):=n^{-1}\sum_{t=1}^{n}\sum_{s=1}^{n}\text{Cov}\{\dot{\rho}_{\alpha}(y_{t}-\mu(\alpha)),\dot{\rho}_{\alpha}(y_{s}-\mu(\alpha))\}\,\mathbf{x}_{jnt}\mathbf{x}_{kns}^{\top},\text{ }\text{ }j,k=1,…,q.
	\end{equation}

	\item[A6.] A central limit theorem is valid for \(\left[\boldsymbol{\zeta}_{1n}(\alpha),\ldots,\boldsymbol{\zeta}_{qn}(\alpha)\right]^{\top}\), i.e., \(\left[\boldsymbol{\zeta}_{1n}(\alpha),\ldots,\boldsymbol{\zeta}_{qn}(\alpha)\right]^{\top}\xrightarrow{D}N(\mathbf{0},[\mathbf{V}_{jk}(\alpha)]_{j,k=1}^q)\) as \(n\to\infty\), where $\boldsymbol{\zeta}_{jn}(\alpha)$ is defined in (\ref{zta}).\footnote{Condition A6 is a relaxation condition, see \textbf{Appendix A.4} for details.}
	
\end{itemize}

\noindent
\textbf{Lemma 1} (Expectile Regression). 
{\it If (A1) - (A6) are satisfied, we have:
$$
\sqrt{n}\left[\hat{\boldsymbol{\beta}}_{1n}(\alpha)-\boldsymbol{\beta}_{0}(\alpha),\ldots,\hat{\boldsymbol{\beta}}_{qn}(\alpha)-\boldsymbol{\beta}_{0}(\alpha)\right]^{\top} \xrightarrow{D}N(\mathbf{0},[\mathbf{\Lambda}^{-1}_{j}(\alpha)\mathbf{V}_{jk}(\alpha)\mathbf{\Lambda}_{k}^{-1}(\alpha)]_{j,k=1}^q),
$$
where  
$$
\hat{\bm{\beta}}_{jn}(\alpha) := \arg\min_{\bm{\beta}_j \in \mathbb{R}^p} \sum_{t=1}^{n} \rho_\alpha(y_t - \mathbf{x}_{jnt}^{\top} \bm{\beta}_j),
$$
$\boldsymbol{\beta}_{0}(\alpha):=[\mu(\alpha),0,\ldots,0]^{\top}\in\mathbb{R}^{p}$, and 
$$
\mathbf{\Lambda}_j(\alpha):=\lim_{n\to\infty}\mathbf{\Lambda}_{jn}(\alpha):=\lim_{n\to\infty} \eta^{-1}(\alpha)\, \mathbf{D}_{jn}   =\eta^{-1}(\alpha)\, \mathbf{D}_j. 
$$}

\noindent
{\bf Proof}. The proof follows the general strategy which was used to establish a similar result for quantile regression \citep{knight1998limiting,koenker2001quantile,li2008laplace,li2012quantile}. First, consider the
case of $q = 1$ for which we drop the subscript $j$ in the notation for simplicity.

Let \( U_t := y_t - \mu(\alpha) = y_t - \mathbf{x}_{nt}^{\top} \bm{\beta}_0(\alpha) \) and \( c_{nt}(\bm{\delta}) := n^{-1/2} \mathbf{x}_{nt}^{\top} \bm{\delta} \) with \( \bm{\delta} \in \mathbb{R}^p \), consider the random function
\[
Z_n(\bm{\delta}) := \sum_{t=1}^n \{ \rho_\alpha(U_t - c_{nt}(\bm{\delta})) - \rho_\alpha(U_t) \}.
\]
Note that by reparameterizing \( \bm{\delta} := \sqrt{n} (\bm{\beta} - \bm{\beta}_0(\alpha)) \) as a function of \( \bm{\beta} \in \mathbb{R}^p \), one can write
$$
Z_n(\bm{\delta}) = \sum_{t=1}^n \{ \rho_\alpha(y_t - \mathbf{x}_{nt}^{\top} \bm{\beta}) - \rho_\alpha(U_t) \}.
$$
Therefore, the minimizer of \( Z_n(\bm{\delta}) \) over \( \bm{\delta} \in \mathbb{R}^p \) takes the form
\[
\hat{\bm{\delta}}_n := \arg \min_{\bm{\delta} \in \mathbb{R}^p} Z_n(\bm{\delta}) = \sqrt{n} (\hat{\bm{\beta}}_n(\alpha) - \bm{\beta}_0(\alpha)).
\]
We aim to prove
\begin{equation}\label{delta}
	\hat{\bm{\delta}}_n = \bm{\Lambda}_n^{-1} \bm{\zeta}_n + o_P(1).
\end{equation}
Here we drop the argument \( \alpha \) from \( \bm{\Lambda}_n \) and \( \bm{\zeta}_n \) for simplicity of notation, as \( \alpha \) is a fixed number. In order to arrive at (\ref{delta}), we first would like to prove that
\begin{equation}\label{zn}
	Z_n(\bm{\delta}) = \tilde{Z}_n(\bm{\delta}) + o_P(1),
\end{equation}
for any fixed \( \bm{\delta} \), where \( \tilde{Z}_n(\bm{\delta}) := -\bm{\delta}^{\top} \bm{\zeta}_n + (1/2) \bm{\delta}^{\top} \bm{\Lambda}_n \bm{\delta} \).  If this is true, then, due to the convexity of \( Z_n(\bm{\delta}) \) and \( \tilde{Z}_n(\bm{\delta}) \) as functions of \( \bm{\delta} \), and following the last part of the proof of Theorem 1 in \cite{pollard1991asymptotics}, we can show that \( \hat{\bm{\delta}}_n \), as the minimizer of \( Z_n(\bm{\delta}) \), is \( o_P(1) \) away from the minimizer of \( \tilde{Z}_n(\bm{\delta}) \), which equals \( \bm{\Lambda}_n^{-1} \bm{\zeta}_n \). This completes the proof of (\ref{delta}).

To prove (\ref{zn}), we observe that for any \( u,c \in \mathbb{R} \), we can write
\[
\rho_{\alpha}(u-c) - \rho_{\alpha}(u) = 
\begin{cases} 
	-2\alpha uc + \alpha c^2 & \text{if } u \geq 0 \text{ and } u-c \geq 0, \\
	(1-2\alpha)u^2 - 2(1-\alpha)uc + (1-\alpha)c^2 & \text{if } u \geq 0 \text{ and } u-c < 0, \\
	-(1-2\alpha)u^2 - 2\alpha uc + \alpha c^2 & \text{if } u < 0 \text{ and } u-c \geq 0, \\
	-2(1-\alpha)uc + (1-\alpha)c^2 & \text{if } u < 0 \text{ and } u-c < 0.
\end{cases}
\]
\[
\hspace{3.68cm}= 
\begin{cases} 
	-\dot{\rho}_{\alpha}(u)c + \alpha c^2 & \text{if } u \geq 0 \text{ and } u-c \geq 0, \\
	-\dot{\rho}_{\alpha}(u)c + (1-2\alpha)(u^2 - 2uc) + (1-\alpha)c^2 & \text{if } u \geq 0 \text{ and } u-c < 0, \\
	-\dot{\rho}_{\alpha}(u)c - (1-2\alpha)(u^2 - 2uc) + \alpha c^2 & \text{if } u < 0 \text{ and } u-c \geq 0, \\
	-\dot{\rho}_{\alpha}(u)c + (1-\alpha)c^2 & \text{if } u < 0 \text{ and } u-c < 0.
\end{cases}
\]
Substituting \( U_t \) and \( c_{nt} := c_{nt}(\bm{\delta}) \) for \( u \) and \( c \) yields
\begin{equation}\label{zn2}
	Z_n(\bm{\delta}) = -\bm{\delta}^{\top} \boldsymbol{\zeta}_n + \sum_{t=1}^n R_{nt},
\end{equation}
where
\begin{equation}\label{rnt}
	R_{nt} := 
	\begin{cases}
		\alpha c_{nt}^2 & \text{if } U_t \geq 0 \text{ and } U_t - c_{nt} \geq 0, \\
		(1-2\alpha)(U_t^2 - 2U_t c_{nt}) + (1-\alpha)c_{nt}^2 & \text{if } U_t \geq 0 \text{ and } U_t - c_{nt} < 0, \\
		-(1-2\alpha)(U_t^2 - 2U_t c_{nt}) + \alpha c_{nt}^2 & \text{if } U_t < 0 \text{ and } U_t - c_{nt} \geq 0, \\
		(1-\alpha)c_{nt}^2 & \text{if } U_t < 0 \text{ and } U_t - c_{nt} < 0.
	\end{cases}
\end{equation}
Here, we drop the argument \(\bm{\delta}\) from \( R_{nt} \) and \( c_{nt} \) for simplicity of notation as \(\bm{\delta}\) is fixed in the remainder of the proof.  (\ref{zn2}) is valid if
\begin{equation}\label{rnt2}
	\sum_{t=1}^n R_{nt} = \frac{1}{2}\bm{\delta}^{\top} \boldsymbol{\Lambda}_n \bm{\delta} + o_P(1).
\end{equation}
This expression can be established by proving that
\begin{equation}\label{ernt}
	\operatorname{E}\left\{\sum_{t=1}^n R_{nt}\right\} = \frac{1}{2}\bm{\delta}^{\top} \boldsymbol{\Lambda}_n \bm{\delta} + o(1),
\end{equation}
and
\begin{equation}\label{vrnt}
	\text{Var}\left\{\sum_{t=1}^{n}R_{nt}\right\}=o(1).
\end{equation}
Note that the leading term in (\ref{rnt2}) is \(O(1)\) for fixed \(\boldsymbol{\delta} \neq \mathbf{0}\) under assumption (A3).

To prove (\ref{ernt}), we observe that
\begin{equation}\label{ernt2}
	\begin{aligned}
		\mathrm{E}\{R_{nt}\} &= \alpha c_{nt}^{2}\mathrm{E}\{I(U_{t} \geq \max(0,c_{nt}))\} \\
		&\quad + \mathrm{E}\left\{[(1-2\alpha)(U_{t}^{2}-2U_{t}c_{nt})+(1-\alpha)c_{nt}^{2}]I(0 \leq U_{t} < c_{nt})\right\} \\
		&\quad + \mathrm{E}\left\{[-(1-2\alpha)(U_{t}^{2}-2U_{t}c_{nt})+\alpha c_{nt}^{2}]I(c_{nt} \leq U_{t} < 0)\right\} \\
		&\quad + (1-\alpha)c_{nt}^{2}\mathrm{E}\{I(U_{t} < \min(0,c_{nt}))\}. 
	\end{aligned}
\end{equation}
A necessary condition for the second term in (\ref{ernt2}) to be nonzero is \(c_{nt} \geq 0\). When this is the case, the absolute value of this term can be bounded above by
\[
\left\{|1-2\alpha|(c_{nt}^{2}+2c_{nt}^{2})+(1-\alpha)c_{nt}^{2}\right\}\Pr\{0 \leq U_{t} < c_{nt}\}.
\]
Under assumption (A2), \(c_{nt} = o(1)\). These, together with assumption (A1), imply that
\[
\Pr\{0 \leq U_{t} < c_{nt}\} = F(\mu(\alpha) + c_{nt}) - F(\mu(\alpha)) = o(1).
\]
Therefore, the second term in (\ref{ernt2}) can be written as \(o(c_{nt}^{2})\). This expression is also valid for the third term in (\ref{ernt2}) which can be nonzero when \(c_{nt} < 0\). Similarly, when \(c_{nt} \geq 0\), the first term in (\ref{ernt2}) can be written as
\[
\begin{aligned}
	\alpha c_{nt}^{2}\Pr\{U_{t} \geq c_{nt}\} &= \alpha c_{nt}^{2}\Pr\{U_{t} \geq 0\} - \alpha c_{nt}^{2}\Pr\{0 \leq U_{t} < c_{nt}\} \\
	&= \alpha c_{nt}^{2}\Pr\{U_{t} \geq 0\} + o(c_{nt}^{2}) \\
	&= \alpha c_{nt}^{2}\{1 - F(\mu(\alpha))\} + o(c_{nt}^{2}).
\end{aligned}
\]
And, when \(c_{nt}<0\), the first term in (\ref{ernt2}) becomes simply \(\alpha c_{nt}^{2}\Pr\{U_{t}\geq 0\}=\alpha c_{nt}^{2}\{1-F(\mu(\alpha))\}\). By a similar argument, we can write the fourth term in (\ref{ernt2}) as
\[
(1-\alpha)c_{nt}^{2}F(\mu(\alpha))+o(c_{nt}^{2})
\]
when \(c_{nt}<0\) and as \((1-\alpha)c_{nt}^{2}F(\mu(\alpha))\) when \(c_{nt}\geq 0\). Under assumption (A2), \(o(c_{nt}^{2})=o(n^{-1})\). Combining these results yields
\[
\mathrm{E}\{R_{nt}\}=\{\alpha(1-F(\mu(\alpha)))+(1-\alpha)F(\mu(\alpha))\}c_{nt}^{2}+o(n^{-1}).
\]
Substituting \(c_{nt}=n^{-1/2}\mathbf{x}_{nt}^{\top}\boldsymbol{\delta}\) in this expression proves (\ref{ernt}).

Furthermore, for any \(0<m<n\), we split the variance in (\ref{vrnt}) into two terms:
\begin{equation}\label{vrnt2}
	\mathrm{Var}\left\{\sum_{t=1}^{n}R_{nt}\right\}=\left(\sum_{(t,s)\in D_{m}}+\sum_{(t,s)\in D_{m}^{\prime}}\right)\mathrm{Cov}(R_{nt},R_{ns}), 
\end{equation}
where \(D_{m}:=\{(t,s):|t-s|\leq m,1\leq t,s\leq n\}\) and \(D_{m}^{\prime}:=\{(t,s):|t-s|>m,1\leq t,s\leq n\}\). The expression (\ref{vrnt}) is obtained if both terms in (\ref{vrnt2}) can be shown to take the form \(o(1)\).

Consider the first term in (\ref{vrnt2}). Under assumption (A2), for any fixed \(\boldsymbol{\delta}\), there exists a constant \(c_{0}:=c_{0}(\boldsymbol{\delta})>0\) such that \(c_{nt}^{2}\leq c_{0}\,n^{-1}\) for all \(t=1,\ldots,n\). Therefore, in the first and fourth cases of (\ref{rnt}), we have, respectively, \(|R_{nt}|=\alpha c_{nt}^{2}\leq\alpha c_{0}\,n^{-1}\) and \(|R_{nt}|=(1-\alpha)c_{nt}^{2}\leq(1-\alpha)c_{0}\,n^{-1}\). If the second case of (\ref{rnt}) is true, we would have \(c_{nt}>0\) and \(0\leq U_{t}<c_{nt}\), which implies
\[
\begin{aligned}
	|R_{nt}| &\leq |1-2\alpha|(c_{nt}^{2}+2c_{nt}^{2})+(1-\alpha)c_{nt}^{2} \\
	&= (3|1-2\alpha|+(1-\alpha))c_{nt}^{2} \\
	&\leq (3|1-2\alpha|+(1-\alpha))c_{0}\,n^{-1}.
\end{aligned}
\]
Similarly, if the third case of (\ref{rnt}) is true, we would have \(c_{nt}<0\) and \(|U_{t}|\leq|c_{nt}|\), which implies
\[
\begin{aligned}
	|R_{nt}| &\leq |1-2\alpha|(c_{nt}^{2}+2c_{nt}^{2})+\alpha c_{nt}^{2} \\
	&= (3|1-2\alpha|+\alpha)c_{nt}^{2} \\
	&\leq (3|1-2\alpha|+\alpha)c_{0}\,n^{-1}.
\end{aligned}
\]
Combining these results yields
\begin{equation}\label{absrnt}
	|R_{nt}|\leq\kappa\,n^{-1}
\end{equation}
for some constant \(\kappa := \kappa(\boldsymbol{\delta}, \alpha) > 0\). Owing to (\ref{absrnt}), we obtain
\[
\text{Var}\{R_{nt}\} \leq \mathrm{E}\{R_{nt}^2\} \leq \kappa^2 n^{-2}.
\]
This implies that
\begin{equation}\label{covrnt}
	| \text{Cov}(R_{nt}, R_{ns}) | \leq \sqrt{\text{Var}\{R_{nt}\} \text{Var}\{R_{ns}\}} \leq \kappa^2 n^{-2}.
\end{equation}
Observe that the number of elements in \(D_m\) is bounded above by \((2m+1)n\). Combining this result with (\ref{covrnt}) yields
\[
\left| \sum_{(t,s) \in D_m} \text{Cov}(R_{nt}, R_{ns}) \right| \leq \sum_{(t,s) \in D_m} | \text{Cov}(R_{nt}, R_{ns}) | \leq (2m+1) \kappa^2 n^{-1}.
\]
If \(m\) is chosen to satisfy \(m \to \infty\) and \(m/n \to 0\) as \(n \to \infty\), then we can write
\begin{equation}\label{sumcovrnt}
	\sum_{(t,s) \in D_m} \text{Cov}(R_{nt}, R_{ns}) = o(1).
\end{equation}
This completes the evaluation for the first term in (\ref{vrnt2}).

To evaluate the second term in (\ref{vrnt2}), we observe that \(\{U_t\}\) is a stationary process with the same strong mixing property as \(\{y_t\}\) under assumption (A4). Because \(R_{nt}\) is a function of \(U_t\) and \(R_{ns}\) is a function of \(U_s\), both bounded by \(\kappa n^{-1}\) according to (\ref{absrnt}), citing the mixing inequality \citep{billingsley2013convergence} yields
\[
| \text{Cov}(R_{nt}, R_{ns}) | \leq 4 \kappa^2 n^{-2} a_{|t-s|}.
\]
Therefore, we have
\[
\begin{aligned}
	\left| \sum_{(t,s) \in D_m'} \text{Cov}(R_{nt}, R_{ns}) \right| &\leq \sum_{m < |\tau| < n, 1 \leq s \leq n} | \text{Cov}(R_{n,s+\tau}, R_{ns}) | \\
	&\leq \sum_{m < |\tau| < n} n \times 4 \kappa^2 n^{-2} a_{|\tau|} \\
	&\leq 8 \kappa^2 n^{-1} \sum_{\tau=m+1}^{n} a_{\tau} \\
	&= 8 \kappa^2 \left\{ n^{-1} \sum_{\tau=1}^{n} a_{\tau} - (m/n) m^{-1} \sum_{\tau=1}^{m} a_{\tau} \right\}.
\end{aligned}
\]
Because \(a_{\tau}\to 0\) as \(\tau\to\infty\), it follows from the Stolz-Ces\`{a}ro theorem that \(n^{-1}\sum_{\tau=1}^{n}a_{\tau}\to 0\) as \(n\to\infty\) and \(m^{-1}\sum_{\tau=1}^{m}a_{\tau}\to 0\) as \(m\to\infty\). This implies that
\begin{equation}\label{sumcovrnt2}
	\sum_{(t,s)\in D^{\prime}_{m}}\text{Cov}(R_{nt},R_{ns})=o(1).
\end{equation}
Collecting (\ref{sumcovrnt}) and (\ref{sumcovrnt2}) proves (\ref{vrnt}). Then, (\ref{delta}) is proved, and thus, 
\begin{equation}\label{ba}
	\sqrt{n}(\hat{\boldsymbol{\beta}}_{n}(\alpha)-\boldsymbol{\beta}_{0}(\alpha)) = \boldsymbol{\Lambda}_{n}^{-1}(\alpha)\boldsymbol{\zeta}_{n}(\alpha)+o_{P}(1). 
\end{equation}

Next, we would like to establish the asymptotic normality of the ER solution.
Because \(\mu(\alpha)\) satisfies (\ref{ne}), it follows that
\[
\mathrm{E}\{\boldsymbol{\zeta}_{n}(\alpha)\}=n^{-1/2}\sum_{t=1}^{n}\mathrm{E}\{ \dot{\rho}_{\alpha}(y_{t}-\mu(\alpha))\}\,\mathbf{x}_{nt}=\mathbf{0}.
\]
Moreover, we have
$$
\mathbf{V}_{n}(\alpha):=\text{Cov}\{\boldsymbol{\zeta}_{n}(\alpha)\}=n^{-1}\sum_{t=1}^{n}\sum_{s=1}^{n}\text{Cov}\{\dot{\rho}_{\alpha}(y_{t}-\mu(\alpha)),\dot{\rho}_{\alpha}(y_{s}-\mu(\alpha))\}\,\mathbf{x}_{nt}\mathbf{x}_{ns}^{\top}.
$$
Then, collecting assumptions (A4)-(A6) and (\ref{ba}), we obtain
\[
\mathbf{\Lambda}_{n}(\alpha)^{-1}\boldsymbol{\zeta}_{n}(\alpha)\xrightarrow{D}N(\mathbf{0},\mathbf{\Lambda}^{-1}(\alpha)\mathbf{V}(\alpha)\mathbf{\Lambda}^{-1}(\alpha)). 
\]

For the general case of $q>1$, we re-add the subscript $j$ in the notation, and define 
$$
Z_{jn}(\bm{\delta}_j) = \sum_{t=1}^n \{ \rho_\alpha(y_t - \mathbf{x}_{jnt}^{\top} \bm{\beta}_j) - \rho_\alpha(U_t) \},
$$ 
and 
$$
Z_{n}^*(\bm{\delta}^*) = \sum_{j=1}^q Z_{jn}(\bm{\delta}_j),
$$
where 
$
\bm{\delta}^* := \left[\bm{\delta}_1,\ldots,\bm{\delta}_q\right]^{\top}, 
$
and $\bm{\delta}_j:= \sqrt{n}(\bm{\beta}_j - \bm{\beta}_0(\alpha))$. By a similar argument, it can be shown that the minimizer of $	Z_{n}^*(\bm{\delta}^*)$, which can be expressed as $\hat{\bm{\delta}}^*_n := \left[\hat{\bm{\delta}}_{1n},\ldots,\hat{\bm{\delta}}_{qn} \right]^{\top}$, is $o_P(1)$ away from $\tilde{\bm{\delta}}_{n}^*:= \left[\tilde{\bm{\delta}}_{1n},\ldots,\tilde{\bm{\delta}}_{n}\right]^{\top}:=\left[\mathbf{\Lambda}^{-1}_{1n} \boldsymbol{\zeta}_{1n} ,\ldots,\mathbf{\Lambda}^{-1}_{qn} \boldsymbol{\zeta}_{qn} \right]^{\top}$, where $\hat{\bm{\delta}}_{jn}$ is the minimizer of $Z_{jn}(\bm{\delta}_j)$. Therefore, $\hat{\bm{\delta}}^*_n$ has the same asymptotic distribution as $\tilde{\bm{\delta}}_{n}^*$. Observe that $\mathrm{E}\{\boldsymbol{\zeta}_{jn}\} = \bm{0}$ and $\text{Cov}(\boldsymbol{\zeta}_{jn},\boldsymbol{\zeta}_{kn}) = \mathbf{V}_{jkn}$. Therefore $\hat{\bm{\delta}}_{n}^*\xrightarrow{D}N(\mathbf{0},[\mathbf{\Lambda}^{-1}_{j}(\alpha)\mathbf{V}_{jk}(\alpha)\mathbf{\Lambda}_{k}^{-1}(\alpha)]_{j,k=1}^q)$, and  Lemma 1 is proved.

\subsection*{A.2 Proof of Theorem 1}\label{pt1}
For fixed \(q>1\) and \(0<\lambda_{1}<\cdots<\lambda_{q}<\pi\), let \(\omega_{\nu_{1}},\ldots,\omega_{\nu_{q}}\) be Fourier frequencies satisfying \(\omega_{\nu_{j}}\to\lambda_{j}\) as \(n\to\infty\) for \(j=1,\ldots,q\). The trigonometric regressor $\mathbf{x}_{jnt}  :=[ 1 , \cos(\omega_{\nu_j} t) , \sin(\omega_{\nu_j} t) ]^{\top}$ is is bounded, so (A2) is satisfied. It is easy to verify that when \(n\to\infty\),
$$
\mathbf{D}_{jn}=n^{-1}\sum_{t=1}^{n}\mathbf{x}_{jnt}\mathbf{x}_{jnt}^{\top}\to\mathbf{D}_j=\operatorname{diag}\{1,1/2,1/2\},
$$
so (A4) is satisfied. It follows from (\ref{vnj}) that, when $j =  k$,
\[
\mathbf{V}_{jjn}(\alpha)=\sum_{|\tau|<n}\gamma(\tau,\alpha)\left\{n^{-1}\sum_{t=\max(1,1+\tau)}^{\min(n,n+\tau)}\mathbf{x}_{jnt}\mathbf{x}_{jn,t-\tau}^{\top}\right\}.
\]
For fixed \(\tau\), we have
\[
n^{-1}\sum_{t=\max(1,1+\tau)}^{\min(n,n+\tau)}\mathbf{x}_{jnt}\mathbf{x}_{jn,t-\tau}^{\top}\to\operatorname{diag}\{1,(1/2)\mathbf{S}(\lambda_j)\},
\]
where
\[
\mathbf{S}(\lambda_j) := \begin{bmatrix}
	\cos(\lambda_j\tau) & -\sin(\lambda_j\tau) \\ 
	\sin(\lambda_j\tau) & \cos(\lambda_j\tau)
\end{bmatrix}.
\]
Then,
\begin{equation}\label{valpha}
	\begin{aligned}
		\mathbf{V}_{jj}(\alpha) &:= \lim_{n\to\infty}\mathbf{V}_{jjn}(\alpha)  \\
		&= \sum_{\tau=-\infty}^{\infty}\gamma(\tau,\alpha)\operatorname{diag}\{1,(1/2)\mathbf{S}(\lambda_j)\} \\
		&= \operatorname{diag}\{h(0,\alpha),(1/2)h(\lambda_j,\alpha),(1/2)h(\lambda_j,\alpha)\}.
	\end{aligned}
\end{equation}
When $j \neq k$,  
\begin{equation}\label{valpha2}
	\mathbf{V}_{jk}(\alpha):= \lim_{n\to\infty}\mathbf{V}_{jkn}(\alpha) = \mathbf{0}.
\end{equation}
Based on (\ref{valpha}) and (\ref{valpha2}), we have
$$
\mathbf{V}_{jk}(\alpha) = \delta_{j-k} \operatorname{diag}\{h(0,\alpha),(1/2)h(\lambda_j,\alpha),(1/2)h(\lambda_j,\alpha)\},
$$
where $\delta_s$ denotes the Kronecker delta function, and
$$
[\mathbf{\Lambda}^{-1}_{j}(\alpha)\mathbf{V}_{jk}(\alpha)\mathbf{\Lambda}_{k}^{-1}(\alpha)]_{j,k=1}^q = \operatorname{diag}\{ g(0,\alpha),2g(\lambda_1,\alpha),2g(\lambda_1,\alpha),\ldots,g(0,\alpha),2g(\lambda_q,\alpha),2g(\lambda_q,\alpha)\}. 
$$
Hence, (A5) is satisfied. Moreover, (C1), (C2) and (C3) repeat (A1), (A3) and (A6), respectively. Finally, all assumptions (A1)-(A6) are satisfied. Therefore, by Lemma 1, 
	\[
	\hspace*{-0.5cm}
	\sqrt{n}
	\begin{bmatrix}
		\hat{\boldsymbol{\beta}}_{n}(\omega_{\nu_1},\alpha)-\boldsymbol{\beta}_{0}(\alpha)\\
		\vdots\\
		\hat{\boldsymbol{\beta}}_{n}(\omega_{\nu_q},\alpha)-\boldsymbol{\beta}_{0}(\alpha)
	\end{bmatrix}
	\xrightarrow{D}
	N\!\left(\mathbf{0},
	\operatorname{diag}\{ g(0,\alpha),2g(\lambda_1,\alpha),2g(\lambda_1,\alpha),\ldots,g(0,\alpha),2g(\lambda_q,\alpha),2g(\lambda_q,\alpha)\}
	\right)
	\]

Further, because \(\hat{\boldsymbol{\beta}}_{n}(\omega_{\nu_j},\alpha) = [\hat{\beta}_{1}(\omega_{\nu_j},\alpha),\hat{\beta}_{2}(\omega_{\nu_j},\alpha),\hat{\beta}_{3}(\omega_{\nu_j},\alpha)]^{\top}\), we have
\[
\sqrt{n}[\hat{\beta}_{2}(\omega_{\nu_j},\alpha),\hat{\beta}_{3}(\omega_{\nu_j},\alpha)]^{\top}/\sqrt{2g(\lambda_j,\alpha)} \xrightarrow{D} N(\mathbf{0},\mathbf{I}),
\]
and hence
\[
{\rm EP}_{n}(\omega_{\nu_j},\alpha)/g(\lambda_j,\alpha) = \frac{n}{4}\{\hat{\beta}^{2}_{2}(\omega_{\nu_j},\alpha) + \hat{\beta}^{2}_{3}(\omega_{\nu_j},\alpha)\}/g(\lambda_j,\alpha) \sim \frac{1}{2}(\xi^{2}_{1} + \xi^{2}_{2}) = \chi^{2}_{2},
\]
where \(\xi_{1}\) and \(\xi_{2}\) are i.i.d. \(N(0,1)\) random variables. This implies that  ${\rm EP}_{n}(\omega_{\nu_j},\alpha)$ are asymptotically independent with asymptotic distribution   
$g(\lambda_{j},\alpha)(1/2)\chi^{2}_{2,j}$ for each $j$, and we obtain (\ref{the2}). Finally,  Theorem 1 is proved.

\subsection*{A.3 Proof of Theorem 2}

	First, by (C5)--(C6), averaging across $M_n$ nearby Fourier frequencies reduces the variance by a factor of order $1/(M_n)$ while introducing only $o(1)$ bias because the true expectile spectrum $g(\lambda,\alpha)$ is continuous in $\lambda$ under assumption (C4).  Second, a central limit theorem for quadratic forms of mixing processes \citep{brillinger2001time, dahlhaus1988empirical, davidson1994stochastic} implies that
\begin{equation}\label{q4}
\sqrt{M_n}\big(	\hat{g}(\omega_{\nu}, \alpha)  -  g(\lambda, \alpha)\big) \overset{d}{\longrightarrow} N(0, \|K^*\|_2^2 2g(\lambda,\alpha)).
\end{equation}
Since both bias and variance vanish under (C4)--(C6), we obtain the consistency in (\ref{p3}) and the asymptotic distribution in \eqref{q4}.

\subsection*{A.4 Substitution of (A6) and (C3)}\label{A7}
Let us take a closer look at assumption (A6) (as well as (C3)). Using the Wald device, one can establish the central limit theorem in assumption (A6) by showing that
\[
S_{n} := \boldsymbol{\lambda}^{\top}\boldsymbol{\zeta}_{n}(\alpha)/\sqrt{\boldsymbol{\lambda}^{\top}\mathbf{V}_{n}(\alpha)\boldsymbol{\lambda}} \xrightarrow{D} N(0,1).
\]
for any fixed \(\boldsymbol{\lambda} \neq \mathbf{0}\). Note that \(S_{n}\) can be written as the sum of a random sequence \(\{w_{nt}\xi_{t}\}\), i.e.,
\[
S_{n} = \sum_{t=1}^{n}w_{nt}\xi_{t},
\]
where \(\xi_{t} := \dot{\rho}_{\alpha}(y_{t} - \mu(\alpha))\) and \(w_{nt} := n^{-1/2}\boldsymbol{\lambda}^{\top}\mathbf{x}_{nt}/\sqrt{\boldsymbol{\lambda}^{\top}\mathbf{V}_{n}(\alpha)\boldsymbol{\lambda}}\). Note that \(\mathrm{E}\{S_{n}\} = 0\) and \(\text{Var}\{S_{n}\} = 1\). A central theorem for this type of random variables is given by \cite{peligrad1997central}, where \(\{\xi_{t}\}\) is assumed to be a possibly nonstationary random sequence with a strong mixing property. In our case, the sequence \(\{\xi_{t}\}\) is stationary under assumption (A4). Therefore, Theorem 2.2(c) of \cite{peligrad1997central} can be simplified as follows.

\noindent
{\bf Proposition 1.} (Peligrad and Utev, 1997).{\it Let \(\{\xi_{t}: t = 1, \ldots, n\}\) be a zero-mean nonzero-variance stationary sequence having the strong mixing property with mixing numbers \(a_{\tau}\) (\(\tau = 1, 2, \ldots\)). Let \(\{w_{nt}: t = 1, \ldots, n\}\) be a triangular array of real numbers. Assume that the following conditions are satisfied.
\begin{itemize}
	\item[P1.] For some \(\delta > 0\), \(\mathrm{E}\{|\xi_{1}|^{2+\delta}\} < \infty\) and \(\sum_{\tau=1}^{n}\tau^{2/\delta}a_{\tau} < \infty\).
	
	\item[P2.] \(\sup_{n}\sum_{t=1}^{n}w_{nt}^{2} < \infty\) and \(\max_{1 \leq t \leq n}|w_{nt}| \to 0\) as \(n \to \infty\).
\end{itemize}
Under these conditions, \(\sum_{t=1}^{n}w_{nt}\xi_{t} \xrightarrow{D} N(0,1)\) as \(n \to \infty\).}

\noindent
{\bf Proof}. When \(\{\xi_{t}\}\) is stationary, we have \(\inf_{t}\text{Var}\{\xi_{t}\} = \text{Var}\{\xi_{1}\} > 0\). If \(\mathrm{E}\{|\xi_{1}|^{2+\delta}\} < \infty\), the stationarity also implies that for any \(\epsilon > 0\) there exists \(c > 0\) such that
\[
\sup_{t} \mathrm{E}\{|\xi_{t}|^{2+\delta}I(|\xi_{t}| \geq c)\} = \mathrm{E}\{|\xi_{1}|^{2+\delta}I(|\xi_{1}| \geq c)\} < \epsilon.
\]
This means that the sequence \(\{|\xi_{t}|^{2+\delta}\}\) is uniformly integrable. Therefore, all conditions in Theorem 2.2(c) of \cite{peligrad1997central} on \(\{\xi_{t}\}\) are fulfilled under (P1) when \(\{\xi_{t}\}\) is stationary. Condition (P2) repeats the assumption (2.1) in \cite{peligrad1997central}.

For the special case with \(w_{nt}=n^{-1/2}\boldsymbol{\lambda}^{\top}\mathbf{x}_{nt}/\sqrt{\boldsymbol{\lambda}^{\top}\mathbf{V}_{n}(\alpha)\boldsymbol{\lambda}}\), condition (P2) is satisfied under the boundedness assumption (A2). Therefore, the following result can be obtained according to Proposition 1.

\noindent
{\bf Lemma 2} (Expectile Regression). {\it In addition to (A3), let \(\{y_{t}\}\) also satisfy the following condition:
\begin{itemize}
	\item[A8.] For some \(\delta>0\), \(\int|y|^{2+\delta}\,dF(y)<\infty\) and \(\sum_{\tau=1}^{n}\tau^{2/\delta}a_{\tau}<\infty\).
\end{itemize} 
Then, assumption (A6) holds if (A2) and (A5) are true.}

\noindent
{\bf Proof}. Under (A3), \(\xi_{t}:=\dot{\rho}_{\alpha}(y_{t}-\mu(\alpha))\) is a zero-mean nonzero-variance strong mixing sequence with mixing numbers \(a_{\tau}\) (\(\tau=1,2,\ldots\)). Under (A8),
\[
\operatorname{E}\{|\xi_{1}|^{2+\delta}\}\leq\operatorname{E}\{|y_{1}|^{2+\delta}\}=\int|y|^{2+\delta}\,dF(y)<\infty.
\]
Therefore, the assertion follows from the Wald device and Proposition 1. Then, one can substitute condition (C3) (also, A6) with (A7).

\end{document}